\documentclass[11pt]{article}
\usepackage{amsfonts}
\usepackage{amsmath,amssymb}
\usepackage{mathrsfs}
\def\un{\underline}
 \def\sc{\scriptscriptstyle \underline}

\makeatletter
\def\eqnarray{\stepcounter{equation}\let\@currentlabel=\theequation
\global\@eqnswtrue
\global\@eqcnt\z@\tabskip\@centering\let\\=\@eqncr
$$\halign to \displaywidth\bgroup\@eqnsel\hskip\@centering
  $\displaystyle\tabskip\z@{##}$&\global\@eqcnt\@ne
  \hfil$\displaystyle{\hbox{}##\hbox{}}$\hfil
  &\global\@eqcnt\tw@ $\displaystyle\tabskip\z@
  {##}$\hfil\tabskip\@centering&\llap{##}\tabskip\z@\cr}
\@addtoreset{equation}{section}
  \def\theequation{\thesection.\arabic{equation}}
\makeatother

\advance \textheight by \topskip
\def\beq{\begin{equation}}
\def\eeq{\end{equation}}
\def\beqa{\begin{eqnarray}}
\def\eeqa{\end{eqnarray}}
\def\barray{\begin{array}}
\def\earray{\end{array}}

\def\cJ{{\cal J}}
\def\parag{\hfil\break} %%%%% paragraph
\def\kikezd{\parag\underbar}

\begin{document}

\title{\textbf{Bosons, fermions  and anyons in the plane, \\
and supersymmetry}}

\author{
{\sf Peter A. Horv\'athy${}^a$}\footnote{e-mails:
horvathy@univ-tours.fr}, {\sf Mikhail S.
Plyushchay${}^b$}\footnote{mplyushc@lauca.usach.cl},
{\sf Mauricio Valenzuela${}^a$}\footnote{valenzue@lmpt.univ-tours.fr}\\
[4pt] {\small \it ${}^a$Laboratoire de Math\'ematiques et
de Physique Th\'eorique,}\\
{\small \it Unit\'e Mixte de Recherche $6083$ du CNRS,
F\'ed\'eration Denis Poisson,
Universit\'e de Tours,}\\
{\small \it Parc de Grandmont,
 F-37200 Tours, France}\\
{\small \it ${}^b$Departamento de F\'{\i}sica, Universidad de
Santiago de Chile}\\
{\small \it Casilla 307, Santiago 2, Chile} }

\date{}%\emph{\textbf{01.01:}} \textbf{Version 27}}

\maketitle

\begin{abstract}
Universal vector  wave equations  allowing for a unified
description of anyons, and also of usual bosons and
fermions in the plane are proposed. The existence of two
essentially different types of anyons, based on unitary and
also on non-unitary infinite-dimensional half-bounded
representations of the (2+1)D Lorentz algebra is revealed.
Those associated with non-unitary representations
interpolate between bosons and fermions. The extended
formulation of the theory includes the previously known
Jackiw-Nair (JN) and Majorana-Dirac (MD) descriptions of
anyons as particular cases, and allows us to compose bosons
and fermions from entangled anyons. The theory admits a
simple supersymmetric generalization, in which the JN and
MD systems are unified in $N=1$ and $N=2$ supermultiplets.
Two different non-relativistic limits of the theory are
investigated. The usual one generalizes L\'evy-Leblond's
spin $1/2$ theory to  arbitrary spin, as well as  to
anyons. The second, ``Jackiw-Nair'' limit (that corresponds
to In\"on\"u-Wigner contraction with both anyon spin and
light velocity going  to infinity), is generalized to
boson/fermion fields and interpolating anyons. The
resulting exotic Galilei symmetry is studied in both the
non-supersymmetric and supersymmetric cases.
\end{abstract}

\newpage\null\newpage

\tableofcontents

\newpage

%%%%%%%%%%%%%%%%%%%%%
\section{Introduction}\label{introd}
%%%%%%%%%%%%%%%%%%%%%%

According to Wigner \cite{Wigner},  elementary particles in
the plane correspond to irreducible representations of the
planar Poincar\'e group \cite{Binegar}. These are labeled
by two Casimir invariants, namely\footnote{We took $c=1$,
so that our metric is $(-1,+1,+1)$.}
\begin{eqnarray}
     (P_{\mu}P^{\mu}+m^2)\psi=0\,,
     \label{KG}
     \\[5pt]
     (P_{\mu}\cJ^\mu-sm)\psi=0\,,
     \label{PaLu}
\end{eqnarray}
where $m$ is the mass, $s$ the spin, and
$\cJ_\mu$ is the ``spin part'' of the total angular
momentum operator
\begin{equation}
    {\cal M}_\mu=-\epsilon_{\mu\nu\lambda}x^\nu
    P^\lambda+\cJ_\mu. \label{calJ}
\end{equation}
 ${\cal M}_\mu$, together with $P_\mu$, generate
the (2+1)D Poincar\'e group \footnote{The irreducible
representations of the universal covering group of
$SO(2,1)$ are, for finite-dimensional non-unitary
representations, fixed uniquely by the values of the
$\mathfrak{so}(2,1)$ Casimir operator ${\cal J}_\mu{\cal
J}^\mu$. Infinite-dimensional (unitary and
non-unitary) representations require, in addition, to
specify also the concrete series we are considering. On the
other hand, the representations of the Poncar\'e group are
labeled, in the massive case,
 by the Poincar\'e spin that corresponds to the
value of the Casimir operator $P_\mu {\cal
M}^\mu/m=P_\mu{\cal J}^\mu/m$.}.

For a spinless particle, $s=0$, the second equation drops out and
the Klein-Gordon equation, (\ref{KG}), is
sufficient in itself.

For a Dirac particle, or for a topologically massive vector
gauge field of Deser, Jackiw and Templeton (DJT) \cite{DJT},
i.e. for Poincar\'e spins
 $s =\pm 1/2$ or
$s=\pm 1$, the Pauli-Lubanski condition, (\ref{PaLu}),
implies (\ref{KG}) and provides, therefore, a satisfactory
description. In the plane the spin can take arbitrary real
values, however, and in the early nineties two, slightly
different descriptions have been proposed for
fractional-spin particles (called anyons
 \cite{LeiMyr,Wil,anysuper,DeserAny,Plany,deSG,Forte,Ghosh}). Both
of them combine the half-bounded, infinite-dimensional,
 unitary, spin-$\alpha$ representations, $D^{\pm}_\alpha$,
 of the Lorentz group \cite{Barg} with a finite dimensional, non-unitary,
 ``internal''  representation.

Choose, for example, $D^{+}_\alpha$, which carries an infinite-dimensional,
bounded-from-below
representation of the $\mathfrak{so}(2,1)$
algebra of the planar Lorentz group, with generators $J_\mu$.
 Diagonalizing
$J_0$, $J_0|\alpha,n\rangle=(\alpha+n)|\alpha,n\rangle,$
with $\alpha>0$ and $n=0,\dots,\infty$.

Jackiw and Nair (JN) \cite{JN1} combine this with the
spin-$1$ representation of topologically massive gauge
theory  \cite{DJT}. In
fact, they consider the dual to the topologically massive
gauge field strength \emph{vector}
  $F^\mu(x)$, which transforms under the spin-1 matrices
$
(j^\mu)_{\nu \lambda}=-i\epsilon_{\nu\ \ \lambda}^{\ \
\mu}.
$
 A ``Jackiw-Nair'' wave function is, hence,
\beq
F^\mu=\sum_{n=0}^\infty F_n^\mu(x)|\alpha,n\rangle .
\label{1.4}
\eeq
Their wave equation is the Pauli-Lubanski
condition (\ref{PaLu}), with $\cJ^\mu$ denoting
the total spin \cite{JN1},
\beq
    P_{\lambda}(\cJ^\lambda)_{n\mu\,\,n'\nu}
    F^\nu_{n'}-smF_{n\mu}=0\,, \qquad
    \cJ^\mu=J^\mu+j^\mu\,, \qquad s=\alpha-1\, .
    \label{totJNspin}
\eeq
This description is redundant, however, necessitating
subsidiary conditions, which reduce the number of
components from $3$ to $1$, and generate the mass shell condition \cite{JN1}.

An alternative description, proposed by one of us
\cite{Pl1991} and referred to below as the ``Majorana-Dirac''
(MD) approach, also uses the infinite dimensional discrete
type  representations $D^{\pm}_\alpha$, but the finite
dimensional part is, rather, a \emph{spinor},
\beq
    \psi=\sum_{n=0}^\infty \psi_n(x)|\alpha,n\rangle\, .
    \label{JNwf}
\eeq
where the $\psi_n(x)$ are two-component
spinors. The posited field equations are the planar
Majorana equation \cite{Ptor} supplemented by the planar
Dirac equation,
\beqa
    (P_{\mu}J^\mu-\alpha m)\psi&=&0\,,
    \label{PaLu2}
    \\[6pt]
    (P_\mu \gamma^\mu-m)\psi&=&0\,,
    \label{PlDirac}
\eeqa
 where the $\gamma^\mu$ are the planar Dirac (in
fact Pauli) matrices.
 The total spin is still $\cJ^\mu=J^\mu+j^\mu$
 as in (\ref{totJNspin}), except for
 $j^\mu=-\gamma^\mu/2$, which carry
a spin-$1/2$ representation, so that $s=\alpha-1/2$.
Note that in (\ref{PaLu2}) the total spin, $\cJ^\mu$,
has been replaced by the $D^+_{\alpha}$ generator $J^\mu$.

The system of two equations  eliminates the unphysical degrees of freedom,
so that the system (\ref{PaLu2})-(\ref{PlDirac}) does not necessitate
any further subsidiary condition.

Both the JN and MD systems are shown to imply the
Klein-Gordon equation, (\ref{KG}), and carry, therefore, an
irreducible representation of the planar Poincar\'e group.
\vskip2mm

$\bullet$ Our first aim is  to unify the two descriptions
by replacing both of them by a suitable vector equation.
\vskip2mm

To this end we use the vector set of equations proposed
earlier  in \cite{CoPl}. By analyzing the lowest/highest
weight representations of $\mathfrak{so}(2,1)$, we show
that the vector set is appropriate not only to describe
anyons and usual boson and fermion fields, but also
provides us with an \emph{anyon interpolation between
bosons and fermions} of arbitrary spin values shifted by
one half.  Considering an extended formulation of the
vector set of equations, we reveal that the Jackiw-Nair
\cite{JN1} and Majorana-Dirac \cite{Pl1991} descriptions of
anyons are incorporated in it as particular cases. By a
kind of  fine tuning, which results in a destructive
interference, the extended formulation allows us to compose
boson and fermion fields from \emph{entangled} anyons. Both
this possibility and the interpolation mechanism mentioned
above rely on  infinite dimensional non-unitary
half-bounded representations of
$\mathfrak{so}(2,1)$.\vskip2mm

$\bullet$ Our second objective is to construct a
supersymmetric extension of the theory. \vskip2mm

We show that our basic vector set of equations admits a
simple and natural supersymmetric generalization. The
construction is based on $\mathfrak{osp}(1|2)$
representations realized in terms of a special deformation
of the Heisenberg algebra \cite{Wig,para,GreenColor,RDHA}.
As interesting examples, we
consider an $N=1$ supermultiplet that includes the Dirac
spinor and topologically massive DJT vector gauge fields,
and unify the JN and MD anyon fields into $N=1$
and $N=2$ supermultiplets.\vskip2mm

$\bullet$ Finally, two different
non-relativistic limits are investigated
both in the supersymmetric and the non-supersymmetric cases. \vskip2mm

The first, usual, limit yields the L\'evy-Leblond
theory for boson/fermion fields of arbitrary spin
\cite{LLeq,HPPLB}, as well
as its supersymmetric and anyon generalizations. Among
particular examples, we consider the non-relativistic limit
of a topologically massive vector gauge field.

 In the second, ``Jackiw-Nair''
non-relativistic limit \cite{JaNa,DuHor}, spin and light speed both tend
simultaneously to infinity  in a special way such that
the (super) Poincar\'e symmetry is reduced to the exotic (super)
Galilei symmetry \cite{SUSYGal1,SUSYGal2,HPV1,Alvarez}.\vskip2mm

\newpage
%%%%%%%%%%%%%%%%%%%%%%%%%%%%%%%%%%%%%%%%%%%%%%%%%%%%
\section{Wave equations}\label{nonsusy}
%%%%%%%%%%%%%%%%%%%%%%%%%%

Our starting point is to posit the vector set of equations
\begin{equation}\label{VmuEq}
    V_\mu\psi=0, \qquad
    V_\mu= \emph{\ss} P_\mu -
    i\epsilon_{\mu\nu\lambda}P^\nu{\cal
    J}^\lambda +m {\cal J}_\mu,
\end{equation}
where $P_\mu=-i\partial_\mu$ and ${\cal J}_\mu$  generates
 (2+1)D Lorentz transformations,
 $[{\cal J}_\mu,{\cal J}_\nu]=-i\epsilon_{\mu\nu\lambda}{\cal J}^\lambda$.
 Such a system of wave equations was proposed before in
 \cite{CoPl} to describe anyons of mass $m>0$ and spin
 $\emph{\ss}$.

Contracting (\ref{VmuEq}) with $P^\mu$, ${\cal J}^\mu$ and
$-i\epsilon^{\mu\nu\lambda}P_\nu {\cal J}_\lambda$ produces three
equations, namely
\begin{eqnarray}
    \left(\emph{\ss} P^2 +  m\, P
    {\cal J}\right) \psi
    &=&0\, ,
    \label{Eqnscal1}\\
    \left((\emph{\ss}-1) P {\cal J} +  m\,
     {\cal J}^2\right)
    \psi&=&0\, ,
    \label{Eqnscal2}\\
    \left(P^2{\cal J}^2 + (P{\cal J})(m - P{\cal
    J})\right)\psi &=&0\,.
    \label{Eqnscal3}
\end{eqnarray}
If $\emph{\ss}=0$, these equations do not to fix the value
of the (2+1)D Poincar\'e Casimir operator $P^2$. We suppose
therefore that
%\begin{equation}\label{betaneq}
 $   \emph{\ss}\neq 0$.
%\end{equation}
 Then the system  (\ref{Eqnscal1})--(\ref{Eqnscal3}) is
equivalent to
%\begin{equation}\label{PJb}
\beqa
    P^2\left(P^2+m^2\right)\psi=0\,,
    \\
    \left(\emph{\ss} P^2+ m\, P{\cal
    J}\right)\psi=0\,,
    \\
    \left(m^2{\cal J}^2+\emph{\ss}
    (1-\emph{\ss})P^2\right)\psi=0\,.
\eeqa
It follows that the field $\psi$ satisfies either $
P^2\psi=0,\,  P{\cal J}\psi=0,\, {\cal J}^2\psi=0$,
 or
\begin{eqnarray}
    \left(P^2+m^2\right)\psi&=&0\, ,\label{Cas+aux1}
    \\
    \left(P{\cal J}-\emph{\ss} m\right)\psi&=&0\, ,
    \label{Cas+aux2}\\
    \left({\cal J}^2 + \emph{\ss}(\emph{\ss} -1)\right)\psi&=&0\, .
    \label{Cas+aux3}
\end{eqnarray}
The first case corresponds to  the trivial representation of the
(2+1)D Poincar\'e group with $P_\mu=0$ and ${\cal J}_\mu=0$ as
seen in the  frame $P^\mu=(p,p,0)$. Excluding this case, we assume
henceforth that  $\psi$ satisfies Eqns.
(\ref{Cas+aux1})--(\ref{Cas+aux3}) with ${\cal J}_\mu\neq 0$. Our
vector system of equations implies, hence, the Klein-Gordon
\eqref{Cas+aux1} and the Pauli-Lubanski condition
\eqref{Cas+aux2}. The consistency of our vector
 equations (\ref{VmuEq}) then follows from
\begin{equation}\label{VV1}
    [V_\mu,V_\nu]=-i\epsilon_{\mu\nu\lambda}\left(
    mV^\lambda +P^\lambda (P{\cal J}-\emph{\ss}
    m)\right).
\end{equation}
When nontrivial solutions of (\ref{VmuEq}) do exist, it
follows that they describe an \emph{irreducible}
representation of the (2+1)D Poincar\'e group with nonzero
mass $m$ and spin $s=\emph{\ss}$~\footnote{The spin zero
case can also be incorporated into the theory, see below.}.

Our vector equations determine  the representation of the spin
part of the Lorentz transformation; eq. \eqref{Cas+aux3} fixes the
value of the Lorentz Casimir.

Taking into account Eq. (\ref{Cas+aux1}) and passing to the rest frame
$P^\mu=(\varepsilon m, 0,0)$ where $\varepsilon=+1$ or $-1$, we
find that our covariant vector equations, (\ref{VmuEq}), are
equivalent to two independent equations, namely to
\begin{eqnarray}\label{J0}
    \left({\cal
    J}_0-\varepsilon\emph{\ss}\right)\psi&=&0\,,\\
    \left({\cal J}_1-i\varepsilon {\cal
    J}_2\right)\psi&=&0\,.\label{J+-}
\end{eqnarray}
From Eq. (\ref{J+-}) we infer that for $\varepsilon=1$,
${\cal J}_\mu$ should belong to a representation bounded
from below, and for $\varepsilon=-1$ a representation has
to be bounded from above~\footnote{Changing $m\rightarrow
-m$ in  (\ref{VmuEq}) together with intertwining
representation bounded from below and above yields  solutions of the same energy
but of the opposite sign of spin .}. (It can also be bounded
from both sides.)

The universal covering group of the (2+1)D Lorentz group
admits
 half-bounded unitary
infinite-dimensional representations,  namely, the discrete
type series $D^\pm_\alpha$ \cite{Barg,CoPl}. They are
characterized by a diagonal operator and a highest (lowest)
vector,
\begin{equation}\label{J0+-psi}
D^+_\alpha\,:\quad    {\cal J}_0\vert \alpha,
    n\rangle= (\alpha+n)\vert \alpha, n\rangle
    \quad\hbox{and}\quad
    ({\cal J}_1- i{\cal J}_2)\vert \alpha,0\rangle=0\,,
\end{equation}
\begin{equation}\label{Dminus}
D^-_\alpha\,:\quad    {\cal J}_0\vert \alpha,
    n\rangle= - (\alpha+n)\vert \alpha, n\rangle
    \quad\hbox{and}\quad
    ({\cal J}_1 + i{\cal J}_2)\vert \alpha,0\rangle=0\,,
\end{equation}
where $n=0,1,\ldots$. The representations $D^+_\alpha$ and
$D^-_\alpha$ are related via the Lorentz algebra
automorphism ${\cal J}_0\rightarrow -{\cal J}_0$, ${\cal
J}_1\rightarrow {\cal J}_1$, ${\cal J}_2\rightarrow -{\cal
J}_2$. The Casimir operator  takes here the value,
\beq\label{Jal}
    {\cal J}^2=-\alpha(\alpha-1).
\eeq These unitary half-bounded infinite-dimensional
representations $D^\pm_\alpha$ appear in most earlier works on
wave equations for relativistic
\cite{JN1,Pl1991,Ptor,CoPl,HPV1,quartion,SorVol,Spinor,MPAnn,SUSYfail,KPR}
and non-relativistic \cite{HPPLB,HPV1} anyons.

 The universal covering group of the (2+1)D
Lorentz group admits also unbounded representations,
namely, the principal and the supplementary continuous
series. They only allow trivial solutions of the vector set
of equations (\ref{VmuEq}), and will therefore be
discarded.

The non-unitary $(2j+1)$-dimensional representations
$\tilde{D}^j$, $j=\frac{1}{2},1,\ldots$, characterized by
(\ref{Jal}) with $\alpha=-j$ i.e., ${\cal J}^2=-j(j+1)$
\footnote{We reserve the notation $j$ to denote the (half-)
integer values, $1/2,1,3/2,...$.}, admits both highest- and
lowest-spin states,
\begin{equation}\label{Dfinite}
\tilde{D}^j\,:\qquad    {\cal J}_0\vert j,
    \ell \rangle= \ell \vert j, \ell \rangle\,,
    \quad
    ({\cal J}_1- i{\cal J}_2)\vert j,0\rangle=0
    \qquad \hbox{and} \qquad
    ({\cal J}_1 + i{\cal J}_2)\vert j,j\rangle=0\,,
\end{equation}
where $\ell=-j, -j+1, ...,j.$
 Hence, they also can be used in our equation.
 In this case, the vector equation describes
 usual bosons and fermions of arbitrary spin.

Up to an unitary transformation, they are  obtained from
the spin-$j$ representations of $su(2)$, ${\cal
J}_0=\hat{\cal J}^3$, ${\cal J}_k=-i\hat{\cal J}^k$,
$k=1,2$, where $[\hat{\cal J}^a,\hat{\cal
J}^b]=i\epsilon^{abc}\hat{\cal J}^c$, $a,b,c=1,2,3$.

 There is a third type of representation of $\mathfrak{so}(2,1)$
 which plays an important role in our equations, namely
 \emph{half-bounded, infinite-dimensional
{non-unitary} representations}. These representations
 are characterized by relations of the form (\ref{Jal}) and
(\ref{J0+-psi}) or (\ref{Dminus}), with the %spin Casimir
parameter $\alpha$ in (\ref{Jal}) taking
\emph{non-half-integer negative} values, $-\alpha \neq
j=1/2,1,... $. They are denoted here by
$\tilde{D}^\pm_\alpha$,
\begin{equation}\label{Dminusbis}
    \tilde{D}^\pm_\alpha\,:\quad
    {\cal J}_0\vert \alpha,
    n\rangle= \pm (\alpha+n)\vert \alpha,
    n\rangle,
    \qquad
    ({\cal J}_1 \mp i{\cal J}_2)\vert
    \alpha,0\rangle=0, \quad -j<\alpha< -j+ 1/2\,.
\end{equation}

These representations describe anyons whose fractional spin
interpolates between bosons and fermions.

Occasionally, we will also use the unifying notation,
\beq
    \label{tilD}
    {\cal D}^\pm_\alpha
    = \left\{\barray{lll}
    D^\pm_\alpha &\hbox{for} &\alpha>0,
    \\[6pt]
    \tilde{D}^\pm_\alpha &\hbox{for} &\alpha<0,\
    \alpha\neq -j,
    \\[6pt]
    \tilde{D}^j &\hbox{for} &\alpha=-j.
    \earray\right.
\eeq
Choosing ${\cal J}_\mu$ in one of irreducible
representations (\ref{tilD}) and putting
\begin{equation}\label{ssalpha}
    \emph{\ss}=\alpha\,,
\end{equation}
Eq. (\ref{Cas+aux3}) is identically satisfied for all real
$\alpha\neq 0$.

Matrix representations of $D^\pm_\alpha$, $\tilde{D}^j$ and
$\tilde{D}^\pm_\alpha$ are considered, in a unified way, in
Section \ref{highlow} below; in Section \ref{RDHA} they are
treated as Fock-like representations of a certain
deformation of the Heisenberg algebra.

In Section \ref{solutions} we describe explicit solutions of the
vector set of equations  with ${\cal J}_\mu$ chosen in any of the
three irreducible representations (\ref{tilD}), and with parameter
$\emph{\ss}$ coherently fixed by (\ref{ssalpha}). In what follows
we refer to such a realization of (\ref{VmuEq}) as a
\emph{minimal} one.

An \emph{extended} formulation, which combines several irreducible
representations, will be considered in Section \ref{solBFA}.

\subsection{Examples: Dirac and Deser-Jackiw-Templeton fields}

The two simplest finite-dimensional representations
$\tilde{D}^j$ are,

\begin{itemize}
\item
$j={\bf -\alpha}=\frac{1}{2}$, ${\cal
J}_\mu=-\frac{1}{2}\gamma_\mu$, when the linear
differential equation (\ref{Cas+aux2}) reduces to the Dirac
equation,
\begin{equation}\label{Dirac}
    (P\gamma - m)\psi=0\, .
\end{equation}
Conversely, our vector system (\ref{VmuEq}) is the Dirac
equation (\ref{Dirac}), multiplied by
$\frac{1}{2}\gamma_\mu$.

\item
$j={\bf -\alpha}=1$, $({\cal
J}_\mu)^\lambda{}_\nu=-i\epsilon^\lambda{}_{\mu\nu}$, when
(\ref{Cas+aux2}) reduces to the equation for a
\emph{topologically massive vector gauge field} \cite{DJT},
\begin{equation}\label{DJT}
    \left(-i\epsilon^\lambda{}_{\mu\nu}P^\mu+
    m\delta^\lambda_\nu\right)\psi^\nu=0\,,
\end{equation}
where we switched to the notation $\psi^\mu$ instead of
$F^\mu$, used in (\ref{1.4}). Eq. (\ref{DJT})  implies the
transversality condition
\begin{equation}\label{pmupsi}
    P_\mu\psi^\mu=0\,,
\end{equation}
that allows us to view the vector field $\psi^\mu$ as dual
to the $U(1)$ gauge invariant tensor,
$\psi_\mu=\frac{1}{2}\epsilon_{\mu\nu\lambda}F^{\nu\lambda}$,
$F_{\mu\nu}=\partial_\mu A_\nu-\partial_\nu A_\mu$.

Conversely, the vectorial form (\ref{VmuEq}) is obtained
from (\ref{DJT})  by multiplication by
 $-i\epsilon^\rho{}_{\sigma\lambda}$ and addition
to Eq. (\ref{pmupsi}), multiplied by $-\delta^\rho_\sigma$.
\end{itemize}

The values $j=\frac{1}{2}$ and $j=1$ are exceptional in
that in these cases (only)  the linear differential
equation (\ref{Cas+aux2}), i .e (\ref{Dirac}) and
(\ref{DJT}), does already imply the Klein-Gordon equation,
(\ref{Cas+aux1}). This explains why, for $|s|\neq
\frac{1}{2},\,1$, getting an irreducible representation of
the Poincar\'e group requires positing a vector set of
equations  for anyons, and for for usual fields of integer
and half-integer spin with $|s|=j$, $j>1$ \footnote{Any two
of three equations (\ref{VmuEq}) generate the third one as
consistency condition, see Section \ref{solutions} below;
the complete set of three equations provides us with an
explicitly covariant formulation.}.

%%%%%%%%%%%%%%%%%%%%%%%%%%%%%%%%%%%%%%%%%%%%%%%%%%%%%%%%%%%%%%%%%%%%%%

%%%%%%%%%%%%%%%%%%%%%%%%%%%%%%%%%%%%%%%%%%
\subsection{Example: Majorana-Dirac anyon
field}\label{MajDirSub}
%%%%%%%%%%%%%%%%%%%%%%%%%%%%%%%%%%%%%%%%%%

Both the Dirac and DJT models are associated with finite
dimensional representations. We can also combine them with
fractional spin, however. Consider, in fact, the
representation of the spin part of the Lorentz generators
as~\footnote{The assumed tensor product,
$\mathcal{J}_\mu=J_\mu\otimes 1-1\otimes \frac12
\gamma_\mu$, is not indicated for the sake of notational simplicity.},
\begin{equation}\label{Maj-Dir}
    \mathcal{J}_\mu=J_\mu-\frac12 \gamma_\mu\,, \qquad J_\mu \in
    D^+_\alpha,\quad  -\frac12 \gamma_\mu \in
    \tilde{D}^{1/2}\,,
    \quad
    \alpha\neq 1/2\,,
\end{equation}
and put
\begin{equation}\label{ssa12}
    \emph{\ss}=\alpha-1/2\,.
\end{equation}
Eq. \eqref{Cas+aux3} allows us to infer,
\begin{equation}\label{Jgal}
    \Pi\psi=0\,,\quad
    {\rm where}\quad \Pi=J\gamma+\alpha\,.
\end{equation}
 The operator $\Pi$ is a projector~:
$\Pi^2=(2\alpha-1)\Pi$. Multiplying (\ref{Jgal}) from the
left by the  [on-shell invertible, cf. (\ref{Cas+aux1})]
operator $P\gamma$, we find that the field also has to
satisfy
\begin{equation}\label{PJge}
    \Upsilon\psi=0,\quad
    {\rm where}\quad \Upsilon= PJ-\alpha P\gamma -\Lambda,\quad
    \Lambda=
    i\epsilon_{\mu\nu\lambda}P^\mu J^\nu \gamma^\lambda\,.
\end{equation}
Multiplication of (\ref{Jgal}) by $-\Lambda$ yields
$\left((\alpha-1)\Upsilon-\Lambda\right)\psi=0$, and we
find, hence, that $\psi$ also has to satisfy the equation
\begin{equation}\label{Lambda}
    \Lambda\psi=0\,.
\end{equation}
In the present case,  Eq. (\ref{Cas+aux2}) reads
$\left(PJ-\alpha m)-\frac{1}{2}(P\gamma-m)\right)\psi=0$.
Combining it  with  Eq. (\ref{PJge}) and taking into
account Eq. (\ref{Lambda}),  we find that our field
$\psi^n_b$, that carries an index $n=0,1,\ldots$ of the
representation $D^+_\alpha$ as well as a spinor index $b$,
has to satisfy, separately, the $(2+1)D$ Majorana and Dirac
equations,
\begin{equation}\label{MajDirac}
    \left(PJ-\alpha m\right)\psi=0\,,\qquad
    \left(P\gamma-m\right)\psi=0\,.
\end{equation}
This system of two equations was proposed in \cite{Pl1991}
to describe $(2+1)D$ anyons; the corresponding field was
called  a ``Majorana-Dirac field".

 It is easy to check that these two equations imply (\ref{Jgal}) and
(\ref{Lambda}) as consistency (integrability) conditions.
Moreover, we have the following remarkable property:  the
full set of four equations (\ref{MajDirac}), (\ref{Jgal})
and (\ref{Lambda}) is generated by positing any two
equations out of the four, (one of which should involve the
mass parameter). It follows that the vector system
(\ref{VmuEq}), with (\ref{Maj-Dir}) and (\ref{ssa12})
describes
 an irreducible representation of the $(2+1)D$
Poincar\'e group of mass $m$ and spin
$s=\alpha-\frac{1}{2}\neq 0$.

%%%%%%%%%%%%%%%%%%%%%%%%%%%%%%%%%%%%%%%
\subsection{Example: Jackiw-Nair anyon field}\label{JNSub}
%%%%%%%%%%%%%%%%%%%%%%%%%%%%%%%%%%%%%%%

Similarly, consider now,
\begin{equation}\label{JNj1}
    \mathcal{J}_\mu=J_\mu+j_\mu\,, \qquad J_\mu \in
    D^+_\alpha,\quad  (j_\mu)^\lambda{}_\nu=
    -i\epsilon^{\lambda}{}_{\mu\nu} \in
    \tilde{D}^{1},\quad
    \alpha\neq 1,
\end{equation}
and
\begin{equation}\label{ssaJN}
    \emph{\ss}=\alpha-1,
\end{equation}
for which the Pauli-Lubanski condition (\ref{Cas+aux2})
becomes,
\begin{equation}\label{JackNa}
    \left(P_\mu J^\mu \delta^\lambda_\nu-
    i\epsilon^{\lambda}{}_{\mu\nu}P^{\mu}-
    (\alpha-1)m\delta^\lambda_\nu\right)\psi^\nu=0\,,
\end{equation}
that is  the anyon equation put  forward by Jackiw and Nair
\cite{JN1} \footnote{Both in (\ref{JackNa}) and in
(\ref{J1split}) the index $n$ of $\psi^n_\mu$ has been
suppressed.}.
 Eq. (\ref{Cas+aux3}) reads in turn
\begin{equation}\label{J1split}
    \left(-i\epsilon^{\lambda}{}_{\mu\nu}J^\mu-
    \alpha\delta^\lambda_\nu\right)\,
    \psi^\nu=0\, .
\end{equation}

 Now we turn to the subsidiary conditions in
\cite{JN1}. Contracting  (\ref{J1split}) from the left with
$J_\lambda$ yields
\begin{equation}\label{Jpsi=0}
    J_\mu\psi^\mu=0\,.
\end{equation}
Next, contraction with the vector $P_\lambda$
 produces
 $\left(-i\epsilon_{\mu\nu\lambda} P^\nu J^\lambda
-\alpha P_\mu\right)\psi^\mu=0$, while contraction with
$-i\epsilon_{\lambda\rho\sigma}P^\rho J^\sigma$ produces an
equation of the same form, but with the sign before
$\alpha$ reversed. In addition to (\ref{Jpsi=0}), the field
also has to satisfy therefore,
\begin{equation}\label{Ppsi=0}
    P_\mu\psi^\mu=0\,,
\quad\hbox{and}\quad
    \epsilon_{\mu\nu\lambda}P^\mu J^\nu \psi^\lambda=0\,,
\end{equation}
which are precisely the subsidiary conditions  of Jackiw
and Nair in Ref.
 \cite{JN1} \footnote{The subsidiary condition (4.12a)
of \cite{JN1} is a linear combination of (\ref{Jpsi=0}) and
of the first equation from (\ref{Ppsi=0}), $(P_\mu
J_\nu-J_\mu P_\nu)\psi^\nu=0$.}.

Contracting  Eq. (\ref{J1split}) with
$i\epsilon^\rho{}_{\sigma\lambda} P^\sigma$ and taking into
account Eq. (\ref{Ppsi=0}), yields
\begin{equation}\label{JackNa*}
    \left((PJ-\alpha m) \delta^\lambda_\nu
    +\alpha(-i\epsilon^{\lambda}{}_{\mu\nu}P^{\mu}+
    m\delta^\lambda_\nu)\right)\psi^\nu=0\,.
\end{equation}
Then, using Eq. (\ref{JackNa})  shows that Eq.
(\ref{J1split}) splits  the JN equation (\ref{JackNa}) into
a Majorana equation, supplemented with the DJT equation of
a topologically massive gauge vector field. Note however
that, unlike in the Majorana-Dirac case (\ref{MajDirac}),
imposing the pair of Majorana and DJT equations alone on
$\psi^n_\mu$ does not imply the condition
 (\ref{J1split}) that guarantees the irreducibility
equation (\ref{Cas+aux3}) as well as the other necessary
subsidiary conditions in (\ref{Ppsi=0}). \vskip3mm

 We shall see in Section \ref{solBFA} that Majorana-Dirac
and Jackiw-Nair descriptions of anyon fields are included
as particular cases into a broader, \emph{extended}
realization of our basic set of equations
(\ref{VmuEq}).

%%%%%%%%%%%%%%%%%%%%%%%%%%%%%%%%%%%%%%%%%%%%%%%%%%%%%%%%%%%%%%%%%%%%%%%%%%%%%
%%%%%%%%%%%%%%%%%%%%%%%%%%%%%%%%%%%%%%%%%%%%%%%%%%%%%%%%%%%%%%%%%%%%%%%%%%
\section{$\mathfrak{so}(2,1)$ lowest/highest weight representations}\label{highlow}
%%%%%%%%%%%%%%%%%%%%%%%%%%%%%%%%%%%%%%%%%%%%%%%%%%%%%%%%%%%%%%%%%%%
%%%%%%%%%%%%%%%%%%%%%%%%%%%%%%%%%%%%%%%%%%%%%%%%%%%%%%%%%%%%%%%%%%%%%%%%%%%%%%%%%%%%%%%
%%%%%%%%%%%%%%%%%%%%%%%%%%%%%%%%%%%%%%%%%%%%%%%%%%%%%%%%%%%%%%%%%%%%%

Now  we discuss some aspects of the representation theory
of $\mathfrak{so}(2,1)$, which will be used
 in the next two Sections to derive explicit solutions.
This will allow us to see, in particular, how the non-unitary
infinite dimensional representations $
\tilde{D}^+_{\alpha}$ (or $ \tilde{D}^-_{\alpha}$) produce
an anyon interpolation between bosons and fermions, and how
bosons and fermions can be composed from anyons. \vskip2mm

Let ${\cal J}_\mu$ generate the
$\mathfrak{so}(2,1)$ algebra, $[{\cal J}_\mu,{\cal
J}_\nu]=-i\epsilon_{\mu\nu\lambda}{\cal J}^\lambda$. In terms of
the ladder operators ${\cal J}_\pm={\cal J}_1\pm i{\cal J}_2$,
\begin{equation}\label{AA1}
    [{\cal J}_-,{\cal
    J}_+]=2{\cal J}_0\,,\qquad
    [{\cal J}_0,{\cal J}_\pm]=\pm{\cal J}_\pm\,.
\end{equation}
{}Eq. (\ref{AA1}) implies
\begin{equation}\label{AA2}
    [{\cal J}_-,{\cal J}_+^n]=2n{\cal J}_+^{n-1}\left(
    \frac{1}{2}(n-1)+{\cal J}_0\right),\qquad
    n=0,1,2,\ldots .
\end{equation}

{}From now on, for the sake of definiteness, we consider the
lowest weight representations built over the state
$|0\rangle$,
\begin{equation}\label{AA3}
    {\cal J}_0|0\rangle=\alpha|0\rangle\,,\qquad
    {\cal J}_-|0\rangle=0\,,\qquad \langle 0|0\rangle=1\,,
\end{equation}
where $\alpha$ is real \footnote{Highest weight
representations can be analyzed in the same way
and
can also be obtained directly from the lowest weight
representations via the Lorentz algebra automorphism ${\cal
J}_0\rightarrow -{\cal J}_0$, ${\cal J}_\pm\rightarrow
{\cal J}_\mp$.}.

 The non-normalized states
\begin{equation}\label{AA4}
    \widetilde{|n\rangle}={\cal J}_+^n|0\rangle\,,
    \qquad
    n=0,1,\ldots,
\end{equation}
are eigenvectors of ${\cal J}_0$,
\begin{equation}\label{AA5}
    {\cal J}_0\widetilde{|n\rangle}=
    (\alpha+n)\widetilde{|n\rangle}\,,
\end{equation}
on which the $\mathfrak{so}(2,1)$ Casimir ${\cal
C}_{\mathfrak{so}(2,1)}=-{\cal
    J}_0^2+\frac{1}{2}({\cal J}_+{\cal J}_-+{\cal J}_-{\cal
    J}_+)$ takes the value
\begin{equation}\label{AA6}
    {\cal C}_{\mathfrak{so}(2,1)}\,
    \widetilde{|n\rangle}=-\alpha(\alpha-1)
    \widetilde{|n\rangle}\,.
\end{equation}
{}From relations (\ref{AA2})--(\ref{AA4}) it follows that
the vectors $\widetilde{|n\rangle}$ form an orthogonal
basis,
\begin{equation}\label{AA7}
    \widetilde{\langle n\,}|\widetilde{n'\rangle}=
    \delta_{nn'}\,{\cal C}_{\alpha,n}^2\,,\quad
    {\cal C}_{\alpha,0}=1\,,\quad
    {\cal C}_{\alpha,n}^2=n!\,
    \prod_{k=0}^{n-1}\left(2\alpha+k\right)\,,
\end{equation}
where ${\cal C}_{\alpha,n}^2$ is a squared norm of
$\widetilde{|n\rangle}$.

For $\alpha\neq -j$, $j=0,1/2,1,\ldots$, the numbers ${\cal
C}_{\alpha,n}^2$ take nonzero values for any
$n=0,1,\ldots$; the corresponding representations are
infinite-dimensional  and irreducible.

For $\alpha= -j$, only the first $2j+1$ numbers ${\cal
C}_{\alpha,n}^2$ with $n=0,\ldots,2j$ take nonzero values,
while ${\cal C}_{\alpha,n}^2=0$ for  $n=2j+1,2j+2,\ldots$.
This happens if $\widetilde{|n\rangle}=0$ for
$n=2j+1,\ldots$, and then we have a $(2j+1)$-dimensional irreducible representation.

There is also another possibility, though~: the states
$\widetilde{|n\rangle}$ with  $n=2j+1,2j+2,\ldots$ may be
nonzero. Curiously, they have zero norm.

 We first consider \emph{irreducible}
representations, where no zero norm states of the form
(\ref{AA4}) are assumed to appear, and then discuss how to
treat representations with zero norm states.

%%%%%%%%%%%%%%%%%%%%%%%%%%%%%%%%%%%%%%%%%%%%%%%%%%%%%%%%%%%
%%%%%%%%%%%%%%%%%%%%%%%%%%%%%%%%%%%%%%%%%
\subsection{Irreducible representations}\label{irreps}
%%%%%%%%%%%%%%%%%%%%%%%%%%%%%%%%%%%%%%%%%
%%%%%%%%%%%%%%%%%%%%%%%%%%%%%%%%%%%%%%%%%%%%%%%%%%%%%%%%%%%

To analyze the solutions of the wave equations, it is
convenient to work with normalized states, denoted by
$|n)$,
\begin{eqnarray}
    |0)=|0\rangle,\qquad
    &
     (\mathcal{J}_+)^n |0)=\left(\Pi_{k=0}^{n-1}
     C^\alpha_k\right) |n)=\widetilde{|n\rangle}\,
     , \label{slrep} &\label{Jn0}\\[6pt]
    &C^\alpha_n=\sqrt{(2\alpha+n)(n+1)}\,.
    &\label{Coeff}
\end{eqnarray}
Note that ${\cal
C}_{\alpha,n}^2=\prod_{k=0}^{n-1}C^\alpha_k$,  cf.
(\ref{AA7}), and that
\begin{equation}
    \mathcal{J}_0\vert n)=(\alpha+n)\vert n)\,,\quad
    \mathcal{J}_+\vert n)=C^\alpha_n\vert n+1)\,,\quad
    \mathcal{J}_-\vert n)=C^\alpha_{n-1}\vert n-1)\,.
    \label{D+}
\end{equation}

Due to the structure of the squared norm ${\cal
C}_{\alpha,n}^2$, the following cases have to be
distinguished.

\begin{itemize}

\item
For $\alpha=0$, we have $C^0_0=1$ [$C^0_n=0$,
$n=1,\ldots$], yielding the \emph{trivial} one-dimensional spin-0
representation, ${\cal J}_\mu|0)=0$.

 \item  $D^+_\alpha$ where $\alpha>0$. The coefficients ${\cal
C}_{\alpha,n}^2$ are positive for any $n$. The $|n)=({\cal
C}_{\alpha,n})^{-1}\widetilde{|n\rangle}$ are just those
$|\alpha,n\rangle$ in (\ref{J0+-psi}), and satisfy the
orthonormality relation $(n|n')=\delta_{nn'}$. This
corresponds to the \emph{infinite-dimensional unitary
half-bounded representations}.

\item $\tilde{D}^j$ with $\alpha=-j$, a negative (half)integer. We have
 both a lowest weight vector, ${\cal J}_-|0)=0$,
 and a highest weight vector, ${\cal J}_+|2j)=0\,$;
 the coefficient $C^{-j}_{2j}$ in (\ref{D+}) vanishes. This is the \emph{usual
$(2j+1)$-dimensional non-unitary representation}.

By  (\ref{Coeff}), the coefficients $C^{-j}_n$,
$n=1,\ldots,2j-1$, are imaginary. Put
\begin{equation}\label{C-j}
    C^{-j}_n=iC'^{-j}_n,\qquad
    C'^{-j}_n=\sqrt{(2j-n)(n+1)}\,.
\end{equation}
Making use of the $\mathfrak{so}(2,1)$  commutation
relations, one finds that the states $|n)$,
$n=0,1,\ldots,2j$, satisfy $(n|n')=(-1)^n\delta_{nn'}$.
The corresponding metric in the vector space $\tilde{D}^j$,
\begin{equation}\label{etametric}
    \eta_{nn'}=diag\, (1,-1,
    \ldots, (-1)^{2j-1}, (-1)^{2j}),
\end{equation}
is indefinite, consistently  with the non-unitary character of
such representations. Redefining the generators as ${\cal
J}_\pm\rightarrow \pm i{\cal J}_\pm$ yields, instead of
(\ref{D+}),
\begin{equation}
    \mathcal{J}_+\vert n)=C'^{-j}_n\vert n+1)\,,\quad
    \mathcal{J}_-\vert n)=-C'^{-j}_{n-1}\vert n-1)\,,\label{C'D+}
\end{equation}
where $C'^{-j}_n$ is given by (\ref{C-j}). Such  redefined
operators ${\cal J}_+$ and ${\cal J}_-$ are mutually
conjugate, $(\Psi|{\cal J}_+\Phi)^*=(\Phi|{\cal J}_-\Psi)$,
w.r.t.  the indefinite  scalar product
$(\Psi,\Phi)=\bar{\Psi}_n\Phi^n$,
$\bar{\Psi}_n=\Psi^{*k}\hat{\eta}_{kn}$, where
$\Phi^n=(n|\Phi)$, and matrix $\hat{\eta}$ is given by
(\ref{etametric}), while $(\Psi|{\cal
J}_0\Phi)^*=(\Phi|{\cal J}_0\Psi)$.

\item
$\tilde{D}^+_\alpha$ with $\alpha$ negative such that
$\alpha\neq -j$. In this case the ${\cal C}_{\alpha,n}^2$
alternate. This corresponds to the
\emph{infinite-dimensional
half-bounded non-unitary representations}.

Let us discuss this case in some detail. For
\begin{equation}\label{alpha<-j}
    -j<\alpha<-j+\frac{1}{2}\,,\qquad j=1/2,1,3/2,\ldots,
\end{equation}
 all
coefficients (\ref{Coeff}) are nonzero. Those with index
$n<2j$ have an imaginary factor $i$, while those with
$n\geq 2j$ are real.  The metric is indefinite,
\begin{equation}\label{etainf}
    \eta_{nn'}=diag\,
    (1,-1,\ldots,(-1)^{2j-1},(-1)^{2j},(-1)^{2j},(-1)^{2j}
    \ldots)\,.
\end{equation}
In the first $(2j+1)$ positions the metric alternates as in
the finite-dimensional representation $\tilde{D}^j$, and
after that it takes the same (constant) value $(-1)^{2j}$
up to infinity.

Both the finite and infinite non-unitary representations can  be
considered in a unified way by putting $\alpha=-(j+\epsilon)$ in
(\ref{D+}).  Redefining ${\cal J}_\pm\rightarrow \pm i{\cal
J}_\pm$, we get
\begin{equation}
    \mathcal{J}_0\vert n)=(-j-\epsilon+n)\vert n)\,,\quad
    \mathcal{J}_+\vert n)=C'^{-(j+\epsilon)}_n\vert n+1)\,,\quad
    \mathcal{J}_-\vert n)=-C'^{-(j+\epsilon)}_{n-1}\vert n-1)\,,\label{C''D}
\end{equation}
where,
\begin{equation}\label{C-j+e}
    C'^{-(j+\epsilon)}_n=\sqrt{(2(j+\epsilon)-n)(n+1)}\,.
\end{equation}
Here the parameter $-1/2\leq \epsilon \leq 0$  interpolates
between
 $-1/2$ and $0$, which
correspond to the finite dimensional representations
 $D^{j-1/2}$ and $D^{j}$, respectively.

\end{itemize}

In all the described cases, the structure of the metric is
consistent with the properties of the  ${\cal
C}_{\alpha,n}^2$ in (\ref{AA7}).

%%%%%%%%%%%%%%%%%%%%%%%%%%%%%%%%%%%%%%%%%%%%%%%%%%%%%
%%%%%%%%%%%%%%%%%%%%%%%%%%%%%%%%%%%%%%%%%%%%%%%%%%%%%%%%%%%%
\subsection{Representations with zero norm
states}\label{ZeroN}
%%%%%%%%%%%%%%%%%%%%%%%%%%%%%%%%%%%%%%%%%%%%%%%%%%%%%%%%%%%%
%%%%%%%%%%%%%%%%%%%%%%%%%%%%%%%%%%%%%%%%%%%%%%%%%%%%%

Consider now the \emph{infinite-dimensional}
representation with $\alpha=-j$, which we denote here by
$\widetilde{D^+_{-j}}$. The necessity to treat such
peculiar representations comes from their appearance in the
extended realization which will be
considered in Section \ref{solBFA}.

The representation $\widetilde{D^+_{-j}}$ is characterized by
the presence of an infinite number of zero norm states of
the form (\ref{AA4}) with $n=2j+1+k$, $k=0,1,\ldots$. In
such a representation, there is a peculiar \emph{zero norm}
state $\widetilde{|2j+1\rangle}={\cal
J}_+^{2j+1}|0\rangle$, which by (\ref{AA2})
and  (\ref{AA5}) satisfies
\begin{equation}\label{AA8}
    {\cal J}_-\widetilde{|2j+1\rangle}=0\,,\qquad
    {\cal
    J}_0\widetilde{|2j+1\rangle}=(j+1)
    \widetilde{|2j+1\rangle}.
\end{equation}
Due to (\ref{AA8}), the infinite set of zero norm states
$\widetilde{|2j+1+k\rangle}={\cal
J}_+^k\widetilde{|2j+1\rangle}$, $k=0,1\ldots$, span a
space invariant under the action of $\mathfrak{so}(2,1)$,
i.e., these states form an irreducible infinite-dimensional
null subspace which we denote by ${\cal D}^{+0}_{j+1}$.
All the infinite-dimensional representation
$\widetilde{D^+_{-j}}$ spanned by the states (\ref{AA4})
with $n=0,1,\ldots$, is, therefore, reducible.

Notice that the invariant null subspace ${\cal
D}^{+0}_{j+1}$ may appear here because the value
$-(-j)((-j)-1)=-j^2-j$ of the Casimir operator (\ref{AA6})
at $\alpha=-j$ admits, by (\ref{AA8}), also the alternative
factorization, $-j^2-j=-(j+1)((j+1)-1)$.

Since the zero norm states $\widetilde{|2j+1+k\rangle}$,
$k=0,1,\ldots$ are orthogonal to all states
$\widetilde{|n\rangle}$, $n=0,1,\ldots$, we can work with
equivalence classes, viewing
$|\Psi\rangle$ and $|\Psi\rangle+|\varphi\rangle$ where
$|\Psi\rangle\in \widetilde{D^+_{-j}}\,\,$ and
$|\varphi\rangle\in {\cal D}^{+0}_{j+1}$  as
equivalent \footnote{Mathematicians call  $\widetilde{D^+_{-j}}$
a ``Verma module''.}.
 In particular,  we consider any state
$|\varphi\rangle$ equivalent to the zero state. In such
a way, we reduce the peculiar infinite-dimensional
representation $\widetilde{D^+_{-j}}$ to the
$(2j+1)$-dimensional non-unitary irreducible representation
$\tilde{D}^j$ described above, i.e., we get
$\widetilde{D^+_{-j}}/{\cal D}^{+0}_{j+1}=\tilde{D}^j$.

%%%%%%%%%%%%%%%%%%%%%%%%%%%%%%%%%%%%%%%%%%%%%%%%%%%%%%%%%%%%%%%%%%%%%%%%%%%%%%%%%%%%
%%%%%%%%%%%%%%%%%%%%%%%%%%%%%%%%%%%%%%%%%%%%%%%%%%%%%%%%%%%%%%%%%%%
\section{Solutions of the wave equations. Anyon interpolation
between bosons and fermions}\label{solutions}
%%%%%%%%%%%%%%%%%%%%%%%%%%%%%%%%%%%%%%%%%%%%%%%%%%%%%%%%%%%%%%%%%%%
%%%%%%%%%%%%%%%%%%%%%%%%%%%%%%%%%%%%%%%%%%%%%%%%%%%%%%%%%%%%%%%%%%%%%%%%%%%%%%%%%%%%%

Having in mind the three essentially different types of
irreducible representations, we analyze first the solutions
of our vector system in the minimal realization
$\emph{\ss}=\alpha$,  assuming that the wave function
$\psi$ carries irreducible representation ${\cal
D}^+_\alpha$, whose type is defined by the value of the
parameter $\alpha$.

The wave function in \eqref{VmuEq} can be decomposed into partial waves,
\begin{equation}
    \psi(x)=\sum_{n=0}^r
    \psi_n(x)\vert n)\,,\label{wave}
\end{equation}
where $r=\infty$ for the representations $D^+_\alpha$ and
$\tilde{D}^+_\alpha$, and $r=2j$ for the
finite-dimensional representation $\tilde{D}^j$
($\alpha=-j$).

Having chosen an irreducible representation,
(\ref{Cas+aux3}) is satisfied identically.  The  three
equations in \eqref{VmuEq} are not independent: choosing
any two of them, the third one follows as a consequence.

  It is convenient to choose equations
\eqref{VmuEq} with $\mu=1,2$. Their complex linear combinations
$V_\pm\psi=0$, $V_\pm=V_1\pm i V_2$,
 give then the system of two
equations
\begin{equation}\label{V+-}
     \left[(\alpha - \mathcal{J}_0 )P_+
     +(m+ P_0) \mathcal{J}_+\right]\psi=0\,,\qquad
     \left[(\alpha + \mathcal{J}_0 )P_-
     +(m- P_0) \mathcal{J}_-\right]\psi=0\,,
\end{equation}
where $P_\pm=P_1\pm iP_2$. Consistently with (\ref{VV1}),
these two equations generate the
  one with $\mu=0$ as an integrability
condition,
\begin{equation}\label{V0eq}
  \left[\alpha P_0+m{\cal J}_0+\frac{1}{2}
  (P_-{\cal J}_+ -P_+{\cal J}_-)\right]\psi=0\,.
\end{equation}
 With (\ref{wave}),  (\ref{Coeff}) and (\ref{D+}),
from (\ref{V+-}) we get the equivalent system
\begin{eqnarray}
    &\sqrt{n+2\alpha}\,(m-P^0)\psi_n-
    \sqrt{n+1}\,P_+\psi_{n+1}=0\,,&\label{HP1}
    \\[6pt]
    &\sqrt{n+2\alpha}\,P_-\psi_n+\sqrt{n+1}\,
    (m+P^0)\psi_{n+1}=0\,.&\label{HP2}
\end{eqnarray}
Then some algebraic manipulations yield the component
form of Eq. (\ref{V0eq}),
\begin{equation}\label{V0comp}
    \left(\alpha P_0+m(\alpha+n)\right)\psi_n
    +\frac{1}{2}\left(\sqrt{(2\alpha +n-1)n}\,P_-\psi_{n-1}-
    \sqrt{(2\alpha+n)(n+1)}\,P_+\psi_{n+1}\right)=0.
\end{equation}

Eqns (\ref{HP1}), (\ref{HP2}) are conveniently
analyzed in
the momentum representation. For positive energy, $P^0>0$,
the operator $P^0+ m\neq 0$ is invertible. Then \eqref{HP2}
allows us to write all partial wave components $\psi_n$
in terms of $\psi_{0}$. From \eqref{HP2} one obtains, first,
\begin{equation}
    \psi_{n+1}=-\frac{\sqrt{n+2\alpha}}{\sqrt{n+1}}\,
    \frac{P_-}{P^0+m}\,\psi_n,\qquad
    P^0>0,\label{n+1n+}
\end{equation}
and then iteratively,
\begin{equation}
    \psi_{n}=\sqrt{{\cal C}_n}\,
    \left(\frac{-P_-}{P^0+m}\right)^n\psi_0\,,
    \label{npsi}
\end{equation}
\begin{equation}\label{Cnbeta}
    {\cal C}_n=
    \frac{2\alpha \,(2\alpha+1)\ldots (2\alpha+n-1)}{n!}=
    (nB(2\alpha,n))^{-1}\,,
\end{equation}
where $B(x,y)=\Gamma(x)\Gamma(y)/\Gamma(x+y)$ is Euler's
beta function.

Note for further reference that if $2\alpha$ is a negative
integer, then  the coefficients  $C_n$ vanish when
$n\geq1-2\alpha$~: the infinite tower of the $\psi$'s
terminates, and the representation space becomes
\emph{finite dimensional}. Alternatively,
 the beta function in (\ref{Cnbeta}) has
poles at negative integers.

The mass-shell condition follows from Eqns.
(\ref{V+-}), and so the component $\psi_0$ has the form
\begin{equation}\label{psi0P}
    \psi_0=C\,
    \delta(P^0-\sqrt{\vec{p}{\,}^2+m^2})\,\delta(\vec{P}-
    \vec{p})\,,
\end{equation}
where $C$ is a (normalization) constant and $\vec{p}$ is the
momentum of the state.

Similarly, for negative energy, ${P^0-m}$ is invertible,
allowing us
 to express $\psi_{n}$ in terms of
$\psi_{n+1}$,
\begin{equation}%\label{}
    \psi_{n}=\frac{\sqrt{n+1}}{\sqrt{n+2\alpha}}\,
    \frac{P_+}{m-P^0}\,\psi_{n+1},
    \qquad P^0<0. \label{n+1n-}
\end{equation}
All components can be expressed in terms of the highest
spin state component. The highest spin state exists,
however,  \emph{only} for the finite dimensional
representations $\tilde{D}^j$, in which case
$\beta=\alpha=-j$, and we get
\begin{equation}
    \psi_{2j-n}=i^{-n}\sqrt{\frac{2j(2j-1)\ldots
    (2j-n+1)}{n!}}\,
    \left(\frac{P_+}{P^0-m}\right)^n \,
    \psi_{2j}\,,
    \label{npsibis}
\end{equation}
where
\begin{equation}\label{psi2jP}
    \psi_{2j}=C\,
    \delta(P^0+\sqrt{\vec{p}{\,}^2+m^2})\,\delta(\vec{P}-
    \vec{p})\,.
\end{equation}

Let us now analyze the positive--energy solutions given by
Eqns. (\ref{npsi}), (\ref{Cnbeta}), (\ref{psi0P}). In the
rest frame $\vec{p}=0$, only the lowest component
$\psi_0$ is nontrivial. For $\vec{p}\neq 0$, every
subsequent component includes an additional kinematical
factor
$P_-/(P^0+m)$.

 For the infinite dimensional
representations $D^+_\alpha$ (and in the generic case of
non-unitary representations $\tilde{D}^+_\alpha$), the numerical
coefficient ${\cal C}_n$ is of order $1$ when $n$ tends to infinity.

 However, in the case of non-unitary
representations $\tilde{D}^+_\alpha$ with  $\alpha$  close
to a negative (half)integer $j$, the nature of coefficients
(\ref{Cnbeta}) is essentially different. When $\alpha$,
supposed to satisfy the relation (\ref{alpha<-j}), tends to
$-j$ from above, the first coefficients with
$n=1,\ldots,2j$ tend to the values of those that correspond
to the finite dimensional case $\tilde{D}^j$. In
particular, ${\cal C}_{2j}\sim j+\alpha \rightarrow 0$ when
$\alpha\rightarrow -j$. In such a case, all higher
components are suppressed by the same factor
$(j+\alpha)^{1/2}\rightarrow 0$.

 With this picture, having also in mind the  metric
 described in Section \ref{irreps},
we conclude that \emph{ the infinite-dimensional
half-bounded non-unitary representations
$\tilde{D}^+_\alpha$ interpolate between boson and fermion
states} of positive energy. This is what corresponds,
intuitively, to an \emph{anyon}.

 It is
worth noting that when $\alpha$ is between $-j$ and
$-j+\frac{1}{2},\, j=1/2,1,3/2$ cf.
(\ref{alpha<-j}) and approaches $-j+\frac{1}{2}$ from
below, the structure
 with Poincar\'e spin $s=-j+\frac{1}{2}$
 is recovered~: it is described by the
finite-dimensional non-unitary representation
$\tilde{D}^{j-1/2}$ and by the  indefinite metric
(\ref{etametric}), with $j$ changed into $j-1/2$.

The spin zero case $\emph{\ss}=\alpha=0$, excluded so far,
 can be incorporated  as the
limit $\alpha\rightarrow 0$ of  ${\cal
J}_\mu\in\tilde{D}_\alpha$, $\emph{\ss}=\alpha<0$.

%%%%%%%%%%%%%%%%%%%%%%%%%%%%%%%%%%%%%%%%%%%%%%%%%%%%%%%%%%%%%%%%%%%%%%%%%%%%%%%%%%%
%%%%%%%%%%%%%%%%%%%%%%%%%%%%%%%%%%%%%%%%%%%%%%%%%%%%%%%%%%%%%%%%%%%
\section{Extended formulation. Bosons and fermions from anyons}\label{solBFA}
%%%%%%%%%%%%%%%%%%%%%%%%%%%%%%%%%%%%%%%%%%%%%%%%%%%%%%%%%%%%%%%%%%%
%%%%%%%%%%%%%%%%%%%%%%%%%%%%%%%%%%%%%%%%%%%%%%%%%%%%%%%%%%%%%%%%%%%%%%%%%%%%%%%%%%%%%

 In the general case ${\cal J}_\mu$ can be taken
as the sum of the generators of the $(2+1)D$ Lorentz group in
representations that correspond to the series
$D^\pm_\alpha$,  $\tilde{D}^\pm_\alpha$, and to
$\tilde{D}^j$. The examples of Majorana-Dirac and
Jackiw-Nair anyon fields considered in Sections
\ref{MajDirSub} and \ref{JNSub} are just particular cases
of such  an \emph{extended} formulation (or, realization)
of the vector set of equations.

Taking, for simplicity,
 just two representations,
\begin{equation}\label{J1J2}
    {\cal J}^\mu=J_\mu+J'_\mu,\quad
    J^2 =-\alpha(\alpha-1),\quad
    J'^2 =-\alpha'(\alpha'-1)\,,
\end{equation}
and put
\begin{equation}\label{ssalpha2}
    \emph{\ss}=\alpha+\alpha'
\end{equation}
where, for a finite-dimensional representation, the
parameter $\alpha$ and/or $\alpha'$ is $-j$, and
(\ref{ssalpha2}) such that $\emph{\ss}\neq 0$. Then Eq.
(\ref{Cas+aux3}) transforms into
\begin{equation}\label{JJpsi}
    \left(JJ'+\alpha\alpha'\right)\psi=0\,.
\end{equation}
Eq. (\ref{JJpsi}) is a non-dynamical, subsidiary equation
for the field $\psi$ [which carries two indices not shown
explicitly]. It, in turn, implies that our composite system
 (\ref{J1J2}), (\ref{ssalpha2}) carries an
\emph{irreducible} representation of the
$\mathfrak{so}(2,1)$-spin $\emph{\ss}=\alpha+\alpha'$,
consistently
 with Eq. (\ref{Cas+aux3}).
This means that the extended formulation, in fact, recovers
the previously considered minimal realization. It provides
us, however, with a new possibility: in this formulation
usual  \emph{boson and fermion fields of arbitrary spin can
be composed from anyons}.\vskip2mm

Before considering concrete  examples, some
general comments are in order. \vskip2mm

$\bullet$ The operator of the non-dynamical equation
(\ref{JJpsi}) commutes with the total vector spin generator
${\cal J}_\mu$. Solutions of (\ref{JJpsi}) then can be
given by the eingestates $\widetilde{|k\rangle}$ of the
compact $\mathfrak{so}(2,1)$ generator ${\cal
J}_0=J_0+J'_0$, ${\cal
J}_0\widetilde{|k\rangle}=(\emph{\ss}+k)\widetilde{|k\rangle}$,
$k=0,\ldots$, presented as a certain linear combinations of
the simultaneous $J_0$ and $J'_0$ eigenstates, $|n)|n')$.
Particularly, the state ${|0\rangle}=|0)|0')$ is a solution
that satisfies relations of the form (\ref{AA3}) with
$\alpha$ changed into $\emph{\ss}=\alpha+\alpha'$
[ for the sake of definiteness we assume
here that $J_\mu\in {\cal
D}^+_\alpha $, $J'_\mu\in {\cal D}^+_{\alpha'}$].
${\cal J}_\mu$ commutes with the scalar operator
$JJ'+\alpha\alpha'=-J_0J_0+\frac{1}{2}(J_+J'_-+J_-J'_+)+\alpha\alpha'$;
all other solutions $\widetilde{|k\rangle}$ of Eq.
(\ref{JJpsi}) are obtained therefore by subsequent
application of ${\cal J}_+$ to the lowest weight state
${|0\rangle}$ cf. Section \ref{highlow}.

For $J_\mu\in \tilde{D}^j$ and $J'_\mu\in\tilde{D}^{j'}$,
Eq. (\ref{JJpsi}) singles out irreducible finite
dimensional representation  $\tilde{D}^{j+j'}$.

For $\emph{\ss}\neq -j$,  Eq. (\ref{JJpsi}) separates an
irreducible bounded from below infinite dimensional
representation of the unitary or non-unitary type  depending
on $\emph{\ss}>0$ or $\emph{\ss}<0$.

When $\emph{\ss}=-j$ and at least one of the
$\mathfrak{so}(2,1)$ spins is in an infinite dimensional
representation, Eq. (\ref{JJpsi}) gives a reducible
representation $\widetilde{D^+_{-j}}$ with invariant
infinite dimensional null subspace ${\cal D}^{+0}_{j+1}$.
As it is explained in Section \ref{ZeroN},
working modulo
zero norm states $|\varphi\rangle \in{\cal D}^{+0}_{j+1}$,
solutions of (\ref{JJpsi}) gives rise to the irreducible
finite-dimensional non-unitary representation
$\tilde{D}^j=\widetilde{D^+_{-j}}/{\cal D}^{+0}_{j+1}$.

As a result, the solution of the vector system of equations
in the extended formulation, for all possible choices of the
 $\mathfrak{so}(2,1)$ representations  in
(\ref{J1J2}) is reduced to the minimal
realization\footnote{The spin zero case $s=0$ is
incorporated into the extended formulation via a limit
$\emph{\ss}\rightarrow 0$ if representations in
(\ref{J1J2}) are chosen in such a way that
$-1/2<\emph{\ss}<0$.}. \vskip2mm

$\bullet$ When ${\cal J}_\mu$ is  composed as in
(\ref{J1J2}), the vector set of equations (\ref{VmuEq})
splits, for each Lorentz index, into two vector sets.

Consider, for instance, $J_\mu\in D^+_\alpha$, $J'_\mu\in
\tilde{D}^j$, assuming, as in the Majorana-Dirac and
Jackiw-Nair cases, (\ref{Maj-Dir}) and (\ref{JNj1}), that
$\emph{\ss}\neq 0$. In the rest frame the system (\ref{J0})
-- (\ref{J+-}) becomes
\begin{equation}\label{JJJ0}
    \left((J_0-\varepsilon \alpha)+(J'_0+\varepsilon
    j)\right)\psi=0\,,
\end{equation}
\begin{equation}\label{JJJ12}
    \left((J_1-i\varepsilon J_2)+(J'_1-i\varepsilon
    J'_2)\right)\psi=0\,.
\end{equation}
Eq. (\ref{JJJ12}) has nontrivial solutions if and only if
$\left(J_1-i\varepsilon J_2\right)\psi=0$. Such a solution,
proportional to the lowest state $\vert \alpha,0\rangle$
for the representation $D^+_\alpha$, is nontrivial only for
$\varepsilon=+1$. Then Eq. (\ref{JJJ12}) splits into
\begin{eqnarray*}%\label{JJJD+}
    \left(J_1-iJ_2\right)\psi=0\,,
\quad\hbox{and}\quad
%\begin{equation}\label{JJJDj}
    \left(J'_1-i J'_2\right)\psi=0\,,
\end{eqnarray*}
splitting Eq. (\ref{JJJ0}) into
\begin{eqnarray*}%\begin{equation}\label{JJJ0D+}
    \left(J_0-\alpha\right)\psi=0\,,
\quad\hbox{and}\quad
    \left(J'_0+ j\right)\psi=0\,.
\end{eqnarray*}%\end{equation}
Then one finds that a solution of (\ref{VmuEq}) has to be a
simultaneous solution of two sets of equations of the form
(\ref{VmuEq}), in one of which ${\cal J}_\mu$ is  ${
J}_\mu\in D^+_\alpha$, and in the other ${\cal J}_\mu$ is
${J}'_\mu\in \tilde{D}^j$. The corresponding solution has
positive energy, mass $m$ and spin $s=\alpha-j\neq 0$.
 This means that the system of equations,
\begin{equation}\label{VV}
    V_\mu^{(\alpha)} \psi=0,\qquad V_\mu^{\alpha'}
    \psi=0,\qquad J J'+\alpha \alpha'=0,
\end{equation}
is equivalent to \eqref{VmuEq} with $\emph{\ss}= \alpha + \alpha'$.
In other words,  spins simply add.
This additive property allows us to create particles of spin $\emph{\ss}$
built from one of spin $\alpha$ and another one of spin $\alpha'$.
Intuitively, the non-dynamical constraint  \eqref{JJpsi}
 ``entangles'' the component systems.

In the Majorana-Dirac and Jackiw-Nair
examples discussed in Section \ref{nonsusy}, the
entangling equation (\ref{JJpsi}) takes the form
(\ref{Jgal}) and (\ref{J1split}), respectively.
\vskip2mm

$\bullet$ Choosing $J_\mu$ in a unitary
infinite-dimensional representation $D^+_\alpha$ with
non-(half)integer $\alpha$, and $J'_\mu$ in a
finite-dimensional non-unitary representation $\tilde{D}^j$
in such a way that $\emph{\ss}=\alpha-j<0$, the resulting
irreducible representation is an infinite-dimensional
half-bounded \emph{non-unitary} representation of
$\mathfrak{so}(2,1)$. Such infinite-dimensional non-unitary
representations belong to the Majorana-Dirac an
Jackiw-Nair descriptions, no attention
was paid to the corresponding interpolating anyons,
in the original References \cite{JN1,Pl1991}
though. \vskip2mm
\goodbreak

$\bullet$ The choices $J_\mu\in \tilde{D}^+_\alpha$,
$J'_\mu\in \tilde{D}^+_{\alpha'}$ or $J'_\mu\in
{D}^+_{\alpha'}\,$, provide us with another interesting
case~: if the parameters $\alpha$ and $\alpha'$ add up to a
negative (half)-integer $\emph{\ss}=-j$,  our extended
formulation, based on two half-bounded infinite-dimensional
(i.e. anyon-like) representations, describes a usual
relativistic finite-component field of spin $j$. In
particular, from two anyon-like representations given by
$\alpha=\alpha'=-1/4$ [considered below],
the extended formulation reproduces the theory of the Dirac
particle with spin $s=-1/2$.

%%%%%%%%%%%%%%%%%%%%%%%%%%%%%%%%%%%%%%%%%%%%%%%%%
%%%%%%%%%%%%%%%%%%%%%%%%%%%%%%%%%%%%%%%%%%%%%%%%%%%%%%%%%%%%%
%\subsection
\kikezd{Examples}\label{ExtEx}
%%%%%%%%%%%%%%%%%%%%%%%%%%%%%%%%%%%%%%%%%%%%%%%%%%%%%%%%%%%%%
%%%%%%%%%%%%%%%%%%%%%%%%%%%%%%%%%%%%%%%%%%%%%%%%%
\vskip1mm

Consider some representative particular
examples.\vskip0.2cm

$\bullet$ Let $J_\mu,\, J'_\mu\in \tilde{D}^{1/2}$\,,
$\alpha=\alpha'=-1/2$, $\emph{\ss} =-1$. The solutions of Eq.
(\ref{JJpsi}) are
\begin{equation}\label{1-21-2}
    |0)=|0)|0')\,,\qquad
    |1)=\frac{1}{\sqrt{2}}\left[|0)|1')+|1)|0')\right]\,,
    \qquad
    |2)=|1)|1')\,.
\end{equation}
They are eigenstates of ${\cal J}_0$ with eigenvalues $-1$,
$0$ and $+1$, respectively, and satisfy the relations
\begin{eqnarray}\label{spin1}
    &{\cal J}_-|0)={\cal J}_+|2)=0\,,\qquad
    {\cal J}_+|0)=i\sqrt{2}\,|1)\,,\qquad
    {\cal J}_+|1)=i\sqrt{2}\,|2)\,,& \\
    &(n|n')=(-1)^n\delta_{nn'}\,,\qquad n,\,n'=0,1,2\,,&\label{spin1'}
\end{eqnarray}
[where we have used Eq. (\ref{Coeff})]  which correspond to the
representation $\tilde{D}^1$. The vector set of equations
\eqref{VmuEq} describes, in this case, a topologically massive
vector gauge field of mass $m$ and spin $s=-1$, which can
be treated as composed of two massive particles
of spin $s=-1/2$ each.

$\bullet$  Let us take
$(J_\mu)^\lambda{}_\nu=-i\epsilon^\lambda{}_{\mu\nu}\in
\tilde{D}^{1}$, $\alpha=-1$,
$J'_\mu=-\frac{1}{2}\gamma_\mu\in \tilde{D}^{1/2}$,
$\alpha'=-1/2$, and $\emph{\ss} =-3/2$. This
corresponds to the Rarita-Schwinger field of spin
$s=-3/2$~\footnote{The chosen representation for
$\alpha=-1$  differs from the matrix representation
of Section \ref{highlow} by a unitary
transformation, see Eqns. (\ref{finiteJ2}), (\ref{U1}) and
(\ref{s=1usual}) in  Appendix \ref{DDJTsusy}.}. Having in
mind the splitting of our vector equations, we find that the
vector-spinor field $\psi^\mu$ [spinor
indices are not display explicitly] satisfies the Dirac and DJT equations,
supplemented with the entangling equation (\ref{JJpsi}),
\begin{eqnarray}
    &(P\gamma - m)\psi^\mu =0\, ,&\label{RS1}
    \\
     &\left(-i\epsilon^\mu{}_{\nu\lambda}P^\nu+
    m\delta^\mu_\lambda\right)\psi^\lambda
    =0\,,&\label{RS2}
    \\
    &
    \Big(i\epsilon^\mu{}_{\nu\lambda}\gamma^\nu
    +\delta^\mu_\lambda\Big)\psi^\lambda
    =0\,.&
    \label{RS3}
\end{eqnarray}
Taking into account the identity
$\gamma_\mu\gamma_\nu=-\eta_{\mu\nu}+i\epsilon_{\mu\nu\lambda}\gamma^\lambda$,
simple algebraic manipulations show that the system
(\ref{RS1})--(\ref{RS3}) is equivalent to the
Rarita-Schwinger equations
\begin{equation}\label{gappapsim}
    (P\gamma - m)\psi^\mu =0\,,\qquad
    \gamma_\mu \psi^\mu=0\,.
\end{equation}
Thus the system (\ref{gappapsim}) generates, in itself, the
DJT,  and also the entangling, (\ref{RS3}), equations.
Notice a similarity between the second equation in
(\ref{gappapsim}) and Eq. (\ref{Jpsi=0}).

$\bullet$ Consider now  the case $J_\mu\in
\tilde{D}^+_{-1/4}$, $J'_\mu\in \tilde{D}^+_{-1/4}\,$,
$\alpha=\alpha'=-1/4$, $\emph{\ss} =-1/2$. The two lowest
normalized eigenstates of ${\cal J}_0$ are given formally
by the same relations as in (\ref{1-21-2}).
It should be
remembered, though, that the states which appear on
the right hand sides of the relations  belong to  two copies of the
infinite dimensional representation $\tilde{D}^+_{-1/4}$. The
 normalization is given by the indefinite metric
described in Section \ref{highlow}.
 These states $|0)$ and
$|1)$ have ${\cal J}_0$-eigenvalues $-1/2$ and $+1/2$,
respectively,  and satisfy the relations
\begin{equation}\label{s1-2}
    (n|n')=(-1)^n\delta_{nn'}\,,\quad
     n,n'=1,2\,, \qquad
     {\cal J}_+|0)=i|1)\,.
\end{equation}
Application of the ladder operator ${\cal J}_+$ to $|1)$
does not annihilate it, but produces a zero-norm state,
$$
    {\cal J}_+|1)=\frac{1}{\sqrt{2}}\left[|0)|2')+|2)|0')
    \right]+i|1)|1')=\widetilde{|2\rangle}\,,\quad
    \widetilde{\langle 2}|\widetilde{2\rangle}=0\,,
$$
which is orthogonal to $|0)$ and $|1)$, and is annihilated
by ${\cal J}_-$.

 Other solutions of Eq. (\ref{JJpsi}) are
obtained by applying, subsequently, ${\cal J}_+$ to the state
$\widetilde{|2\rangle}$, which, together with the latter,
span the null space ${\cal D}^{+0}_{1/2}$\,. Factoring out
the states $|\varphi\rangle\in {\cal D}^{+0}_{1/2}$\,, we
get the two-dimensional non-unitary representation
$\tilde{D}^{1/2}$. In such a way two anyons of identical $\mathfrak{so}(2,1)$
spin $-1/4$, yield, in our extended
realization, a fermion field of spin $-1/2$\,. By
this reason, in our picture, these anyons \emph{can} be called
\emph{semions}.\vskip2mm

The picture we have just described can be reinterpreted in
the following way. Consider a pair of positive energy
states (particles) with Poincar\'e spins $-1/4$.
Separately, each such particle can be described within
the minimal realization with $J_\mu\in \tilde{D}^+_{-1/4}$,
$\emph{\ss}=\alpha=-1/4$. The wave function of each
particle picks up a phase, $e^{-i\pi/2}=-i$, under a $2\pi$
spatial rotation generated by $J_0$, and the system can be
interpreted as a pair of \emph{semions}.

When switching to the
extended formulation with $J_\mu\in \tilde{D}^+_{-1/4}$,
$J'_\mu\in \tilde{D}^+_{-1/4}\,$, $\alpha=\alpha'=-1/4$,
$\emph{\ss} =-1/2$, the vector system of equations splits,
as discussed above, into two vector systems of equations,
each describing a semion field. These two fields are not more
independent, however. The vector system implies the non-dynamical
equation (\ref{JJpsi}) that guarantees the irreducibility
of the Poincar\'e representation realized on our composed,
two-semion system. This equation introduces a kind of
entanglement between the two semion states, that provides
us, as a result, with a spin $s=-1/2$ fermion. Each
independent semion is described by an infinite-component
field in a frame different from the rest frame [with
higher components suppressed by a kinematical factor, see
Eq. (\ref{npsi})]. Equation (\ref{JJpsi}) then introduces a
kind of destructive interference between the semion wave
functions, with an infinite number of components
associated with the null subspace, with the exception of two
surviving independent wave components, which, due to
the entanglement, describe a two-component positive-energy fermion state.

Notice here that if we choose instead two unitary infinite
dimensional representations, $J_\mu\in {D}^+_{+1/4}$,
$J'_\mu\in {D}^+_{+1/4}\,$, $\alpha=\alpha'=+1/4$,
$\emph{\ss} =+1/2$, our vector set of wave equations would
describe a positive-energy particle of mass $m$  and spin
$+1/2$. Although its wave function  picks up a phase $-1$
under a $2\pi$ rotation, we do not get a usual fermion,
since its wave function has infinite number of components
when $\vec{p}\neq 0$, cf. the previous Section. Anyons
described by the unitary representations ${D}^+_{+1/4}$ can
not be referred to, therefore, as semions~\footnote{The
name `semions' cannot be applied therefore to the
supersymmetric system constructed in \cite{SorVol}, based
on the two infinite dimensional representations
${D}^+_{+1/4}$ and ${D}^+_{+3/4}$\,; the term `quartions'
used alternatively seems to be more appropriate.}.

In the same way, if we add two spins corresponding to
infinite dimensional anyon representations such that
$\alpha+\alpha'=\emph{\ss}=-j$, we can treat the resulting
boson or fermion field as composed of two anyons.

$\bullet$ Finally, we consider shortly another
particular interesting case, $J_\mu\in {D}^+_{1/2}$,
$J'_\mu\in \tilde{D}^1$, $\alpha=1/2$, $\alpha'=-1$,
$\emph{\ss}=-1/2$. In our framework, it describes a
usual fermion field~\footnote{This case formally
corresponds to the Jackiw-Nair scheme, no attention
was paid to it in  \cite{JN1}, though.}.

According to Eq. (\ref{Coeff}), the nonzero coefficients
are  $C^{-1}_0=C^{-1}_1=i\sqrt{2}$, and $C^{1/2}_n=n+1$.
The $-1/2$ and $+1/2$ eigenstates of ${\cal J}_0$ are
\beq
|0)=|0)|0')
\quad\hbox{and}\qquad
 |1)={\cal
J}_+|0)=|1)|0')+i\sqrt{2}\,|0)|1'),
\eeq
 which satisfy the
relations (\ref{s1-2}). The lowest zero-norm state is given
by
\beq
{\cal J}_+|1)=\widetilde{|2\rangle}
\quad\hbox{where}\quad
\widetilde{|2\rangle}=2\left[ |2)|0')+i\sqrt{2}\,
|1)|1')-|0)|2')\right].
\eeq
${\cal
J}_-\widetilde{|2\rangle}=0$.
Higher zero norm states
of the null subspace ${\cal D}^{+0}_{1/2}$ are produced by
applying the ladder operator ${\cal J}_+$ to
$\widetilde{|2\rangle}$.

%%%%%%%%%%%%%%%%%%%%%%%%%%%%%%%%%%%%%%%%%%%%%%%%%%%%%%%%%%%%%%%%%%%%%%%%%%%%%%%%%%%
%%%%%%%%%%%%%%%%%%%%%%%%%%%%%%%%%%%%%%%%%%%%%%%%%%%%%%%%%%%%%%%%%%
\section{Deformed oscillator representation of
$\mathfrak{osp}(1|2)$}\label{RDHA}
%%%%%%%%%%%%%%%%%%%%%%%%%%%%%%%%%%%%%%%%%%%%%%%%%%%%%%%%%%%%%%%%%%%%%%%%%%%%%%%%%%%%
%%%%%%%%%%%%%%%%%%%%%%%%%%%%%%%%%%%%%%%%%%%%%%%%%%%%%%%%%%%%%%%%%%%

To get a concrete realization of the supersymmetric
extensions of our model we shall discuss later, we use the
reflection-deformed Heisenberg algebra (RDHA). The latter
was (implicitly) introduced by Wigner \cite{Wig} and led
subsequently to parastatistics \cite{para}, that condensed,
finally, into QCD color \cite{GreenColor}. To make our
discussion self-contained, we outline below those
representations of the RDHA which are necessary for our
purposes. See  Ref. \cite{RDHA}\footnote{In the context of
quantum mechanical supersymmetry this algebra was used in
\cite{MPAnn,P_para_susy}.} for more details.

Let us consider the reflection-deformed Heisenberg algebra
of the harmonic oscillator
\begin{equation}\label{DHAR}
    \big[a^-,a^+\big]=1+\nu R\, ,  \qquad  R^2=1\,,
    \qquad \{a^{\pm},R\}=0\,,
\end{equation}
where $\nu$ is a real deformation parameter and  $R$ is the
reflection operator, $R =(-1)^{\cal N}=\exp(i\pi {\cal
N})$. The number operator
\begin{equation}\label{Naa}
    {\cal N}=\frac{1}{2}\{a^+,a^-\}-\frac{1}{2}(\nu+1)\, ,
    \qquad [{\cal N},a^\pm]=\pm
    a^\pm\, ,
\end{equation}
defines the Fock space,
\begin{equation}\label{calF}
    \mathcal{ F}=\{|n\rangle, n=0,1,2,...\}\,,\quad
    {\cal N}|n\rangle=n|n\rangle\,,
\end{equation}
where $|n\rangle=C_n(a^+)^n|0\rangle$, $n=0,1,\ldots,$
$a^-|0\rangle=0$, $\langle 0|0\rangle=1$,
\begin{equation}\label{Cn!}
    C_n=([n]_\nu!)^{-1/2}\,,\quad[0]_\nu !=1\,,
    \quad [n]_\nu !=\prod_{l=1}^n
    [l]_\nu\,,\quad  n\geq 1\,,\quad
    [l]_\nu=l+\frac{1}{2}(1-(-1)^l)\nu\, .
\end{equation}
The explicit  form of the coefficients (\ref{Cn!})
indicates that for $\nu>-1$ the algebra has
infinite-dimensional, parabosonic-type unitary irreducible
representations; for negative-odd values $\nu=-(2r+1)$,
$r=1,2,\ldots$, it has $(2r+1)$-dimensional, non-unitary
parafermionic-type representations, see below\footnote{The
Fock space construction here is similar to that in Section
\ref{highlow}. We refrain from repeating  similar technical
details related to the appearance of  null subspaces at
$\nu=-(2r+1)$.}. For $\nu<-1$, $\nu\neq -(2r+1)$, it has
infinite dimensional non-unitary representations of the
$\tilde{D}^\pm_\alpha$ type.

The quadratic operators
\begin{eqnarray}
    &{\cal J}_0=\frac{1}{4}\{a^+,a^-\},\qquad {\cal J}_\pm={\cal J}_1\pm
    i{\cal J}_2=\frac{1}{2}(a^\pm)^2,&\label{J0,Ji}
\end{eqnarray}
together with linear operators
\begin{equation}\label{Lalpha}
    \mathcal{L}_a=\left(\,\frac{a^++a^-}{\sqrt{2}},\,\,
    i\,\frac{a^+-a^-}{\sqrt{2}}\right),\qquad a=1,2\,,
\end{equation}
generate the $\mathfrak{osp}(1|2)$ superalgebra,
\begin{eqnarray}\label{osp12}
    &[{\cal J}_\mu,{\cal J}_\nu]=-i\epsilon_{\mu\nu\lambda}J^\lambda\,,
    &\label{Lorentz 2+1}\\[6pt]
    &\{\mathcal{L}_a,\mathcal{L}_b\}=4i ({\cal J}\gamma)_{ab}\,,\qquad
    [{\cal J}_{\mu},\mathcal{L}_{a}]= \frac{1}{2}(\gamma_{\mu})_{a}^{\
    b}\mathcal{ L}_{b}\,.&\label{Lorentz 2+1 a}
\end{eqnarray}
The Lorentz operators ${\cal J}_\mu$ are even, and
$\mathcal{L}_a$ are odd supercharges with respect to a
$\mathbb{Z}_2$-grading provided by the reflection operator,
$ [R,{\cal J}_\mu]=0$, $\{R,\mathcal{L}_a\}=0$. The $(2+1)$
dimensional Dirac matrices are in the Majorana
representation,
\begin{equation}\label{Dmatrices}
    (\gamma^0)_{a}^{\ b}= -(\sigma_2)_{a}^{\ b}, \qquad
    (\gamma^1)_{a}^{\ b}= i(\sigma_1)_{a}^{\ b}, \qquad
    (\gamma^2)_{a}^{\ b}= i(\sigma_3)_{a}^{\ b},
\end{equation}
which satisfy,
\begin{equation}
 (\gamma_\mu)_a{}^\rho(\gamma_\nu)_\rho{}^b=
    -\eta_{\mu\nu}\epsilon_{a}{}^b+i\epsilon_{\mu\nu\lambda}
    (\gamma^\lambda)_a{}^b,
    \qquad
\gamma^\mu_{ab}=\gamma^\mu_{b a}, \qquad
\gamma^{\mu\dagger}_{ab}=-\gamma^\mu_{ab}.\label{Cliff}
\end{equation}
The antisymmetric tensor, $\epsilon^{ab}=-\epsilon^{b a}$
($\epsilon^{12}=1$), plays the role of a spinor metric,
rising and lowering indices, $A^a=\epsilon^{ab}A_b$,
$A_a=A^b\epsilon_{b a}$. The scalar product has the
property $A^{a} B_a=-A_a B^a $ for any spinors $A_a$ and
$B_a$. Then the Lorentz generators can be written  as
\begin{equation}\label{J mu}
    {\cal J}_\mu=\frac{i}{4} \mathcal{L}^a (\gamma_\mu)_a{}^b \mathcal{
    L}_b\,,\quad \mu=0,1,2\,.
\end{equation}
The representation of $\mathfrak{osp}(1|2)$ built in this
way is irreducible, characterized by the super-Casimir
operator
\begin{equation}
     \mathcal{ C}={\cal J}_\mu {\cal J}^\mu-\frac{i}{8}
     \mathcal{L}^a \mathcal{L}_a=\frac{1}{16}(1-\nu^2)\,.
\end{equation}
The representation of the Lorentz subalgebra is, however,
reducible, due to  supersymmetry. This is reflected
by the
eigenvalues of the Casimir of the Lorentz subalgebra,
\begin{equation}\label{JJ}
    {\cal J}_\mu {\cal J}^\mu=-\hat{\alpha}(\hat{\alpha}-1)\,
\quad\hbox{with}\quad
    \hat{\alpha}=\frac{1}{4}(2+\nu)-\frac{1}{4}R\,.
\end{equation}
The irreducible components are obtained by projecting to
the eigenspaces of $R$,
\begin{eqnarray}
    &R\,\mathcal{F}_\pm=\pm\mathcal{F}_\pm\,,&
    \label{calF+-}
    \\[6pt]
    &\mathcal{F}_+=\big\{|n\rangle_+=|2n\rangle, n=0,1,2,...\big\}\,,
    \quad
    \mathcal{F}_-=\big\{|n\rangle_- =|2n+1\rangle, n=0,1,2,...\big\}\,.  &
    \label{FFpm}
\end{eqnarray}
On these subspaces, the $\mathfrak{so}(2,1)$ Casimir
\eqref{JJ} takes the values
\begin{eqnarray}\label{J^2}
    {\cal J}_\mu {\cal J}^\mu\mathcal{ F}_\pm=-\alpha_\pm(\alpha_\pm-1)\mathcal{
    F}_\pm\,,\quad\hbox{where}\quad \alpha_+=\frac{1+\nu}{4}\,,\quad
    \alpha_-=\alpha_+ + \frac{1}{2}\,.
\end{eqnarray}
The irreducible representations of the Lorentz algebra are
extracted by the projectors,
$\mathcal{F}_\pm=\Pi_\pm\mathcal{F}$,
\begin{eqnarray}
    &J^{(\pm)}_\mu= {\cal J}_\mu \Pi_\pm\,,\qquad
    [J^{(\pm)}_\mu,J^{(\pm)}_\nu]=-i
    \epsilon_{\mu\nu}{}^\lambda J^{(\pm)}_\lambda\,,
    \qquad [J^{(+)}_\mu,J^{(-)}_\nu]=0\,,&
    \label{AJ+-}
    \\[6pt]
    &J^{(\pm)}_\mu J^{(\pm)}{}^\mu=-\alpha_\pm(\alpha_\pm-1){
    \Pi}_\pm\,,&
\end{eqnarray}
\begin{equation}\label{projector}
    \Pi_\pm=\frac{1}{2}(1\pm R)\,,
    \qquad \Pi_\pm^2=\Pi_\pm\,, \quad \Pi_+\Pi_-=0\,,\quad
    \Pi_++\Pi_-=1\,.
\end{equation}
The representation of ${\cal J}_\mu$ in \eqref{J0,Ji} is
therefore a direct sum, ${\cal J}_\mu=J^{(+)}_\mu +
J^{(-)}_\mu$. {}Consistently with \eqref{J0,Ji} and
(\ref{Naa}), ${\cal J}_0$  has eigenvalues $n+\alpha_\pm$\
,
\begin{equation}
    {\cal J}_0 |n\rangle_\pm =(n+\alpha_\pm)|n\rangle_\pm\,,
    \qquad |n\rangle_\pm  \in  \mathcal{F}_\pm\,.\label{J0n+-}
\end{equation}
For future use, it is convenient to introduce the
shifted number operator,
\begin{equation}
    \mathbf{N}={\cal J}_0-\hat{\alpha}\,,
    \qquad \mathbf{N}|n\rangle_\pm =n|n\rangle_\pm\,,
    \qquad \hat{\alpha}|n\rangle_\pm=
    \alpha_\pm|n\rangle_\pm\,.
\label{J0N+-}
\end{equation}

The ladder operators ${\cal J}_\pm={\cal J}_1\pm i {\cal
J}_2$ act as
\begin{eqnarray}
     &{\cal J}_+\vert n\rangle_\pm=C^{\alpha_\pm}_n\vert n+1\rangle_\pm\,,\quad
    {\cal J}_-\vert n\rangle_\pm=C^{\alpha_\pm}_{n-1}\vert n-1\rangle_\pm \,,
    \quad |n\rangle_\pm \in \mathcal{F}_\pm\,,\label{CnJ+-}&\\[6pt]
    & C^{\alpha_\pm}_n=\sqrt{(2\alpha_\pm+n)(n+1)}\,,\label{Cn++}
\end{eqnarray}
cf. (\ref{Coeff}). The rising/lowering operators
interchange the $\mathcal{F}_+$ and $\mathcal{F}_-$
subspaces, $a^\pm:\mathcal{F}_+ \leftrightarrow
\mathcal{F}_-$,
\begin{eqnarray}
 &a^+\vert n\rangle_+=\sqrt{2(n+2\alpha_+)}\, \vert n\rangle_-\,,\quad
    a^+\vert n\rangle_-=\sqrt{2(n+1)}\,\vert n+1\rangle_+\,,&\label{aaevod+}
    \\[6pt]
    &a^-\vert n\rangle_+=\sqrt{2n}\,\vert n-1\rangle_-\,,\quad
    a^-\vert n\rangle_-=\sqrt{2(n+2\alpha_+)}\,\vert n
    \rangle_+\, .&\label{aaevod-}
\end{eqnarray}
On the subspaces  $\mathcal{F}_\pm$ the $\mathfrak{so}(2,1)$ spin is shifted by
\beq
\alpha_--\alpha_+=1/2;
\label{halfshift}
\eeq
it is therefore legitimate to interpret the $a^\pm$ as
odd supercharges.

The deformed oscillator provides us with infinite-, and
also finite, dimensional representations of
$\mathfrak{osp}(1|2)$. Observe that, in \eqref{aaevod+} and
\eqref{aaevod-}, the coefficient $n+2\alpha_+$ vanishes
when $n$  takes the value $n=-2\alpha_+=-(1+\nu)/2$. This
happens for odd-integer negative values of the deformation
parameter, $\nu=-(2r+1),\, r=1,2,\ldots,$ for which $n=r$.
Then the Fock space becomes finite,
\begin{eqnarray}
    &\mathcal{F}=\{|0\rangle,
    |1\rangle,\dots,|2r\rangle\}=\mathcal{F}_+ + \mathcal{F}_-\,,&\\[6pt]
     &\mathcal{F}_+=\{|0\rangle_+=
     |0\rangle,\dots,
     |r\rangle_+=|2r\rangle\}\,,\quad \mathcal{F}_-=
     \{|0\rangle_-=|1\rangle,\dots,|r-1\rangle_-=|2r-1\rangle\}\,
     .&\nonumber
\end{eqnarray}
These spaces are characterized by the equations
$$
a^{-(2r+1)}=a^{+(2r+1)}=0, \quad a^-|0\rangle=0, \quad
a^+|2r\rangle=0, \quad dim\,(\mathcal{F})=|\nu|=2r+1.
$$
  The
spins carried by the subspaces $\mathcal{F}_+$ and
$\mathcal{F}_-$  are $\alpha_+=-r/2$ and
$\alpha_-=-r/2+1/2$,  respectively.

For $\nu>-1$, the spectrum of ${\cal J}_0$ is positive
definite and bounded from below, and the spin $\alpha_\pm$
takes only positive values. This produces the discrete
series ${D}^+_{\alpha_\pm}$ of $\mathfrak{so}(2,1)$. Every
series is generated by the operators \eqref{AJ+-}. The
discrete series bounded from above, ${D}^-_{\alpha_\pm}$,
of the Lorentz algebra are obtained by the external
automorphism $({\cal J}_0,{\cal J}_1,{\cal J}_2)\rightarrow
(-{\cal J}_0,{\cal J}_1,-{\cal J}_2)$, and projection to
the corresponding subspaces $\mathcal{F}_\pm$.

For $\nu<-1$, $\nu\neq -(2r+1)$, the RDHA provides
us also with infinite dimensional half-bounded non-unitary
irreducible representations $\tilde{D}^\pm_{\alpha_\pm}$ of $\mathfrak{so}(2,1)$.

For the finite dimensional representation, the operators
$a^+$ and $a^-$ are conjugate,
$(\Psi,a^-\Phi)^*=(\Phi,a^+\Psi)$, w.r.t. the scalar
product,
\begin{equation}
(\Psi,\Phi)=\bar{\Psi}_{n}\Phi^n, \quad
\bar{\Psi}_n=\Psi^{*k}\hat{\eta}_{kn}, \label{a4}
\end{equation}
provided by the matrix $\hat{\eta}=diag(+1, -1, -1, +1, +1,\ldots,
(-1)^{r},
    (-1)^{r})$,
where $\Phi^n=\langle n|\Phi\rangle$.

The two simplest examples of finite-dimensional
representations of the reflection-deformed Heisenberg
algebra (\ref{DHAR}), namely those of
$\alpha_+=-1/2,\,\alpha_-=0$, and of
$\alpha_+=-1,\,\alpha_-=-1/2$, are summarized in the
Appendix.

Note that for $\nu=-1$,  the algebra
(\ref{DHAR}) has a one-dimensional trivial representation, in
which $R=1$ and $a^\pm=0$. This corresponds
to the spin-$0$ representation of $\mathfrak{so}(2,1)$.

The algebra (\ref{DHAR}) admits a nonlinear realization
\cite{RDHA1},
\begin{equation}
    a^+ =\sqrt{1+\frac{\nu}{\tilde{\mathcal{N}}}}
    \;\,\tilde{a}^+\tilde{\Pi}_+ +
    \tilde{a}^+\tilde{\Pi}_-\,,
    \qquad a^- =
    \sqrt{1+\frac{\nu}{\tilde{\mathcal{N}}+1}}
    \;\,\tilde{a}^-\tilde{\Pi}_-+\tilde{a}^-
    \tilde{\Pi}_+\,, \label{a+a-F}
\end{equation}
in terms of the usual, non-deformed  Heisenberg algebra of a bosonic oscillator,
\begin{equation}
    [\tilde{a}^-,\tilde{a}^+]=1\,,
    \qquad [\tilde{\mathcal{N}},\tilde{a}^\pm]=
    \pm \tilde{a}^\pm\,.\label{tildea+-}
\end{equation}
Here $\tilde{\mathcal{N}}=\tilde{a}^+\tilde{a}^-$ is the number
operator, and $\tilde{\Pi}_\pm=\frac{1}{2}\left( 1\pm
\tilde{R}\right)$,  $\tilde{R}=(-1)^{\tilde{\mathcal{N}}}$, are
 projectors, $[\tilde{R},\tilde{a}^\pm]=0$. Consistently with
(\ref{a+a-F}) and (\ref{Naa}), the number operators of the deformed,
(\ref{DHAR}), and the non-deformed, (\ref{tildea+-}), algebras
coincide, $\mathcal{N}=\tilde{\mathcal{N}}$.

%%%%%%%%%%%%%%%%%%%%%%%%%%%%%%%%%%%%%%%%%%%%%%%%%%%%%%%%%%%%%%%%%%%%%%
\section{$N=1$ supersymmetric
generalization}\label{susyN1gen}
%%%%%%%%%%%%%%%%%%%%%%%%%%%%%%%%%%%%%%%%%%%%%%%%%%%%%%%%%%%%%%%%%%%%%%
Having written the various spin cases in a unified form,
now we proceed to unify them into a supersymmetric theory.
Consider in fact the following modification of Eq.
\eqref{VmuEq},
\begin{equation}\label{Vsusy}
    V_\mu^{(\hat{\alpha})}\psi(x)=0\,,\qquad
    V_\mu^{(\hat{\alpha})}=\hat{\alpha}P_\mu
    -i\epsilon_{\mu \nu \lambda}P^\nu
    {\cal J}^\lambda+
    m {\cal J}_\mu\,.
\end{equation}
Here ${\cal J}_\mu$ is the direct sum of irreducible
representations of $\mathfrak{so}(2,1)$, and $\hat{\alpha}$
is  not more a $c$-number,  but a (diagonal) \textit{operator},
which takes the corresponding $\mathfrak{so}(2,1)$-spin
value, $\alpha$, on each $\mathfrak{so}(2,1)$-irreducible
subspace,
\begin{equation}\label{JcalAl}
    {\cal J}_\mu{\cal J}^\mu=
    -\hat{\alpha}(\hat{\alpha}-1)\,,
\end{equation}
cf. Eq. (\ref{Jal}). The vector system (\ref{Vsusy}) yields
equations of the form (\ref{Cas+aux1})--(\ref{Cas+aux3}) with
$\emph{\ss}$ replaced by $\hat{\alpha}$. Note that the analog of
Eq. (\ref{Cas+aux3}) is automatically  satisfied here, due to
(\ref{JcalAl}). Eq. (\ref{Vsusy})  describes a multiplet of
particles of the same mass, and the values of the Poincar\'e spins are
 given by the diagonal elements of $\hat{\alpha}$.

As found before (but in a less general framework) \cite{HPV1},
choosing the direct sum of just two $\mathfrak{so}(2,1)$
representations shifted by one half,
\begin{equation}\label{hata}
    \hat{\alpha}=diag\,
    \left(\alpha,\alpha+1/2\right),
\end{equation}
provides us with an $N=1$ supersymmetric system. To
identify the supersymmetric structure of
(\ref{Vsusy}), (\ref{hata}), we realize the
$\mathfrak{so}(2,1)$-generators quadratically in terms of
the creation-annihilation operators of the RDHA, see
(\ref{J0,Ji}). Consistently with (\ref{JcalAl}), they
satisfy the relation (\ref{JJ}).
% with
%$\hat{\alpha}$ given by (\ref{hataR}).
 They commute with
the reflection operator $R$, identified as the
$\mathbb{Z}_2$-grading operator of the super-Poincar\'e
structure we are looking for. The Fock space representation
of the RDHA depends on  the deformation parameter, and
provides us with infinite- ($\nu\neq -(2r+1)$), and
$(2r+1)$-dimensional ($\nu=-(2r+1)$, $r=1,2,\ldots$)
irreducible representations of the $\mathfrak{osp}(1\vert
2)$ superalgebra. Each such representation is the direct
sum of two irreducible $\mathfrak{so}(2,1)$ representations
with spin values shifted by $1/2$,
\begin{equation}\label{a+a-R}
    \alpha_+=\frac{1}{4}(1+\nu),\qquad
    \alpha_-=\alpha_++\frac{1}{2}\,,
\end{equation}
cf. (\ref{halfshift}), and consistently also with
(\ref{hata}). These
 are just the values taken by
 the operator $\hat{\alpha}$ in (\ref{JJ}),
 when restricted to the subspaces
$\mathcal{F}_\pm$. The latter are invariant under the
action of ${\cal J}_\mu$, and are eigensubspaces of the
reflection operator,  see (\ref{calF+-}), (\ref{FFpm}). The
wave function is expanded as
\begin{equation}\label{psi+psi-}
    \psi(x)=\psi^+(x) +\psi^-(x)\,,\qquad \psi^\pm(x)= \sum_{n=0}
    \psi^\pm_n(x)|n\rangle_\pm\,, \qquad |n \rangle_\pm \in
    \mathcal{F}_\pm\,,
\end{equation}
and Eq. (\ref{Vsusy}) is reduced to two independent
equations,
\beq
    V^{(\alpha_+)}_\mu\psi^+(x)=0
    \qquad\hbox{and}\qquad
    V^{(\alpha_-)}_\mu\psi^-(x)=0\,,\label{Vpsi+-}
\eeq
 whose solutions form
a supermultiplet with spins $\alpha_+$ and
$\alpha_++1/2=\alpha_-$, respectively. The spinor
supercharge of such an $N=1$-supersymmetric system is given
by
\begin{equation}\label{QL}
    Q_a=\frac{i}{\sqrt{2m(1+\nu)}}\,\left((P^\mu \gamma_\mu)_{a}{}^b
    -mR
    \delta_a{}^b\right){\cal
    L}_b\,,\qquad a,b=1,2\,,
\end{equation}
where the $\gamma_\mu$'s are the Dirac matrices
(\ref{Dmatrices}) in the Majorana representation. It is an
operator-valued linear combination of the odd
$\mathfrak{osp}(1\vert 2)$ generators (\ref{Lalpha}). To
see that the (\ref{QL}) is indeed the supercharge, we
calculate the commutator
\begin{equation}\label{VaQ}
    [V_\mu^{(\hat{\alpha})}, Q_a]=
    \frac{i}{\sqrt{2m(1+\nu)}}
    \left( (D_\mu)_a{}^b {\cal D}_b \Pi_+  -
    \frac{1}{2}(\gamma_\mu)_a {}^b \mathcal{L}_b
    (P^2+m^2)\right)\,,
\end{equation}
where
\begin{equation}
    (D_\mu)_a{}^b= P_\mu \delta_a ^b -m (\gamma_\mu)_a
    {}^b,\quad {\cal D}_b=(P_\mu (\gamma^\mu)_a {}^b -m \delta_a
    {}^b\,)\mathcal{L}_b\,,
    \label{susycons}
\end{equation}
and $\Pi_+$ is the projector on the even subspace
$\mathcal{F}_+$, see  (\ref{projector}). The action of
(\ref{VaQ}) on the wave functions that satisfy Eq.
(\ref{Vsusy}) produces zero if  ${\cal D}_a\psi^+=0$. Such
a spinor set of equations was studied in \cite{MPAnn},
where it was shown that its solutions describe a particle
of mass $m$ and spin $\alpha_+$.

 Moreover,
$
    V_\mu^{(\alpha_+)}\psi^+(x)=
    \frac{1}{4}\mathcal{L}^{a}(\gamma^\mu
    D_\mu)_a{}^{b}\,
 {\cal D}_{b}\psi^+(x),
$ and any solution of the spinor set of equations in the
even subspace is also a solution of the vector set.

 We conclude, therefore,
that (\ref{QL}) is indeed a supercharge. Together with the
Poincar\'e generators, it yields the off-shell
\emph{nonlinear} superalgebra
\begin{equation}\label{SuperPoi1}
    [P_\mu,P_\nu]=0\,,\qquad
    [{\cal M}_\mu, P_\nu]=-i\epsilon_{\mu\nu\lambda}P^\lambda\,,\qquad
    [{\cal M}_\mu,{\cal M}_\nu]=-i\epsilon_{\mu\nu\lambda}{\cal
    M}^\lambda\,,
\end{equation}
\begin{equation}\label{SuperPoi2}
    [P_\mu,Q_a]=0\,,\qquad
    [{\cal M}_\mu,Q_a]=\frac{1}{2}(\gamma_\mu)_a{}^b
    Q_b\,,
\end{equation}
\begin{eqnarray}
    &\{Q_a,Q_b\}=-2i(P\gamma)_{ab}&\label{QQPab}
    \\[4pt]
    &+\frac{2i}{m(1+\nu)}\left[({\cal J}\gamma)_{ab}(P^2+m^2)-
    2(P\gamma)_{ab}(P{\cal J}-
    m\hat{\alpha})\right]\,.&\label{SuperPoi3}
\end{eqnarray}
The last term   \eqref{SuperPoi3} vanishes on-shell, and
\eqref{SuperPoi1}--\eqref{QQPab}  yield the usual $N=1$
planar super-Poincar\'e  algebra.

The deformed oscillator representation of the Lorentz
algebra produces  both the infinite half-bounded series,
and the usual finite-dimensional series with integer or
half-integer spin. Hence, the system of equations
\eqref{Vsusy}, (\ref{hata})  describes, universally, a
supersymmetric system of massive particles of anyonic spin
when $\nu\neq -(2r+1)$, $r=1,2,\ldots$, and of usual
integer/half-integer spin for $\nu=-3,-5,-7,\ldots$. This
is summarized in Table 1,
 where the notation  (\ref{tilD}) was used.
Note that for
$-3<\nu<-1$, the supermultiplet includes anyons with spins
of opposite signs, $-1/2<\alpha_+<0$ and
$\alpha_-=\alpha_++1/2>0$, which are described in terms of
infinite-dimensional half-bounded non-unitary and unitary
representations, respectively.

\begin{table}[ht]  \label{T1}
\caption{Poincar\'e supermultiplets}
\begin{center}
\begin{tabular}{|c|c|c|c|}
 \hline
  Deformation parameter & Poincar\'e spin supermultiplet
   &  Lorentz rep. & Supermultiplet \\\hline\hline
  $\nu\neq -(2r+1)$ & $(\alpha_+,\alpha_-)$ &
  ${\cal D}%\hat{D}
  _{\alpha_+}^+ \oplus
  {\cal D}%\hat{D}
  _{\alpha_-}^+$ & anyons \\\hline
  $\nu=-(2r+1)$, $r=1,2,\ldots$ & $
  (-r/2,-r/2+1/2)$ & ${\cal D}^+%\tilde{D}^
  _{-r/2}\oplus
  {\cal D}^+%\tilde{D}^
  _{-r/2+1/2}$ & boson/fermion
\\\hline
\end{tabular}
\end{center}
\end{table}

Some examples of interest are listed in Table \ref{T2}.
Notice  that the supermultiplet (a) was studied in a
superfield approach in \cite{SorVol}, see also Footnote 14.
For the supermultiplet (c), the vector operator
$V_\mu^{(\hat{\alpha})}$ vanishes identically on the
spin-$0$ subspace. In this sector the Klein-Gordon equation
should therefore be imposed, assigning to it $|0\rangle_-$
in the Fock space. Then the superpartner-fields are mapped
into each other by the supercharge \eqref{QL}. Another
possibility consists in taking $\nu<-3$, and then
considering the limit $\nu\rightarrow -3$.

The supermultiplet (e) is studied in the next Section; the
supermultiplets (f) and (g) can be considered in the same
way. The analog of the DJT formulation  \cite{DJT} for the
massive graviton was considered in \cite{3TMGGD,3TMGG};
here we have the linearized form of the corresponding field
equations.

\begin{table}[ht]
\caption{Examples}  \label{T2}
\begin{center}
\begin{tabular}{|c|c|c|}
 \hline
Deformation parameter & Spin supermultiplet  &
Supermultiplet \\\hline\hline $\nu=0$  & $(1/4,3/4)$ & (a)
Quartions
\\\hline $\nu=-2$ & $(-1/4,1/4)$ & (b)
Semion/quartion
\\\hline $\nu=-3$ & $(-1/2,0)$ & (c)
Dirac/scalar
\\\hline $\nu=-4$ & $(-3/4,-1/4)$ & (d)
Semions
\\\hline $\nu=-5$ & $(-1,-1/2)$ & (e) DJT/Dirac \\\hline
$\nu=-7$ & $(-3/2,-1)$ & (f) Rarita-Schwinger/DJT
\\\hline $\nu=-9$ & $(-2,-3/2)$ & (g) Massive
graviton/gravitino
\\\hline
\end{tabular}
\end{center}
\end{table}

%%%%%%%%%%%%%%%%%%%%%%%%%%%%%%%%%%%%%%
%%%%%%%%%%%%%%%%%%%%%%%%%%%%%%%%%%%%%%%%%%%%%%%%%%%%%%%%

\subsection{Dirac/Jackiw-Deser-Templeton
supermultiplet}\label{DJDTsusy}

%%%%%%%%%%%%%%%%%%%%%%%%%%%%%%%%%%%%%%%%%%%%%%%%%%%%%%%%%%%%%%
%%%%%%%%%%%%%%%%%%%%%%%%%%%%%%%%%%%%%%%%%%%%%%%%%%%%%%%%%%%%

To illustrate the general theory of the previous Section,
we consider a supermultiplet composed of a \emph{Dirac
field}, paired with the \emph{topological massive gauge
vector field} of Deser, Jackiw and Templeton \cite{DJT}. It
is described by the vector system \eqref{Vsusy}, where the
spin operator $\hat{\alpha}$ has now eigenvalues $-1$ and
$-1/2$, respectively. The deformed algebra
 \eqref{DHAR} with deformation parameter  $\nu=-5$
 provides us with an  $\mathfrak{osp}(1|2)$ representation,
 whose irreducible
 Lorentz components have spin $j=-\alpha_+=1$ and $j=-\alpha_-=1/2$, respectively.
 The superfield $\psi(x)$ is expanded in
  the $|\nu|=5$-dimensional basis \eqref{Fock5},
\begin{equation}
\psi(x)=
 \left(
\begin{array}{c}
\psi^+_0(x)  \\
\psi^-_0(x)  \\
\psi^+_1(x)  \\
\psi^-_1(x)  \\
\psi^+_2(x) %
\end{array}
\right).\label{psi5}
\end{equation}
The $\mathfrak{osp}(1|2)$ generators are $5\times5$
matrices given by Eqns.  \eqref{finiteJ2}, \eqref{5a+a-} and
\eqref{Lalpha}. Using the projectors \eqref{RPI} and
taking into account  \eqref{F-} and \eqref{F+}, we extract
the components with spins $1$ and $1/2$,
\begin{eqnarray}
& j=-\alpha_+=1:\qquad \psi^+(x)=\Pi_+\psi(x)=
 \left(
\begin{array}{c}
\psi^+_0(x)  \\
\psi^+_1(x)  \\
\psi^+_2(x) %
\end{array}
\right)_{\!\!+},
&\label{j1}\\[10pt]
&j=-\alpha_-=1/2:\qquad \psi^-(x)=\Pi_-\psi(x)=
 \left(
\begin{array}{c}
\psi^-_0(x)  \\
\psi^-_1(x)  \\
\end{array}
\right)_{\!\!-}.
&\label{j1/2}
\end{eqnarray}
The operator  $V_\mu^{(\hat{\alpha})}$ in Eq. \eqref{Vsusy},
projected to these components,
\beq
V_\mu^{(\hat{\alpha})}\Pi_\pm \psi(x)=
V_\mu^{(\alpha_\pm)}\psi^\pm(x),
\eeq
 reduces to
\begin{equation}
V_\mu^{(\alpha_\pm)}=\alpha_\pm P_\mu+m
J^{(\pm)}_\mu-i\epsilon_{\mu \nu \lambda}P^\nu
J^{(\pm)}{}^\lambda,
\end{equation}
where,
\begin{eqnarray}
&J^{(-)}_0=-\frac{1}{2}\sigma_ 3,\qquad J^{(-)}_1=
\frac{i}{2}\sigma_1,\qquad J^{(-)}_2=-\frac{i}{2}\sigma_2,
&\\[6pt]
& J^{(+)}_0=\left(
\begin{array}{ccc}
-1 & 0 & 0 \\
0 & 0 & 0 \\
0 & 0 & 1%
\end{array}%
\right), \;\; J^{(+)}_1=\frac{i}{\sqrt{2}}\left(
\begin{array}{ccc}
0 & 1 & 0 \\
1 & 0 & 1 \\
0 & 1 & 0%
\end{array}%
\right), \;\; J^{(+)}_2=\frac{1}{\sqrt{2}}\left(
\begin{array}{ccc}
0 & -1 & 0 \\
1 & 0 & -1 \\
0 & 1 & 0%
\end{array}%
\right),&
\end{eqnarray}
\beq
J^{(-)}_\mu J^{(-)}\,^\mu=-\frac{3}{4},
\qquad
J^{(+)}_\mu
J^{(+)}\,^\mu=-2.
\eeq

 The
Pauli-Lubanski condition (\ref{Cas+aux2})  implied by the
equations \eqref{Vsusy} becomes, on one of the sectors, the
Dirac equation\footnote{Note that the representation of
Dirac matrices, $\gamma'_\mu$  in (\ref{gammapr}), is
related to that in  (\ref{Dmatrices}) by the unitary
transformation (\ref{gprimeg}).}
\begin{equation}\label{gammapr}
(P\gamma'-m)_a{}^b(\psi^-(x))_b=0,\qquad  \gamma'_\mu =
 -2 J^{(-)}_\mu,
\end{equation}
while, on the other sector, it becomes  equivalent to the
topologically massive gauge vector field equation \footnote{The
standard (adjoint) representation of the spin-$1$ Lorentz algebra,
$(J_\mu)^\nu{}_\lambda=-i\epsilon^{\nu}{}_{\mu\lambda}$, is
obtained by the unitary transformation \eqref{U1}.},
$$
(P^\mu J^{(+)}_\mu +
m) \psi^+(x)=0\,.
$$
Then we obtain \eqref{DJT}, with
\begin{eqnarray}
    \left(-i\epsilon^{\lambda}{}_{\mu\nu}
    P^\mu+m\delta^\lambda_\nu\right)
    \psi^\nu=0\,,\quad \psi^\nu= (U^\dagger \psi^+(x))^\nu\,,\quad -
    i\epsilon^{\nu}{}_{\mu\lambda}= (U^\dagger J^{(+)}_\mu
    U)^\nu{}_\lambda\; .
\end{eqnarray}

To see how the supercharge acts on this supermultiplet, it
is convenient to consider the linear combinations

\begin{equation}\label{QQpm}
    Q_\pm=Q_1\mp i Q_2\,,\qquad
    Q_\pm=\frac{i}{\sqrt{m(1+\nu)}}\left[\pm a^\mp P_\pm
    +
    a^\pm (mR\mp P_0)\right],
\end{equation}
where $P_\pm=P_1\pm iP_2$. Using the matrix representation
\eqref{5a+a-} of the rising/lowering operators $a^\pm$, and
the representation \eqref{RPI} of the reflection operator,
the explicit action of the supercharges (\ref{QQpm}) on the
spin-1 ($\psi^+$) and spin-1/2 ($\psi^-$) components of the
supermultiplet is found,
\begin{eqnarray}
 Q_+\psi^-(x)&=&
 \frac{1}{\sqrt{m}}\left(
\begin{array}{c}
iP_+\psi^-_0(x)  \\
\frac{i}{\sqrt{2}}P_+\psi^-_1(x)-\frac{1}{\sqrt{2}}(m+P_0) \psi^-_0(x)  \\
-(m+P_0) \psi^-_1(x) %
\end{array}
\right)_+,\label{Q+-1}\\[8pt]
 Q_-\psi^-(x)&=&
\frac{1}{\sqrt{m}}\left(
\begin{array}{c}
i(P_0-m) \psi^-_0(x)\\
-\frac{1}{\sqrt{2}}P_-\psi^-_0(x)+\frac{i}{\sqrt{2}}(P_0-m) \psi^-_1(x)  \\
 -P_-\psi^-_1(x)%
\end{array}
\right)_+,\label{Q+-2}
\end{eqnarray}

\begin{eqnarray}
Q_+\psi^+(x)&=& \frac{1}{\sqrt{m}} \left(
\begin{array}{c}
\frac{1}{\sqrt{2}}P_+\psi^+_1(x)+i(m-P_0) \psi^+_0(x)  \\
P_+\psi^+_2(x)+\frac{i}{\sqrt{2}}(m-P_0) \psi^+_1(x)
\end{array}
\right)_-,\label{Q+-3}\\[8pt]
Q_-\psi^+(x)&=&\frac{1}{\sqrt{m}}\left(
\begin{array}{c}
-iP_-\psi^+_0(x)+\frac{1}{\sqrt{2}}(m+P_0) \psi^+_1(x)  \\
-\frac{i}{\sqrt{2}}P_-\psi^+_1(x)+(m+P_0) \psi^+_2(x)
 \end{array}
\right)_-.\label{Q+-4}
\end{eqnarray}
Here, it is explicitly shown how the components of the
 transformed spin-$1$ field depend on the untransformed
 spin-$1/2$ components \eqref{j1/2} of the superfield \eqref{psi5}.

 Conversely, the components of the transformed spin-$1/2$
 field are those of the spin $1$ field \eqref{j1}.

%%%%%%%%%%%%%%%%%%%%%%%%%%%%%%%%%%%%%%%%%%%%%%%%%%%%
%%%%%%%%%%%%%%%%%%%%%%%%%%%%%%%%%%%%%%%%%%%%%%%%%%%%%%%%%%%
\subsection{Jackiw-Nair/Majorana-Dirac supersymmetric
system}\label{JNMDN1susy}
%%%%%%%%%%%%%%%%%%%%%%%%%%%%%%%%%%%%%%%%%%%%%%%%%%%%%%%%%%%
%%%%%%%%%%%%%%%%%%%%%%%%%%%%%%%%%%%%%%%%%%%%%%%%%%%%

Now, extending the Dirac--DJT correspondence established in the
previous Section,
 we construct an \emph{$N=1$ supersymmetric unification} of the
Jackiw-Nair and Majorana-Dirac anyonic fields. This
can be done by combining the extended realization
(\ref{J1J2}) and (\ref{ssalpha2}) [that corresponds to the
sum of the $\mathfrak{so}(2,1)$-representations] with the
 $N=1$ supersymmetric scheme of Section
\ref{susyN1gen}. Taking the particular realization of the
latter from Section \ref{DJDTsusy} [i.e. taking the
Dirac/DJT supermultiplet], we get, as a result, a particular,
  $N=1$ supermultiplet of Majorana-Dirac and JN fields.

The generator $J_\mu$ in (\ref{J1J2}) is taken in an
irreducible representation $D^+_\alpha$, while $J'_\mu$ is
realized as a direct sum of two finite-dimensional
$\mathfrak{so}(2,1)$ representations with Lorentz spins
$j=1$ and $j=1/2$. Finite-dimensional representations can
be obtained by means of the RDHA with deformation parameter
$\nu=-5$.
The parameter $\alpha'$ is  promoted to
a diagonal operator with eigenvalues $-1$ and $-1/2$. This
immediately provides us with a supermultiplet of spins
$\alpha-1$ and $\alpha-1/2$. On the corresponding subspaces
with $\beta=\alpha-1$ and $\beta=\alpha-1/2$, our vector
 equations will take the form which  corresponds
to the Jackiw-Nair and Majorana-Dirac fields, respectively.

Explicitly, the $N=1$ JN/MD supersymmetric system is
described by the vector system of equations
\begin{equation}\label{N=1JNMD}
    V_\mu^{(\alpha+\hat{\alpha}')}\psi(x)=0\,,\qquad
    V_\mu^{(\alpha+\hat{\alpha}')}=
    (\alpha+\hat{\alpha}')P_\mu-i\epsilon_{\mu \nu \lambda}P^\nu
    {\cal J}^\lambda+
    m {\cal J}_\mu\,,
\end{equation}
where
\begin{equation}
    {\cal J}_\mu=J_\mu +\mathfrak{J}'_\mu,\qquad
    J_\mu\in D^+_\alpha,\qquad
    \mathfrak{J}'_\mu\in D^{1}\oplus D^{1/2},\qquad
    \hat{\alpha}'=diag\, (-1,-1/2)\,.
\end{equation}
The five-dimensional irreducible representation of
$\mathfrak{osp}(1|2)$ that corresponds to the chosen
direct sum for $\mathfrak{J}'_\mu$ was described in the
previous Section in terms of the RDHA with parameter
$\nu=-5$. A solution of the system (\ref{N=1JNMD}) is given
by the JN field realized on $D^+_\alpha\oplus \tilde{D}^1$,
and by the MD field realized on $D^+_\alpha\oplus
\tilde{D}^{1/2}$. Here $\tilde{D}^1$ corresponds to the
three-dimensional (vector) even subspace of
five-dimensional Fock space
%of the RDHA with $\nu=-5$,
 and $\tilde{D}^{1/2}$ corresponds to its
two-dimensional (spinor) odd  subspace.

The spinor supercharge of such a supersymmetric system
becomes
\begin{equation}\label{QLJNMD}
    Q_a=\frac{1}{2\sqrt{2m}}\,\left((P^\mu \gamma_\mu)_{a}{}^b
    -mR
    \delta_a{}^b\right){\cal
    L}_b\,,\qquad a,b=1,2\,,
\end{equation}
cf. (\ref{QL}). It is implied that (\ref{QLJNMD}) acts
identically on the index $n$
 [not shown explicitly and
corresponding to the infinite-dimensional representation
$D^+_\alpha$] of the wave function $\psi(x)$, while the matrix operator
${\cal L}_b$ acts in
the 5-dimensional irreducible representation of the
$\mathfrak{osp}(1|2)$ superalgebra. The action of this
supercharge is given by Eqns. (\ref{QQpm}),
(\ref{Q+-1})--(\ref{Q+-4}), where, in this case, it is implied
that the components of the wave functions carry also
 [an not displayed] index $n$. Together with ${\cal M}_\mu$ and
$P_\mu$, this supercharge generates a nonlinear
superalgebra of the form
(\ref{SuperPoi1})--(\ref{SuperPoi3}), with ${\cal J}_\mu$
in the last term (\ref{SuperPoi3}) changed to
$\mathfrak{J}'_\mu$, and $\hat{\alpha}$ changed to
$\hat{\alpha}'$.

 It was shown in Section 2 that the vector
sets of equations for JN and MD fields decouple
into Majorana and DJT equations, and Majorana
and Dirac equations,
respectively.  This means that the  factor
$(P\mathfrak{J}'-m\hat{\alpha}')$  disappears on-shell like
(\ref{SuperPoi3}). Therefore, on-shell,
the standard $N=1$ superalgebra given by Eqns.
(\ref{SuperPoi1})--(\ref{QQPab}) is obtained.

%%%%%%%%%%%%%%%%%%%%%%%%%%%%%%%%%%%%%%%%%%%%%%%%%%%%%%%%%%%%%%%%%%%%%%%
%%%%%%%%%%%%%%%%%%%%%%%%%%%%%%%%%%%%%%%%%%%%%%%%%%%%%%%%
\section{$N=2$ supersymmetry}\label{N2susy}
%%%%%%%%%%%%%%%%%%%%%%%%%%%%%%%%%%%%%%%%%%%%%%%%%%%%%%%%%%%%%%%%
%%%%%%%%%%%%%%%%%%%%%%%%%%%%%%%%%%%%%%%%%%%%%%%%%%%%%%%%%%%%%%%%%%%%%%%%%

In Section \ref{susyN1gen} we have shown
 that promoting  the parameter $\alpha$ in (\ref{ssalpha})
 to a (diagonal) operator, $\hat{\alpha}$ in
(\ref{hata}), and realizing the Lorentz generator ${\cal
J}_\mu$ in terms of the RDHA creation-annihilation
operators yields an $N=1$ supersymmetric system of anyons,
or of usual fields of arbitrary integer and half-integer
spin.

In the previous Section we showed in turn that $N=1$
supersymmetry can be obtained, alternatively, if, in the
extended realization of Section \ref{solBFA},  the
second parameter $\alpha'$  is promoted to a diagonal operator, while
$\alpha$ is left a fixed numerical parameter. The
alternative construction allowed us to to unify the
Majorana-Dirac and JN fields in one anyon supermultiplet,
and to get supermultiplets with partner anyon fields
described by infinite-dimensional %non-unitary
representations.

The construction of the previous
section was asymmetric,
and now we argue that the same prescription,
applied to (\ref{ssalpha2}) \emph{symmetrically} in the
parameters $\alpha$ and $\alpha'$,   i.e., promoting
them \emph{both} to operators, provides us with an
$N=2$-supersymmetric system of anyons, or of bosons and
fermions. In such a way we unify, in particular, the
Majorana-Dirac and Jackiw-Nair anyon systems into a
single extended $N=2$ supermultiplet.

Let us take \begin{equation}
    \mathcal{J}^\mu=\mathfrak{J}_{\sc 1}^\mu +
    \mathfrak{J}_{\sc 2}^\mu\,,\qquad
    \mathfrak{J}_{\sc A}{}_\mu \mathfrak{J}_{\sc A}^\mu=
    -\hat{\alpha}_{\sc A} (\hat{\alpha}_{\sc A}-1)\,,\quad
    \un A= \un1,\, \un2\,,
     \label{frakJ}
\end{equation}
[with no summation in $\un A$ implied], and postulate
the vector system of equations
\begin{eqnarray}\label{Vsusy2}
    &V_\mu^{(\hat{\emph{\ss}\;})}\psi(x)=0\,,
    \qquad  V_\mu^{(\hat{\emph{\ss}\;})}=
    \hat{\emph{\ss}\;} P_\mu+m \mathcal{J}_\mu-i
    \epsilon_{\mu \nu \lambda}P^\nu \mathcal{J}^\lambda\,,& \\[6pt]
    &\hat{\emph{\ss}\;}=\hat{\alpha}_{\sc 1}+\hat{\alpha}_{\sc
    2}\,.&\label{sspin}
\end{eqnarray}
Here we assume that each Lorentz generator
$\mathfrak{J}_{\sc A}{}_\mu$, $\un A= \un1,\, \un2$, is
realized quadratically in terms of the
creation-annihilation operators of  two independent
RDHA's,
\begin{equation}\label{DHARa}
    \big[a_{\sc A}^-,a_{\sc A}^+\big]=
    (1+\nu_{\sc A} R_{\sc A})\,, \quad \{a_{\sc A}^{\pm},R_{\sc A}\}=0\,,
    \quad R_{\sc A}^2=1\,,\quad \un A= \un1,\, \un2\,.
\end{equation}
So, here
\begin{eqnarray}
    &\hat{\alpha}_{\sc A}=
    \frac{1}{4}(2+\nu_{\sc A} - R_{\sc A})\,,&\label{alpha12}\\[6pt]
    &\mathfrak{J}_{\sc A}{}_0 =\frac{1}{2}
    N_{\sc A}+ \alpha_{\sc
    A +}\,,
    \qquad \alpha_{\sc A +}= \frac{1}{4}(1+\nu_{\sc
    A})\,.&\label{frakJ0}
\end{eqnarray}
In internal space we have two copies of
$\mathfrak{osp}(1|2)$ superalgebra,
\begin{equation}
    \mathfrak{osp}_{\sc 1}(1|2)\oplus \mathfrak{osp}_{\sc
    2}(1|2)\,,\qquad \mathfrak{osp}_{\sc A}(1|2)=\{
    \mathcal{L}_{\sc A}{}_a,\; \mathfrak{J}_{\sc
    A}^\mu\}\,,\qquad \un A= \un1\,, \un2\,, \label{2osp}
\end{equation}
which provide us with a representation of the Lorentz
generators (\ref{frakJ}) with four irreducible components
[two for each $\mathfrak{osp}_{\sc A}(1|2)$]. The ${\cal
J}_\mu$ act on the tensor product of the Fock spaces,
\begin{equation}
    \mathfrak{F}=\mathcal{F}_{\sc 1} \otimes \mathcal{F}_{\sc 2},
    \quad
    \mathfrak{F}=\big\{ |n_1,n_2\rangle = |n_1\rangle
    \otimes |n_2\rangle\;\big\vert\; |n_1\rangle \in
    \mathcal{F}_{\sc 1},\,
    |n_2\rangle \in \mathcal{F}_{\sc
    2}\big\}\,,\label{FxF}
\end{equation}
where $\mathcal{F}_{\sc A}$ is the Fock space of the
corresponding RDHA. The wave function is expanded on
$\mathfrak{F}$,
\begin{eqnarray}
    &\psi(x)=\sum_{n_1,n_2=0} \psi_{n_1,\,
    n_2}(x)|n_1,n_2\rangle
    \,.&\label{4wave}
\end{eqnarray}
 The subdivision of the Fock
spaces $\mathcal{F}_{\sc A}$ in even and odd subspaces,
$$
    \mathcal{F}_{\sc A +}=\big\{|n_{\scriptscriptstyle A}\rangle_+=
    |2n_{\scriptscriptstyle A}\rangle\big\},
    \qquad
    \mathcal{F}_{\sc A-}=
    \big\{|n_{\scriptscriptstyle A}
    \rangle_-=|2n_{\scriptscriptstyle A}+1\rangle
    \big\},
$$
induces the decomposition of $\mathfrak{F}$,
\begin{equation}\label{totalF}
    \mathfrak{F}= \mathcal{F}_{\sc 1 +} \otimes
    \mathcal{F}_{\sc 2 +} +\mathcal{F}_{\sc 1 +} \otimes
    \mathcal{F}_{\sc 2 -} +\mathcal{F}_{\sc 1 -} \otimes
    \mathcal{F}_{\sc 2 +}+\mathcal{F}_{\sc 1 -} \otimes
    \mathcal{F}_{\sc 2 -}\,.
\end{equation}
The spin operator (\ref{sspin}) takes, on the four
corresponding subspaces in (\ref{totalF}), the values
\begin{equation}\label{ssN2}
    \hat{\emph{\ss}\;}=diag\left(\chi,\chi+\frac{1}{2},
     \chi+\frac{1}{2},\chi+1\right),
     \qquad \chi = \alpha_{\sc 1 +} +
     \alpha_{\sc 2 +} = \frac{1+\nu_{\sc 1}}4
     + \frac{1+\nu_{\sc 2}}4\,.
\end{equation}
Observe that the spin eigenvalue $\chi+\frac{1}{2}$ is
doubly degenerated. The wave function \eqref{4wave} has
therefore four components,
\begin{eqnarray}
    &\psi(x)=\psi^{++}(x)+\psi^{+-}(x)+
    \psi^{-+}(x)+\psi^{--}(x)\,,&
    \label{psidecomp}
    \\[8pt]
    &\psi^{\pm \pm}(x)=\sum_{n_1,n_2=0}
     \psi^{\pm \pm}_{n_1,n_2}(x)|n_1,n_2
     \rangle_{\pm \pm}\,,
     \label{psidecomp2}
     \\[6pt]
     &\psi^{\pm \mp}(x)
     =\sum_{n_1,n_2=0} \psi^{\pm \mp}_{n_1,n_2}(x)
     |n_1,n_2\rangle_{\pm \mp}\,,&
     \label{psidecomp3}
     \\[6pt]
    &|n_1,n_2\rangle_{\pm \pm}\in \mathcal{F}_{\sc 1 \pm}
    \otimes \mathcal{F}_{\sc 2 \pm}\, ,\qquad
    |n_1,n_2\rangle_{\pm \mp}\in \mathcal{F}_{\sc 1 \mp} \otimes
    \mathcal{F}_{\sc 2 \mp}\,.&
\end{eqnarray}
The wave function \eqref{psidecomp}  solves \eqref{Vsusy2} if
each of its component solves, independently,
\begin{equation}\barray{lll}
    V_\mu^{(\chi)}\psi^{++}(x)&=&0,\\[4pt] %\quad
    V_\mu^{(\chi+1/2)}\psi^{+-}(x)&=&0,\\[4pt] %\quad
    V_\mu^{(\chi+1/2)}\psi^{-+}(x)&=&0,\\[4pt] % \quad
    V_\mu^{(\chi+1)}\psi^{--}(x)&=&0.
    \earray\label{Vchi}
\end{equation}
Each component carries therefore an irreducible
representation of the Poincar\'e group with  mass $m$ and
spin indicated by the upper index of the vector operator.

Fields  which differ by one half in their spins can be
connected by supercharges analogous to \eqref{QL}. They are
given by
\begin{eqnarray}
    &Q_{\sc1
    a}=\frac{i}{\sqrt{2m(1+\nu_{\sc1})}}\,\left((P^\mu
    \gamma_\mu)_{a}{}^b
    -mR_{\sc1} \delta_a{}^b\right){\cal L}_{\sc1 b}\,,
    &\label{Q_1N=2}\\[6pt]
    &Q_{\sc2 a}=\frac{i}{\sqrt{2m(1+\nu_{\sc2})}} R_{\sc1}
    \,\left((P^\mu \gamma_\mu)_{a}{}^b -mR_{\sc2}
    \delta_a{}^b\right){\cal L}_{\sc2 b}\,.&\label{Q_2N=2}
\end{eqnarray}
The additional factor $R_{\sc1}$ is included into the
second supercharge in order to make (\ref{Q_1N=2}) and
(\ref{Q_2N=2})  anti-commute, see (\ref{QQanti})
below. The action of the supercharges on the
supermultiplet components is illustrated on Fig.
\ref{fig1}.
\begin{figure}[ht]
\begin{eqnarray}
     [\psi^{+-}(x);\,s=\chi+1/2]\,
     \quad\longleftarrow
     \quad &Q_{\sc 1}& \quad \longrightarrow \quad [\psi^{--}(x);
     \,s=\chi+1]\nonumber\\[6pt]
    \uparrow \qquad \qquad &&\qquad \qquad \uparrow \nonumber\\[6pt]
    \quad \quad Q_{\sc 2} \quad\;\;
    \qquad && \quad\;\;\; \qquad Q_{\sc 2} \nonumber\\[6pt]
    \downarrow \qquad \qquad && \qquad \qquad
    \downarrow  \nonumber\\[6pt]
    [\psi^{++}(x);\,s=\chi]\,\quad \longleftarrow \quad
    & Q_{\sc
    1}& \quad \longrightarrow \quad [\psi^{-+}(x);\,s=\chi+1/2]
\nonumber
\end{eqnarray}
\caption[] {\it The  action of supercharges on the
supermultiplet components. Each supercharge, $Q_{\sc 1}$
and $Q_{\sc 2}$, changes the spin by $1/2$; the first (second)
supercharge acts nontrivially in the first (second) index
of the Fock space decomposition (\ref{4wave}).} \label{fig1}
\end{figure}

To see that (\ref{Q_1N=2}) and (\ref{Q_2N=2}) are physical
operators in that they preserve
 the physical subspace,
  we pass to the rest frame~\footnote{It is implied
  that at least one of the parameters
  $\nu_{\sc A}$, $\un A=1,2,$ corresponds to an
  infinite-dimensional representation of RDHA ($\nu>-1$).},
$P^\mu=(m,0,0)$, in which  the vector system  is equivalent
to $(\mathcal{J}_0 - \hat{\emph{\ss}\;})\psi(x)=0$,
$\mathcal{J}_-\psi(x)=0$, cf. (\ref{J0+-psi}). Observe that
$$
\mathcal{J}_-|0,0\rangle_{\pm\pm}=0,\qquad
\mathcal{J}_-|0,0\rangle_{\pm\mp}=0\,,
$$
and that
$|0,0\rangle_{++}=|0,0\rangle,$
$|0,0\rangle_{+-}=|0,1\rangle,$
$|0,0\rangle_{-+}=|1,0\rangle,$
$|0,0\rangle_{--}=|1,1\rangle$.
 The same set of kets
is annihilated also by the operator $\mathcal{J}_0 -
\hat{\emph{\ss}\;}$.  In the rest frame, the general
solution  of  \eqref{Vsusy2}
 is, therefore, an arbitrary linear combination of the four lower
ket states,
$$
    \psi(x)\propto \psi^{++}_{0,0}|0,0\rangle_{++}
    \;+\;\psi^{+-}_{0,0}|0,0\rangle_{+-}
    \;+\;\psi^{-+}_{0,0}|0,0\rangle_{-+}
    \;+\;\psi^{--}_{0,0}|0,0\rangle_{--}\,,
$$
multiplied by a time-dependent phase [not displayed here].
The coefficients are arbitrary constants. In the
rest frame, the operators (\ref{Q_1N=2}) and
(\ref{Q_2N=2}) are proportional to $a_{\sc
A}^+\Pi_{\sc A +}$ and $a_{\sc A}^-\Pi_{\sc A -}$ with
values of index $\un A=\un 1$ and $\un 2$, respectively.
\begin{figure}[ht]
\begin{eqnarray}
|0,1\rangle \qquad \quad \,&\begin{array}{c}
\longrightarrow a_{\sc 1}^+ \longrightarrow
\\[6pt]
\longleftarrow a_{\sc 1}^- \longleftarrow
\end{array}& \qquad \quad\,
|1,1\rangle \\
%%%%%%%%%%%%%%%%%%%%%%%%%%%%%%%%%%%%%%%%%%%%%%
 a_{\sc 2}^+\uparrow \;\; \downarrow a_{\sc 2}^-
 \qquad & & \qquad  a_{\sc 2}^+\uparrow\;\;
 \downarrow a_{\sc 2}^- \nonumber\\[6pt]
%%%%%%%%%%%%%%%%%%%%%%%%%%%%%%%%%%%%%%%%%%%%%%
|0,0\rangle \qquad \quad \, &\begin{array}{c}
\longrightarrow a_{\sc 1}^+ \longrightarrow
\\[6pt]
\longleftarrow a_{\sc 1}^- \longleftarrow
\end{array}& \qquad \quad\,
|1,0\rangle \nonumber
\end{eqnarray}
\caption[] {\it In the rest frame $P^\mu=(m,0,0)$, the action of
the
supercharges,  shown in Fig. \ref{fig1}, is reduced to
that of creation-annihilation operators.} \label{fig2}
\end{figure}

\noindent Their action on solutions is reduced to that of
the creation and annihilation operators, as shown on Figure
\ref{fig2}. Therefore, the space of solutions is invariant
under the action of (\ref{Q_1N=2}) and (\ref{Q_2N=2}). They
are odd operators with respect to
 $\Gamma=R_{\un 1}R_{\un 2}$, identified as the
$\mathbb{Z}_2$-grading operator of the $N=2$ Poincar\'e
superalgebra. The even part of this superalgebra is given
by Eq. \eqref{SuperPoi1}. The part that involves the
supercharges is
\begin{eqnarray}
    &[P_\mu,Q_{\sc A a}]=0\,,
    \qquad [{\cal J}_\mu,Q_{\sc A \,a}]
    =\frac{1}{2}(\gamma_\mu)_a{}^b Q_{\sc A \,b}\,,&\label{QQant}\\[6pt]
    &\{Q_{\sc A \,a},Q_{\sc B \,b}\}
        =-2i\delta_{\sc{AB}}
    (P\gamma)_{ab}\,,&\label{QQanti}
\end{eqnarray}
where  the anti-commutator of the supercharges is shown in
on-shell form.

When both parameters $\nu_{\sc A}$, $\un A=1,2,$ correspond
to finite-dimensional representations of RDHA, the
equations (\ref{Vsusy2}), (\ref{sspin}) provide us with a
usual, boson/fermion $N=2$ super-multiplet (\ref{ssN2}),
with states of  both signs of the energy.

%%%%%%%%%%%%%%%%%%%%%%%%%%%%%%%%%%%%%%%%%%%%%%%%%%%%%%%%%%%%
%%%%%%%%%%%%%%%%%%%%%%%%%%%%%%%%%%%%%%%%%%%%%%%%%%%%%
%\subsection
\kikezd{Jackiw-Nair/Majorana-Dirac $N=2$
supermultiplet}\vskip2mm %%%%%%%%%%%%%%%%%%%%%%%%%%%%%%%%%%%%%
%%%%%%%%%%%%%%%%%%%%%%%%%%%%%%%%%%%%%%%%%%%%%%%%%%%%%%%%%%%%

The Lorentz generator \eqref{frakJ}, $\mathcal{J}^\mu=
\mathfrak{J}_{\sc 1}^\mu+\mathfrak{J}_{\sc 2}^\mu$, is
constructed in terms of two RDHA algebras, one for
$\mathfrak{J}_{\sc 1}^\mu$ and another for
$\mathfrak{J}_{\sc 2}^\mu$. We  have the freedom to choose
the deformation parameters $\nu_{\sc 1}$ and $\nu_{\sc 2}$
independently. As a result, the system of equations
\eqref{Vsusy2} can describe a variety of $N=2$
supermultiplets (\ref{ssN2}), based on different
representations of the $\mathfrak{so}(2,1)$ algebra, by
mixing ``anyonic" and/or usual integer/half-integer spin
series.

As an example, we consider here the
Jackiw-Nair/Majorana-Dirac $N=2$ supermultiplet. It is
realized by choosing $\nu_{\sc 1}>-1$ and $\nu_{\sc 2}=-5$.
 This implies,
\begin{equation}%\label{}
    \alpha_{\sc 1 +}\,>\,0 ,\qquad
    -\alpha_{\sc 2 +}=j_{\sc
    2}=1 \qquad \Rightarrow
    \qquad \chi= \alpha_{\sc 1 +}-1.
\end{equation}
Hence we get the spectrum described in Table \ref{JNMD}.

\begin{table}[ht]
\caption{$N=2$ JN/MD supermultiplet}  \label{JNMD}
\begin{center}
\begin{tabular}{|c|c|c|c|c|}\hline
Field component & Fine spin structure & Spin $s$ & Internal
space & Field type
\\\hline\hline
$\psi^{++}(x)$ & $\alpha_{\sc 1 +}-1$ & $\alpha_{\sc 1
+}-1$ & $\mathcal{F}_{\sc 1+}^{(\infty)} \otimes
\mathcal{F}_{\sc 2 +}^{\,(3)}$
& JN \\[4pt]\hline
$\psi^{+-}(x)$ & $\alpha_{\sc 1 +}-1/2$ & $\alpha_{\sc 1
+}-1/2$ &  $\mathcal{F}_{\sc 1+}^{(\infty)} \otimes
\mathcal{F}_{\sc 2 -}^{\,(2)}$ &
MD\\[4pt]\hline
$\psi^{-+}(x)$ & $\alpha_{\sc 1 -}-1$ & $\alpha_{\sc 1
+}-1/2$ & $\mathcal{F}_{\sc 1-}^{(\infty)} \otimes
\mathcal{F}_{\sc 2 +}^{\,(3)}$
 &  JN\\[4pt]\hline
$\psi^{--}(x)$ & $\alpha_{\sc 1 -}-1/2$ & $\alpha_{\sc 1
+}$ & $\mathcal{F}_{\sc 1-}^{(\infty)} \otimes
\mathcal{F}_{\sc 2 -}^{\,(2)}$ &  MD\\[4pt]\hline
\end{tabular}
\end{center}
\end{table}

It is interesting to note that, from the viewpoint of the
Poincar\'e spin value $s$ [see the third column in the
Table], the field component $\psi^{--}$ looks like the field
we got in Section 2 using the irreducible representation
$D^+_\alpha$ with $\alpha=\alpha_{\sc 1 +}>0$. The
components $\psi^{-+}$ and $\psi^{+-}$ look like anyonic
fields produced by the Majorana-Dirac system.
Finally, the component $\psi^{++}$ looks like that
described by the Jackiw-Nair equations. {}Having in mind,
however, that the equations for each field component have a
structure that corresponds to the scheme (\ref{J1J2}),
(\ref{ssalpha2}) based on the direct sum of two irreducible
representations of $\mathfrak{so}(2,1)$ (see the second and
fourth columns), each component is interpreted as a field of
the type indicated in the fifth column.

The wave functions have two Lorentz indices, associated to
$\mathcal{F}_{\sc 1}^{(\infty)}=\mathcal{F}_{\sc 1
-}^{(\infty)}+\mathcal{F}_{\sc 1 +}^{(\infty)}$ (infinite
dimensional) and $\mathcal{F}_{\sc 2
}^{\,(5)}=\mathcal{F}_{\sc 2 -}^{\,(2)}+\mathcal{F}_{\sc 2
+}^{\,(3)}$ ($5$-dimensional) representations of RDHA.
Explicitly,
\begin{eqnarray}
    &\psi^{++}(x)=&\sum_{n=0}^\infty\sum_{\lambda'=1,2,3}
    (\psi_n^{++}(x))^{\lambda'}|n,\lambda'\rangle_{++}\, ,
    \qquad |n,\lambda'\rangle_{++} \in
    \mathcal{F}_{\sc 1 +}^{(\infty)}
    \otimes \mathcal{F}_{\sc 2 +}^{\,(3)}\,, \\
    &\psi^{+-}(x)=&\sum_{n=0}^\infty \sum_{a=1,2}
    (\psi_n^{+-}(x))^{a}|n,a\rangle_{+-}\,,
    \qquad |n,a\rangle_{+-} \in
    \mathcal{F}_{\sc 1 +}^{(\infty)}
    \otimes \mathcal{F}_{\sc 2 -}^{\,(2)}\,, \\
    &\psi^{-+}(x)=&\sum_{n=0}^\infty\sum_{
    \lambda'=1,2,3} (\psi_n^{-+}(x))^{\lambda'}
    |n,\lambda'\rangle_{-+}\,,\qquad |n,\lambda'
    \rangle_{-+} \in \mathcal{F}_{\sc 1 -}^{(\infty)}
    \otimes \mathcal{F}_{\sc 2 +}^{\,(3)}\,, \\
    &\psi^{--}(x)=&\sum_{n=0}^\infty \sum_{ a=1,2}
    (\psi_n^{--}(x))^{a}|n,a\rangle_{--}\,,\qquad
    |n,a\rangle_{--} \in \mathcal{F}_{\sc 1 -}^{(\infty)} \otimes
    \mathcal{F}_{\sc 2 -}^{\,(2)}\,.
\end{eqnarray}
Here, $a=1,2,$ is a spinor index in the basis \eqref{F-},
in which the  $\mathfrak{so}(2,1)$ generators are given by
Eq. \eqref{Cliff:DHA}. The vector index $\lambda'$ corresponds
to the basis \eqref{F+}, and transforms under the
$\mathfrak{so}(2,1)$ representation \eqref{finiteJ2}.
Index $n$ transforms under the infinite dimensional
$\mathfrak{so}(2,1)$ representation ($D_{\alpha_{\sc 1
\pm}}^+$), and endows a wave function with the anyonic part
of the spin. Supercharges (\ref{Q_1N=2}), (\ref{Q_2N=2})
transform these fields as shown on Fig. \ref{fig1}.

%\newpage
%%%%%%%%%%%%%%%%%%%%%%%%%%%%%%%%%%%%%%%%%%%%%%%%%%%%%%%%%
%%%%%%%%%%%%%%%%%%%%%%%%%%%%%%
\section{Nonrelativistic limit}\label{NRlimit}
%%%%%%%%%%%%%%%%%%%%%%%%%%%%%%%
%%%%%%%%%%%%%%%%%%%%%%%%%%%%%%%%%%%%%%%%%%%%%%%%%%%%%%%%%%

In this Section we study the non-relativistic limit of our
vector system of equations \eqref{VmuEq}. This will provide
us with a universal non-relativistic description of either
fermion/boson fields of arbitrary half-integer/integer
spin, or of non-unitary anyons interpolating between them,
or of anyons based on unitary representations. As shown
below, the system of linear differential equations we get
implies, for each component,  the Schr\"odinger equation as
integrability condition  -- just like the Klein-Gordon
equation is implied in the relativistic case. Also, the
equations guarantee that the  system carries an irreducible
representation of the corresponding Galilei symmetry.

We analyze two
different types of non-relativistic limits
\cite{HPPLB}.

$\bullet$ The first one is the
usual limit, in which the velocity of light, $c$, tends to
infinity, while the remaining parameters are kept constant.
This generalizes the \emph{L\'evy-Leblond equations}
valid for spin $1/2$ \cite{LLeq} to arbitrary spin.

$\bullet$  In the second, \textit{``exotic''} type of limit
introduced by Jackiw and Nair \cite{JaNa} (see also
\cite{DuHor,HPPLB}) the spin, $s$, also goes to infinity in
such a way that a ratio
\beq
    c\rightarrow \infty\,,\quad
    s \rightarrow \infty\,,\quad s/c^2=\kappa
    \label{exolim}
\eeq
remains constant. Its \textit{``raison
d'\^etre''} is that it yields \emph{exotic Galilei
symmetry} \cite{ExoGal}, which admits, besides the mass,
$m$, also
 a second central charge, $\kappa$,  associated with
the non-commutativity of Galilean boosts \cite{ExoGal,DH}.

Both non-relativistic limits are conveniently
derived from \eqref{VmuEq}, after reinstating the velocity
of light,
 $c$, and substituting $m\rightarrow mc$ and
$\psi\rightarrow e^{-imc^2t}\psi$, which gives
\begin{equation}
    \label{P0mc}
   P^0= \frac{1}{c}\left(i\frac{\partial}{\partial t}+m
    c^2\right)\,.
\end{equation}
Then Eqns. \eqref{HP1} and \eqref{HP2} take the equivalent form,
\begin{eqnarray}
    &\frac{1}{c}\sqrt{n+2\alpha} \left(i\frac{\partial}{\partial
    t}\right){\psi}_n+\sqrt{n+1}\,P_+
    {\psi}_{n+1}=0\,,&\label{1HP1}
    \\[6pt]
    &\sqrt{n+2\alpha}\,P_-{\psi}_n+\sqrt{n+1}\,
    \left(2mc+\frac{1}{c}\left(i\frac{\partial}{\partial
    t}\right)\right){\psi}_{n+1}=0\,,&\label{1HP2}
\end{eqnarray}
while Eq. (\ref{V0comp}) reduces to
\begin{equation}\label{V0compNR}
    \left(-\alpha \frac{1}{c}i\frac{\partial}{\partial t} +mcn)\right)\psi_n
    +\frac{1}{2}\left(\sqrt{(2\alpha +n-1)n}\,P_-\psi_{n-1}-
    \sqrt{(2\alpha+n)(n+1)}\,P_+\psi_{n+1}\right)=0.
\end{equation}
The Klein-Gordon equation,  a consequence of Eqns.
(\ref{1HP1})--(\ref{V0compNR}), takes here the equivalent  form
\begin{equation}\label{NR1+}
    \left(\left(1+\frac{i}{2mc^2}\partial_t\right)
    i\partial_t-\frac{1}{2m}\vec{P}{}^2\right)\psi=0\,.
\end{equation}
We put
\beqa\label{KiNR+}
    K_i&=&-\frac{1}{c}\epsilon_{ij}{\cal M}_j=
    -tP_i+mx_i-\epsilon_{ij}{\cal J}_j+\frac{1}{c^2}x_i\cdot
    i\partial_t\,,
    \\[8pt]
    {\rm J}&=&\mathcal{M}_0=\epsilon_{ij}x_iP_j+{\cal J}_0\,,
\eeqa
and get, for the commutator of the boosts,
\begin{equation}\label{KiKjJ}
    [K_i,K_j]=-i\frac{1}{c^2}\epsilon_{ij}\,{\rm J}\,,
\end{equation}
where $\epsilon_{ij}=-\epsilon_{ji}$, $\epsilon_{12}=1$.

Let us insist that all these formulas are still
relativistic; we simply wrote them in a form where the role
of the speed of light is highlighted.

%%%%%%%%%%%%%%%%%%%%%%%%%%%%%%%%%%%%%%%%%%%%%%%%%%%%%%%%%%%%%%%%%
%%%%%%%%%%%%%%%%%%%%%%%%%%%%%%%%%%%%%%%%%%%%%%%%%%%%%%%%%%%%%%%%%%%%%
\subsection{Usual non-relativistic limit}\label{uslimit}
%%%%%%%%%%%%%%%%%%%%%%%%%%%%%%%%%%%%%%%%%%%%%%%%%%%%%%%%%%%%%%%%%%%%%
%%%%%%%%%%%%%%%%%%%%%%%%%%%%%%%%%%%%%%%%%%%%%%%%%%%%%%%%%%%%%%%%%

From Eq. (\ref{1HP2}) we infer  that, as
$c\rightarrow\infty$, every subsequent component is
\emph{O$(\frac{1}{c})$}-times the previous one. The case of
a boson/fermion of nonzero spin  $j=n/2>0$ is described by a
$(2j+1)$-component field. Considering the non-relativistic
limit,  one could try to preserve all terms up to order
$1/c^n$, $n>1$. Though such an approximation would be
rather natural for boson/fermion fields of corresponding
spin, it will not possess a property of universality.
Indeed, particularly, as it follows from (\ref{NR1+}),  the
Hamiltonian operator in the Schr\"odinger equation then
will contain corrections with terms of the form
$(\vec{P}\,{}^2/mc^2)^k$, for which the highest value of
$k$ will depend on the  value of spin. Also, according to
(\ref{KiKjJ}), the boost generators will commute for
$j=1/2$, but will be non-commuting for $j>1/2$. The
question is then\,:  what kind of symmetry will underlie
the resulting theory?

Investigating the  general question goes beyond our scope
here; we consider, instead, a certain non-relativistic
limit which is characterized by the following properties\,:

\begin{itemize}
\item
It has a universal structure; in particular,  the
dynamics is described by the Schr\"odinger equation
with a Hamiltonian operator having the usual non-relativistic form;

\item
The corresponding Lie-algebraic symmetry is produced by
In\"on\"u-Wigner contraction from Poincar\'e symmetry;

\item
For $j=1/2$, it reproduces the
L\'evy-Leblond theory for a non-relativistic spin one-half field.

\end{itemize}

Let us denote the multicomponent field by $\Psi$, and
consider the (invertible) similarity transformation $\Psi\rightarrow
\Phi$,
\begin{equation}\label{rescaledPhi}
    \Phi = \mathbf{M} \Psi \,, \qquad \mathbf{M}=diag(1,c^1, c^2, ...)\,.
\end{equation}
The dimension of the field-column
\begin{equation}
    \Phi=\left(
    \begin{array}{c}
    \phi_0 \\ \phi_1 \\ \vdots
    \end{array}
    \right),\label{PHI}
\end{equation}
and of diagonal matrix $\mathbf{M}$  depends on the chosen
representation of the Lorentz algebra;
 in the anyon case, it is
infinite. The components of the transformed field $\Phi$,
unlike those of $\Psi$, are of order $c^0=1$ cf. Eq. (\ref{1HP2}).

The transformation (\ref{rescaledPhi}) induces a similarity
transformation of operators~: to any operator
$\mathfrak{O}$ acting on a state $\Psi$, there corresponds
an operator
\begin{equation}\label{similarO}
    \check{\mathfrak{O}}=\mathbf{M}\mathfrak{O}\mathbf{M}^{-1}\,
\end{equation}
that acts on $\Phi$. For the translational invariant (spin)
part of the Lorentz generators,
$\mathfrak{O}=\mathcal{J}_0,$ $\mathcal{J}_+,$
$\mathcal{J}_-$, we get then
\begin{equation}\label{Lorcheck}
    \check{\mathcal{J}}_0=\mathcal{J}_0\,,\qquad
    \check{\mathcal{J}}_+ = c
    {\mathcal{J}}_+\,,\qquad
    \check{\mathcal{J}}_- = \frac{1}{c}
    \mathcal{J}_- \, .
\end{equation}
The components of the vector operator $V_\mu$ of our basic
equations are transformed into
\begin{eqnarray}
    && \check{V}_0=-\frac{1}{c}
    \left( \alpha\, i\partial_t +\frac{1}{2}P_+\mathcal{J}_-
    \right)+ c\left(m(\mathcal{J}_0-\alpha)+\frac{1}{2}P_-
    \mathcal{J}_+\right), \\[6pt]
    && \check{V}_+= -(\mathcal{J}_0-\alpha)P_+ -
    \mathcal{J}_+i \partial_t  \,, \\[6pt]
    && \check{V}_-=(\mathcal{J}_0+\alpha)P_-+2m
    \mathcal{J}_-+\frac{1}{c^2} i \partial_t\,.
\end{eqnarray}
So far we merely deformed the  theory, which is still
relativistic. To obtain the non-relativistic limit of the
equations presented in terms of the rescaled field $\Phi$, it is
necessary to `renormalize' $\check{V}_0$ and consider
\begin{eqnarray}
    && \mathfrak{V}_0=\lim_{c\rightarrow\infty} \frac{1}{c} \check{V}_0 =
    m(\mathcal{J}_0-\alpha)+\frac{1}{2}\mathcal{J}_+P_-\,, \label{VnonR+1}\\[6pt]
    && \mathfrak{V}_+=- \lim_{c\rightarrow\infty} \check{V}_+ = (\mathcal{J}_0-\alpha)P_+
    +\mathcal{J}_+ i \partial_t \, \,, \label{VnonR+2}\\[6pt]
    && \mathfrak{V}_-=\lim_{c\rightarrow\infty}
    \check{V}_- = (\mathcal{J}_0+\alpha)P_-+2m
    \mathcal{J}_-\,.\label{VnonR+3}
\end{eqnarray}
Then we infer our new vector equations
\begin{equation}\label{NRPhieq}
    \mathfrak{V}_0 \Phi= 0\,,\qquad \mathfrak{V}_+
    \Phi= 0\,,\qquad \mathfrak{V}_- \Phi= 0\,.
\end{equation}
The dynamics of the field $\Phi$ is given by the Schr\"odinger
equation, $i\partial_t\Phi=\frac{1}{2m}\vec{P}\,{}^2\Phi$.  The
latter appears in fact as consistency (integrability) condition
for the system (\ref{NRPhieq}).

In component form the last two equations read
\begin{eqnarray}
    \sqrt{n+2\alpha} \left(i\frac{\partial}{\partial t}
    \right) \phi_n+\sqrt{n+1}\,P_+\phi_{n+1}&=&0\,,\label{2HP1}
    \\[6pt]
    \sqrt{n+2\alpha}\,P_-\phi_n+ 2m \sqrt{n+1}\,
    \phi_{n+1}&=&0\,,
    \label{2HP2}
\end{eqnarray}
where $n=0,1,\ldots,\infty$ for anyons ($\alpha>0$, or
$0>\alpha\neq -j$), and $n=0,1,\ldots,2j$ for bosons/fermions
($\alpha=-j$). The component form of the first equation from
(\ref{NRPhieq}) is reduced here to Eq. (\ref{2HP2}). Expressing
$\phi_{n+1}$ from Eq. \eqref{2HP1},
\begin{equation}\label{ulrecf}
    \phi_{n+1}=-\frac1{2m}\sqrt{\frac{n+2\alpha}{n+1}}\,
    \,P_-\phi_n\,.
\end{equation}
Note that the tower of states is automatically finite as it should
 if $2\alpha$ is a negative integer. Inserting (\ref{ulrecf})
into  \eqref{2HP2} shows explicitly that each component satisfies,
separately, the Schr\"odinger equation,
\begin{equation}\label{scheq}
    \left(i\frac{\partial}{\partial t}-
    \frac{\vec{P}{}^2}{2m}
    \right) \phi_n=0\,.
\end{equation}
Iterating (\ref{ulrecf})  allows us to express all
components $\phi_n$ in terms of the lowest one,
\begin{equation}
    \phi_{n}=\left(n B(2\alpha,n )\right)^{-1/2}
    \left(\frac{-P_-}{2m}\right)^{n} \,\phi_0\,,\label{nphi}
\end{equation}
cf. \eqref{npsi}.
 By Eq. (\ref{scheq}), $\phi_0$
is a (superposition of) plane waves,
\begin{equation}\label{psi0nr}
    \phi_{0}=\exp \left\{ -it
    \frac{\vec{p}{\,}^2}{2m}+i \vec{x}\cdot\vec{p} \right\}\,.
\end{equation}
Note that for integer/half-integer spin
$j=-\alpha=1/2,1,3/2,\ldots$, Eqns.
\eqref{ulrecf}-\eqref{nphi} yield $\phi_{n}=0$ for $n>2j$,
consistently with our expectation that spin-$j$ particle
should be described by a $(2j+1)$-component field, cf.
Section 3.

As in the relativistic case,  the infinite-component anyon
fields of spin $s=\alpha$ given in Eq. (\ref{alpha<-j}),
interpolate between the finite-component non-relativistic
fields of spins $j-1/2$ and $j$, respectively. Here,
however, each higher component $\phi_n$ is suppressed by
the additional hidden factor $\frac{1}{c^n}$ w.r.t. the
leading
 component $\phi_0$. When $\alpha$
tends to either of the boundary values $-j+1/2$ or $-j$,
the components $\phi_n$ with $n>2j-1$ or $n>2j+1$ are
suppressed, in addition, by the numerical factor
$(j-\frac{1}{2}+\alpha)^{1/2}$ or $(j+\alpha)^{1/2}$, see
Section 3.  In the case  of anyons based on the unitary
representations $D^+_\alpha$ with $\alpha>0$, no such
additional suppression arises.

Now we show that the system \eqref{2HP1}, \eqref{2HP2} has
a $1$-parameter centrally extended Galilei symmetry,
obtained by In\"on\"u-Wigner contraction from the $(2+1)$D
Poincar\'e algebra,
\begin{equation}\label{Poinc}
    [P_\mu,P_\nu]=0\,,\qquad
    [\mathcal{M}_\mu\,, P_\nu]=
    -i\epsilon_{\mu\nu\lambda}P^\lambda\,,\qquad
    [\mathcal{M}_\mu,\mathcal{M}_\nu]=
    -i\epsilon_{\mu\nu\lambda}\mathcal{M}^\lambda\,.
\end{equation}

 The Galilean boost generators are
defined by
\beq
   \mathcal{K}_i=-\lim_{c\rightarrow\infty}
   \epsilon_{ij}\check{\mathcal{M}}_j/c\,.
\eeq
With taking into account Eqns. (\ref{KiNR+}) and
(\ref{Lorcheck}), we get
\begin{equation}
    \mathcal{K}_\pm=\mathcal{K}_1\pm i\mathcal{K}_2
    =-tP_\pm+m x_\pm+\Delta_\pm\,,
    \qquad
    \Delta_-=0\,,\quad
    \Delta_+=i{\cal J}_+\,.
    \label{simpleboost}
\end{equation}
Note here the presence of the spin-dependent part $\Delta_+$.

The generator of rotations has, by (\ref{KiNR+}),
(\ref{Lorcheck}), the same form as in  relativistic
case,
\begin{equation}
    {\rm J}=\epsilon_{ij}x_iP_j + \mathcal{J}_0\,.
    \label{simplerot}
\end{equation}
Defining also
\begin{equation}\label{Hrenorm+}
    {\cal H}=cP^0-mc^2\,,
\end{equation}
we find that $P_i$, ${\cal K}_i$, ${\rm J}$ and ${\cal
H}=i\frac{\partial}{\partial t}$ are symmetry operators  of
the system \eqref{2HP1}, \eqref{2HP2}, and generate the
extended Galilei (``Bargmann'') algebra with a unique
central charge $m$,
\begin{eqnarray}
    &[{\cal K}_i,P_j]=im\delta_{ij}\,,\quad [P_i,P_j]=0\,,
    \quad [\mathcal{H},P_i]=0\,, \quad
    [\mathcal{K}_i,\mathcal{K}_j]=0\,,
    &\label{eG1}
    \\[6pt]
    &[\mathcal{K}_i,\mathcal{H}]=iP_i\,,\quad
    [{\rm J},P_i]=i\epsilon_{ij}P_j\,,\quad
    [{\rm J},\mathcal{K}_i]=i\epsilon_{ij}
    \mathcal{K}_j\,.
    &\label{eG2}
\end{eqnarray}
Note that by (\ref{KiKjJ}) boosts commute.

The Casimir operators of the algebra (\ref{eG1}),
(\ref{eG2}) are
\begin{equation}\label{Cas}
    \mathcal{C}_1=P_i^2-2m{\cal H}\,,\qquad
    \mathcal{C}_2=m{\rm J}-\epsilon_{ij}\mathcal{K}_iP_j\,.
\end{equation}
On-shell, i.e. on the surface of Eqns. \eqref{2HP1},
\eqref{2HP2}, they take the values
\begin{equation}
    \mathcal{C}_1=0\,, \qquad \mathcal{C}_2=\alpha\, m\,.
    \label{Casvl}
\end{equation}
The non-relativistic equations \eqref{2HP1}, \eqref{2HP2}
describe therefore a massive non-relativistic particle of
spin $\alpha$.

The particular cases of spin $1/2$ and spin $1$ will be discussed in Section
\ref{NRDDJT}.

%%%%%%%%%%%%%%%%%%%%%%%%%%%%%%%%%%%%%%%%%%%%%%%%%%%%%%%%%%%%%%%%%%%%%%%%%
%%%%%%%%%%%%%%%%%%%%%%%%%%%%%%%%%%%%%%%%%%%%%%%%%%%%%%%%%%%%%%%%%

\subsection{Exotic limit}\label{exlimit}

%%%%%%%%%%%%%%%%%%%%%%%%%%%%%%%%%%%%%%%%%%%%%%%%%%%%%%%%%%%%%%%%%
%%%%%%%%%%%%%%%%%%%%%%%%%%%%%%%%%%%%%%%%%%%%%%%%%%%%%%%%%%%%%%%%%%%%%%%%%

For the infinite-dimensional unitary representations
$D^+_\alpha$, $\alpha>0$,  we define the operators,
\begin{equation}
    \mathbf{N}=\mathcal{J}_0-\alpha\,,\qquad
    b^\pm=\frac{\mathcal{J}_\pm}{
    \sqrt{2\alpha}}\,,\label{Nb+b-}
\end{equation}
and put $\alpha \rightarrow \infty.$ In this limit, the operators
(\ref{Nb+b-}) generate the harmonic oscillator algebra,
\begin{equation}
    [\mathbf{N}, b^\pm]=b^\pm\,,
    \qquad [b^-,b^+]=1\,,\label{HANb}
\end{equation}
and the $\mathfrak{so}(2,1)$ representation \eqref{D+}
transforms into a Fock space representation,
\begin{equation}
    \label{Jbn}
    \mathbf{N}|n)=n|n), \qquad b^-|n)=\sqrt{n+1}|n+1),
    \qquad b^+
    |n)=\sqrt{n}|n-1), \qquad n=0,1,2,\ldots\,.
\end{equation}
The exotic non-relativistic limit
$
c\rightarrow \infty\,,\,
\alpha \rightarrow \infty\,,\,
\alpha/c^2=\kappa=const\,,
$
cf. (\ref{exolim}),
 applied to  the equations
\eqref{V+-}--\eqref{V0eq} with the substitution (\ref{P0mc}),
produces
\begin{eqnarray}
    &\Lambda_0\,{\psi}=0\,,\qquad \Lambda_+\,
    { \psi}=0\,,\qquad
    \Lambda_-\,{\psi}=0\,, &\label{lambdas}
    \\[8pt]
    &\Lambda_0=i\frac{\partial}{\partial t}-\frac{v_-P_+}{2} +
    \frac{v_+}{2}\Lambda_-\,, \quad
    \Lambda_+=\kappa v_+ \left( i\frac{\partial}{\partial
    t}-\frac{1}{2} v_- P_+\right)\,,
    \quad \Lambda_-=P_- - mv_-\,
    .\qquad&\label{Lamdet}
\end{eqnarray}
Here, $\Lambda_0$, $\Lambda_-$ and $\Lambda_+$ correspond
to the limit (\ref{exolim}) of the operators $-cV_0 /\alpha$,
$V_- /2\alpha$ and $V_+$, respectively, and we
have introduced the notation
\begin{equation}
    v_\pm =v_1\pm i v_2=-\sqrt{\frac{2}{\kappa}}b^\pm\,.\label{v+-}
\end{equation}
Taking the combination $\Lambda_0-v_+\Lambda_-$, we get a
minimal set of equations,
\begin{equation}
    \left(i\frac{\partial}{\partial
    t}-{H}\right){\psi}=0\,,\qquad \Lambda_-{\psi}
    =0\,, \label{minexeq}
\end{equation}
where
\begin{equation}\label{Hexo}
    H=\vec{P}\vec{v} -
    \frac{m}{2} v_+v_-\,
\end{equation}
is identified as the [linear-in-$P$] \emph{Hamiltonian
operator}.

 Taking into account the second equation
from (\ref{minexeq}), the first one becomes the
Schr\"odinger equation,
\begin{equation}
    \label{Sch0m}
    \left(i\frac{\partial}{\partial t}-
    \frac{1}{2m}\vec{P}{\,}^2\right){\psi}=0\,.
\end{equation}

On the other
hand, decomposing as $\psi=\sum_n \psi_n |n)$, the
equations $\Lambda_\pm\,\psi=0$  read
\begin{eqnarray}
    \sqrt{2\kappa} \left(i\frac{\partial}{\partial
    t}\right)\psi_n+\sqrt{n+1}\,P_+\psi_{n+1}=0\,,
    \\[6pt]
    \sqrt{2\kappa}\,P_-\psi_n+2m\sqrt{n+1}\psi_{n+1}=0\,.
    \label{exlim2}
\end{eqnarray}
These equations  can also be obtained, alternatively, by applying
the exotic limit (\ref{exolim}) to Eqns. \eqref{1HP1},
\eqref{1HP2}.
Using the second equation
from (\ref{exlim2}), the higher components can be
expressed in terms of the lowest one, namely as
$$
    \psi_n=\frac{1}{\sqrt{n!}}
    \left(-\sqrt{\frac{\kappa}{2}}\,\frac{P_-}{m}\right)^n\psi_0\,,
$$
while, according to the first equation, the
dynamics of each component is governed by the Schr\"odinger
equation, (\ref{Sch0m}).

The exotic limit
% (\ref{simpleboost})-(\ref{simpleH})-(\ref{simplerot}),
 with the additional `renormalization'
\beq
\mathcal{M}_0\rightarrow
\mathcal{M}_0-\alpha=\mathcal{M}_0-\kappa c^2,
\eeq
 cf. (\ref{Hrenorm+}), provides us with rotation and
Galilean boosts generators,
\begin{equation}\label{JK}
    {\rm J}=\epsilon_{ij}x_iP_j+\frac{1}{2}\kappa
    v_+v_-\,,\qquad
    \mathcal{K}_i= mx_i-tP_i+\kappa\epsilon_{ij}v_j\,.
\end{equation}
Together with ${\cal H}=i\frac{\partial}{\partial t}$ and momentum
$P_i$, they generate the exotic Galilei symmetry,
\begin{eqnarray}
    &[\mathcal{K}_i,P_j]=im\delta_{ij}\,,\quad
    [P_i,P_j]=0, \quad [\mathcal{H},P_i]=0\,,\quad
    [\mathcal{K}_i,\mathcal{K}_j]=-i\kappa\epsilon_{ij}\,,&\label{EG1}
    \\[8pt]
    &[\mathcal{K}_i,{\cal H}]=iP_i\,,\quad
    [{\rm J},P_i]=i\epsilon_{ij}P_j\,,\quad
    [{\rm J},\mathcal{K}_i]=i\epsilon_{ij}\mathcal{K}_j\,.&\label{EG2}
\end{eqnarray}
In particular, the parameter $\kappa$, which
measures the non-commutativity of boosts,  becomes the second
central charge, highlighting the exotic Galilean symmetry
\cite{ExoGal,DH}. The
Casimir operators are
\begin{equation}\label{CC}
    \mathcal{C}_1=P_i^2-2m{\cal H}\,,\qquad
    \mathcal{C}_2=m{\rm J}-\epsilon_{ij}
    \mathcal{K}_i P_j +\kappa
    \mathcal{H}\,.
\end{equation}
On-shell, they take the values
$\mathcal{C}_1=\mathcal{C}_2=0$. The system of equations
\eqref{minexeq} describes an exotic (2+1)D non-relativistic
particle with mass $m$ and the exotic parameter $\kappa$. The
system may be reinterpreted as a free massive particle on the
non-commutative plane, see \cite{HPPLB}. \vskip0.2cm

 The exotic
limit can be carried out also by letting $\alpha\rightarrow
-\infty$, that involves the non-unitary representations
$\tilde{D}^j$ and $\tilde{D}^+_\alpha$. We reproduce here
the same results as for $\alpha\rightarrow +\infty$.
Putting $\alpha=-j$, $j\rightarrow \infty$, the
limit can be applied, in particular, to  usual boson/fermion fields.

Starting  from Eqns. (\ref{C''D}), (\ref{C-j+e}), we  now
consider the exotic non-relativistic limit (\ref{exolim}),
i.e., $ c\rightarrow \infty\,,\ j \rightarrow \infty\,,\
j/c^2=\kappa=const\,. $ The equations  \eqref{lambdas},
with corresponding operators (\ref{Lamdet}), are obtained
by applying this limit to
$$
(\kappa c)^{-1} V_0\psi=0,
\qquad
V_+\psi=0,
\qquad
-\kappa
(2\kappa c^2)^{-1} V_-\psi=0,
$$
 respectively.  The
oscillator number and creation-annihilation operators
are found by applying the limit $j\rightarrow\infty$ to
\begin{equation}
    \mathbf{N}=\mathcal{J}_0+j+\epsilon\,,\qquad
    b^\pm=\pm
    \frac{\mathcal{J}_\pm}{\sqrt{2(j+\epsilon)}}\,.\label{b+-def}
\end{equation}
Note that while the nature of the operators (\ref{C''D})
requires an indefinite metric, due to
the sign in the definition of $b^\pm$ in (\ref{b+-def}),
we have now $\mathbf{N}^\dagger=\mathbf{N}$,
$(b^+)^\dagger=b^-$ with respect to the usual, positive
definite metric, and reproduce the relations (\ref{Jbn}).

%%%%%%%%%%%%%%%%%%%%%%%%%%%%%%%%%%%%%%%%%%%%%%%%%%%%%%%%%%%%%%%%%%%%%%%%%
%%%%%%%%%%%%%%%%%%%%%%%%%%%%%%%%%%%%%%%%%%%%%%%%%%%%%%%%%%%%%%%%%%%%%%%%%%%%%%%%%%%%%

\section{Non-relativistic supersymmetry}\label{secNRSUSY}

%%%%%%%%%%%%%%%%%%%%%%%%%%%%%%%%%%%%%%%%%%%%%%%%%%%%%%%%%%%%%%%%%%%%%%%%%%%%%%%%%%%%%
%%%%%%%%%%%%%%%%%%%%%%%%%%%%%%%%%%%%%%%%%%%%%%%%%%%%%%%%%%%%%%%%%%%%%%%%%

Studying the non-relativistic limit of the vector equation
 \eqref{VmuEq}, we derived first order equations for
 bosons and fermions, and for anyons. The symmetry of
 these systems is the mass-extended Galilei symmetry, or the exotic-Galilei
 symmetry, depending on the type of limit.
 They correspond to different In\"on\"u-Wigner
 contractions of the Poincar\'e group.

In Sections \ref{susyN1gen} and  \ref{N2susy}, the relativistic
wave equation \eqref{VmuEq} was generalized to \eqref{Vsusy}, or
\eqref{Vsusy2}, respectively. They describe
$N=1$ or $N=2$ Poincar\'e
supermultiplets. It is natural to expect, therefore,
that the non-relativistic limits of these equations will describe
(exotic) Galilei supermultiplets. It is also expected that the
symmetries of these systems will be the In\"on\"u-Wigner
contraction of the super-Poincar\'e algebra.

In the case of $N=1$ supersymmetry, for instance, the
supermultiplet is described by \eqref{Vsusy}.
Projected to the corresponding subspaces, it yields Eqns.
\eqref{Vpsi+-}, $V^{(\alpha_+)}_\mu\psi^+=0$ and
$V^{(\alpha_-)}_\mu\psi^-=0$. The fields $\psi^+$ and
$\psi^-$ have spin $\alpha_+$ and $\alpha_-$, respectively.
The procedure described in
Section~\ref{NRlimit} to obtain the nonrelativistic
limit can be repeated in each subspace.
Having in mind that the fields $\psi^\pm$ are expanded on
the even (odd) Fock subspaces of the RDHA, see Eqns.
\eqref{psi+psi-} and \eqref{calF+-},
$\psi^\pm(x)=\sum_{n=0} \psi^\pm_n(x)|n\rangle_\pm$, we
first obtain  \eqref{1HP1}, \eqref{1HP2}.
Then, taking the limit, yields Eqns. \eqref{2HP1} and
\eqref{2HP2} with parameters $\alpha_+$ and
$\alpha_-=\alpha_++1/2$, respectively. Each set of non-relativistic
fields will be Galilei-symmetric, and the fields are
interchanged by the non-relativistic supercharge,
 obtained as the non-relativistic limit of the relativistic supercharge \eqref{QL}.

The extended $N=2$ supersymmetry is derived by a similar
procedure, starting from
 equations \eqref{Vsusy2} and applying
 In\"on\"u-Wigner contraction.
We present this in detail in the following subsections.

%%%%%%%%%%%%%%%%%%%%%%%%%%%%%%%%%%%%%%%%%
\subsection{Usual  $N=1$ non-relativistic supermultiplet}
%%%%%%%%%%%%%%%%

In this case we get two copies of the non-relativistic
equations \eqref{2HP1}, \eqref{2HP2},
\begin{eqnarray}
    \sqrt{n+2\alpha_\pm} \left(i\frac{\partial}{\partial t}\right)
     \phi^\pm_n+\sqrt{n+1}\,P_+\phi^\pm_{n+1}=0,
     \label{HP+-1}
     \\[6pt]%\qquad
     \sqrt{n+2\alpha_\pm}\,P_-\phi^\pm_n+ 2m \sqrt{n+1}\,
     \phi^\pm_{n+1}=0.\label{HP+-2}
\end{eqnarray}
Defining $\phi^\pm_n =c^n\psi^\pm_n$ and taking the limit
$c\rightarrow \infty$ as in \eqref{2HP1}, \eqref{2HP2}, we get,
  cf. \eqref{nphi}, \eqref{psi0nr},
\begin{equation}
    \phi^\pm_{n}=\left(n B(2\alpha_\pm,n )\right)^{-1/2}
     \left(\frac{-P_-}{2m}\right)^{n} \,\phi_0,\qquad
      \phi^\pm_{0}=A^\pm\exp
      \left\{ -it \frac{\vec{p}{\,}^2}{2m}+i
      \vec{p}\vec{x}
    \right\}\,,
\end{equation}
where the $A^\pm$ are constants. The Galilean generators are
defined as in \eqref{simpleboost}, (\ref{simplerot}),
(\ref{Hrenorm+}). Augmented with the translation generators $P_i$,
they span the algebra \eqref{eG1}, \eqref{eG2}.
 Here  the reducible representations
\eqref{J0,Ji} of the Lorentz algebra  (see also \eqref{AJ+-}) are
used for ${\cal J}_+$ and ${\cal J}_0$  in the
boost and rotation operators,
${\cal K}_+$ and ${\rm J}$, respectively. The
operator $\mathcal{C}_2$, defined as in \eqref{Cas}, is therefore
multi-valued. On-shell, we have
\begin{equation}
    \mathcal{ C}_2=m\hat{\alpha}\,,\qquad \mathcal{C}_2
    \Phi^\pm=\mathcal{C}^\pm_2 \Phi^\pm\,, \qquad
    \mathcal{C}^\pm_2=m\alpha_\pm\,.
    \label{C2susy}
\end{equation}
The states $\Phi^\pm$ are vectors of the form
 \eqref{PHI}, with components $\phi^\pm_n$.
The operator $\mathcal{C}_1$ vanishes in both subspaces.

The supercharges that interchange the fields $\Phi^\pm$
are the nonrelativistic limits of those in
 \eqref{QL}. To identify them, we note that
the multi-component field $\Psi$ of our $N=1$ supermultiplet has the
structure $\Psi^T=(\psi_0^+,\psi^-_0,\psi_1^+,\psi^-_1,\ldots)$,
where $T$ denotes transposition. The
rescaled field $\Phi$  has  analogous structure. They are  related by the transformation of
the form (\ref{rescaledPhi}) with a matrix $\mathbf{M}$
\begin{equation}\label{MatrixRescSUSY}
    \mathbf{M}=diag (1,1,c,c,c^2,c^2,\ldots)=\mathbf{M}_+
    +\mathbf{M}_-,\qquad \mathbf{M}_\pm=\mathbf{M}\Pi_\pm\,,
\end{equation}
where $\Pi_+$ and $\Pi_-$ are the projectors (\ref{projector}).
The similarity transformation (\ref{similarO}), applied to the RDHA
creation and annihilation operators, gives
\begin{equation}\label{Resca+a-}
    \check{a}^+=ca^+\Pi_-+a^+\Pi_+,\qquad
    \check{a}^-=c^{-1}a^-\Pi_++a^-\Pi_-\,.
\end{equation}
From  \eqref{QQpm} and (\ref{Resca+a-})
 we infer the non-relativistic supercharges,
\begin{equation}\label{q+-def}
    \mathcal{ Q}_\pm={\lim}_{c\rightarrow \infty}
    \frac{\check{Q}_\pm}{c}\,,
\end{equation}
\begin{equation}
    \mathcal{Q}_+=2i\,\sqrt{\frac{m}{1+\nu}}\,a^+\Pi_+\,,\qquad
    \mathcal{Q}_-=-2i\,\sqrt{\frac{m}{1+\nu}}\left(a^-+\frac{1}{2m}P_-a^+\right)\Pi_-
    \,,\label{q+-}
\end{equation}
where $\nu\neq -1$. The only  nontrivial anticommutator is
\begin{equation}
    \{\mathcal{ Q}_+,\mathcal{ Q}_-\}=\frac{8}{1+\nu}\,
    \mathcal{S}\,,
    %\qquad \mathcal{S}=m\left(\mathcal{J}_0+
    %\frac{R}{4}+\frac{\nu}{4}\right)\,.\label{Q+Q-cNR}
\end{equation}
where
\begin{equation}
    \mathcal{S}=\mathcal{ C}_2-
    m\hat{\alpha}+m\frac{1+\nu}{2}=\mathfrak{V}_0+\frac{1}{2}m(1+\nu)\,,
\end{equation}
with $\mathfrak{V}_0$ defined in (\ref{VnonR+1}). The operator
$\mathcal{S}$ commutes with all Galilei generators.
 Unlike $\mathcal{ C}_2$, $\mathcal{S}$
commutes also with the supercharges $\mathcal{ Q}_\pm$. It
can be identified [up to the factor $m$], therefore, with
the superspin Casimir operator. Taking into account
\eqref{C2susy} [or the first equation from
(\ref{NRPhieq})],  we get on-shell
$\mathcal{S}=m\frac{1+\nu}{2}.$ The supercharges $\mathcal{
Q}_\pm$ extend the Galilei algebra \eqref{eG1}-\eqref{eG2}
with on-shell (anti)commutation relations,
\begin{eqnarray}
    &[{\rm J},\mathcal{Q}_\pm]=\pm
    \frac{1}{2}\mathcal{Q}_\pm\,,
    \qquad
     \{\mathcal{Q}_+,\mathcal{Q}_-\}=4m\,,&\label{JQQ1}
     \\[6pt]
     & [\mathcal{K}_i,\mathcal{ Q}_\pm]=
     [P_i,\mathcal{Q}_\pm]=[\mathcal{H},\mathcal{
    Q}_\pm]=0\,,\qquad
    \mathcal{ Q}_\pm^2=0\,,&\label{JQQ2}
\end{eqnarray}
giving rise to the $N=1$ super-extension of the Galilei
algebra \cite{GalSUSY} \footnote{ The $N=1$ super-Galilei
algebra \cite{GalSUSY} has, besides the $\mathcal{Q}_\pm$,
one more supercharge, namely the square root of $H$. It is,
however, \emph{not} obtained as the non-relativistic limit
of relativistic SUSY. The possibility of extending the
Galilei symmetry conformally, yielding a generalized
Schr\"odinger symmetry from $s=1/2$ \cite{DHP} to any spin,
as well as its supersymmetric extensions, \cite{SchSUSY},
are, in general, open questions. (See, however,
\cite{HPVLett}). The extension of exotic symmetry  appears
to be non-linear \cite{Alvarez,HMS}.}.

%%%%%%%%%%%%%%%%%%%%%%%%%%%%%%%%%%%%%%%%%%%%%%%%%%%%%%%%
\subsection{Exotic $N=1$ non-relativistic supermultiplet}
%%%%%%%%%%%%%%%%%%%%%%%%%%%%%%%%%%%%%%%%%%%%%%%%%%%%%%%%

Based on the deformed-oscillator representation of the
 Lorentz algebra, we define now the operators,
\begin{equation}
    \mathbf{N}=\mathcal{J}_0-\hat{\alpha}\,,\qquad
    b^\pm=\sqrt{\frac{2}{\nu}}\,\mathcal{J}_\pm\,,\label{Nb+b-2}
\end{equation}
and take the limit $\nu \rightarrow \infty.$
%Making use of  the equations \eqref{JJ}, \eqref{J0N+-}, \eqref{CnJ+-} %and \eqref{Cn},
 In this limit we reproduce the usual harmonic
oscillator algebraic relations of the form (\ref{HANb}),
\begin{equation}
    [\mathbf{N}, b^\pm]=b^\pm\,,
    \qquad [b^-,b^+]=1\,.\label{HANb2}
\end{equation}
However, here we get the direct sum of two irreducible
representations of the Heisenberg algebra, realized in the
subspaces $\mathcal{F}_\pm$ of the total Fock space
$\mathcal{F}$ of the RDHA (see \eqref{calF+-},
\eqref{FFpm}),
$$
    \mathbf{N}|n\rangle_\pm=n|n\rangle_\pm\,,
    \quad
    b^-|n\rangle_\pm=\sqrt{n+1}|n+1\rangle_\pm\,,
    \quad b^+
    |n\rangle_\pm=\sqrt{n}|n-1\rangle_\pm\,,
    \quad n=0,1,2,\ldots\,.
$$
Generalizing the exotic limit (\ref{exolim}), we define,
cf. (\ref{exolim}),
\begin{equation}
    \label{exolim2+}
     c\rightarrow \infty\,,\qquad
      \nu \rightarrow \infty\,,\qquad
       \frac{\nu/4}{c^2}=\kappa=const\,.
\end{equation}
This limit produces $\mathcal{J}_\pm/c
\rightarrow \sqrt{2\kappa}\,b^\pm$ and,
\begin{equation}
    \frac{\hat{\alpha}}{c^2}
    \rightarrow \kappa\,.\label{kappa2}
\end{equation}
It follows that applying
 \eqref{exolim2+} to the supersymmetric equations \eqref{Vsusy}
 with the substitution (\ref{P0mc}) yields the
 \emph{same} equations
 \eqref{lambdas} and \eqref{minexeq} as in the non-supersymmetric case.
 There, the
 $v_\pm$ operators are defined as in \eqref{v+-}, in terms
 of the operators \eqref{Nb+b-2}.

The Poincar\'e reducibility of the system is revealed
by a nontrivial integral,  namely the
reflection operator $R$. We have
\begin{equation}
    \left(i\frac{\partial}{\partial
    t}-{H}\right)\psi^\pm=0\,,\qquad \Lambda_-\psi^\pm
    =0\,, \qquad R\psi^\pm=\pm \psi^\pm\,,
    \label{minexeq2}
\end{equation}
where the operators $H$ and $\Lambda_-$, which commute with
$R$, are formally given by the same relations as in
(\ref{Hexo}), (\ref{Lamdet}).

 The finite part of the operator  $\mathcal{M}_0$ is
now
\begin{equation}
    {\rm J}=(\epsilon_{ij}x_iP_j+\mathcal{J}_0)-
    \frac{\nu}{4}=\epsilon_{ij}x_iP_j+ \mathbf{N}+
    \frac{1}{2}-\frac{1}{4}R\,.\label{JSex}
\end{equation}
The exotic Galilean generators are given by
\begin{equation}\label{HJK}
    {\rm J}=
    \epsilon_{ij}x_iP_j+\frac{1}{2}\kappa
    v_+v_-+\frac{1}{2}-\frac{R}{4}\,,\qquad
    \mathcal{K}_i= mx_i-tP_i+\kappa\epsilon_{ij}v_j\,.
\end{equation}
Together with time and space translations, ${\cal
H}=i\frac{\partial}{\partial t}$,  $P_i$, they generate
on-shell the algebra \eqref{EG1}, \eqref{EG2}. The Casimir
operator $\mathcal{C}_1$, defined in \eqref{CC},
vanishes on-shell. In turn,  the operator
 $\mathcal{C}_2$, \eqref{CC}, takes different eigenvalues
when  acts on the $\psi^\pm$ wave functions,
\begin{equation}
    \mathcal{C}_2=m\frac{2-R}{4}\,,\qquad
     \mathcal{C}_2\psi^\pm=\mathcal{C}^\pm_2\psi^\pm\,,
     \qquad \mathcal{C}^\pm_2=
     m\left(\frac{1}{2}\pm \frac{1}{4}\right).
    \label{C2NR}
\end{equation}

The supercharges associated with the system
\eqref{minexeq2} are obtained from the exotic limit of
\eqref{QQpm},
\begin{equation}
    \mathcal{ Q}_\pm={\rm lim}_{c,\nu\rightarrow \infty}\,
    \frac{1}{\sqrt{c}}Q_\pm= \pm 2i \sqrt{m} \left(
     {\rm lim}_{\nu \rightarrow \infty}\, \Pi_\mp
     \frac{a^\pm}{\sqrt{\nu}}  \right) .\label{q+-E}
\end{equation}
On account of the realization \eqref{a+a-F}  of the RDHA generators
in terms of the bosonic oscillator operators, we get
\begin{equation}
    {\rm lim}_{\nu\rightarrow \infty}
    \frac{a^+}{\sqrt{\nu}} =\frac{1}{\sqrt{\tilde{\mathcal{N}}}}
    \;\tilde{a}^+\tilde{\Pi}_+\equiv c^+\,,\qquad
    {\rm lim}_{\nu\rightarrow \infty}
    \frac{a^-}{\sqrt{\nu}} =\frac{1}{\sqrt{1+\tilde{\mathcal{N}}}}
    \;\tilde{a}^-\tilde{\Pi}_-\equiv c^-\,.\label{c+-}
\end{equation}
The operators $c^\pm$ and $\tilde{R}=R$ satisfy the same
algebra as the Pauli matrices
$\sigma_\pm=\frac{1}{2}(\sigma_1\pm i\sigma_2)$ and
$\sigma_3$,
\begin{equation}
    \{c^+,\,c^-\}=1,\qquad c^\pm{}^2=0\,,
    \quad [c^+,\,c^-]=R\,,\quad
    \{c^\pm,R\}=0\,,\quad
    R^2=1\,.\label{}
\end{equation}
We obtain therefore
\begin{equation}
    \mathcal{ Q}_\pm= \pm 2i
    \sqrt{m} \,c^\pm\,.\label{q+-E1}
\end{equation}
The supercharges $\mathcal{ Q}_\pm$ extend the exotic
Galilei algebra \eqref{EG1}-\eqref{EG2} with
(anti)commutation relations identical to those in
\eqref{JQQ1}-\eqref{JQQ2}. The only difference
 with the
usual non-relativistic limit of the previous subsection is
the non-commutativity of Galilean boosts.

The Casimir operator operator (\ref{C2NR}) is generalized
to
$$
    \mathcal{S}=\mathcal{C}_2+\frac{1}{16}
    [\mathcal{ Q}_+,\mathcal{ Q}_-]\,,
$$
that commutes, unlike $\mathcal{C}_2$, with all
superalgebra generators, including the supercharges
$\mathcal{ Q}_\pm$, and on-shell takes the value
$\frac{1}{2}m$. It is identified [up to the factor $m$] as
the superspin operator.

%%%%%%%%%%%%%%%%%%%%%%%%%%%%%%%%%%%%%%%%%%%%%%%%%%%%%%%%%%%%%%%%%%%%%%%%%%%%%%%%%%%%%%%%%%%%%%%
\subsection{Nonrelativistic Dirac/DJT
supermultiplet}\label{NRDDJT}
%%%%%%%%%%%%%%%%%%%%%%%%%%%%%%%%%%%%%%%%%%%%%%%%%%%%%%%%%%%%%%%%%%%%%%%%%%%%%%%%%%%%%
%%%%%%%%%%%%%%%%%%%%%%%%%%%%%%

$\bullet$ For spin one-half ($-\alpha=1/2$), putting
\begin{equation}\label{s12LL}
     {\phi}_0=\psi_0\,,\quad {\phi}_1=c\psi_1,
\end{equation}
the general theory of Section \ref{NRlimit} reduces, for
$n=0$, to the non-relativistic analogs of the $(2+1)$D
Dirac equation due to the  L\'evy-Leblond
 \cite{LLeq,DHP}, which take here a form
\begin{equation}
\barray{lll}
    i\displaystyle\frac{\partial}{\partial t}
    {\phi}_0 + -iP_+{\phi}_1&=&0\,,
     \\[8pt]
      P_-{\phi}_0-2im \phi_1&=&0\,.
     \earray
\label{LLequations}
\end{equation}
Eliminating one component yields the Schr\"odinger equation
for the other one. Eqns. (\ref{2HP1})--(\ref{2HP2}) for
$n=1$ and then for any $n>1$ yield $\phi_n=0$.

The boosts operators (\ref{simpleboost}) involve nontrivial spin \cite{LLeq,DHP},
\begin{equation}
    \mathcal{K}_\pm=\mathcal{K}_1\pm i\mathcal{K}_2
    =-tP_\pm+m x_\pm+\Delta_\pm\,,
    \qquad
    \Delta_-=0\,,\qquad
    \Delta_+=\left(%
\begin{array}{cc}
  0 & 0 \\
  -1 & 0 \\
\end{array}%
\right)\,.
    \label{simple1/2boost}
\end{equation}

The angular momentum, (\ref{simplerot}), reads in turn,
\begin{equation}
    {\rm J}=\epsilon_{ij}x_iP_j + \mathcal{J}_0\,,
    \qquad
    \mathcal{J}_0=\left(\barray{cc}
    -1/2&0\\0&1/2\earray\right).
    \label{simple1/2rot}
\end{equation}

%%%%%%%%%%%%%%%%%%%%%%%%%%%%%%%%%%%%%%%%%%%
%\subsection{Nonrelativistic Deser-Jackiw-Templeton equation}
%%%%%%%%%%%%%%%%%%%%%%%%%%%%%%%%%%%%%%%%%%%

$\bullet$  For $-\alpha=j=1$ one gets instead the
non-relativistic analog of the DJT vector equation for a
topological massive gauge field.
Introducing the redefined fields
\begin{equation}\label{phipsi}
   \varphi_0=\psi_0\,,\qquad
    \varphi_1=c\psi_1\,,\qquad
    \varphi_2=c^2\psi_2\,,
\end{equation}
yields, putting $n=0$ into (\ref{2HP1})-(\ref{2HP2}) and
multiplying by $-i$, the pair of equations
$$
\begin{array}{lll}
\sqrt{2}\,i\frac{\partial}{\partial t}\varphi_0 -i
P_+\varphi_1&=&0\,,
\\[8pt]
\displaystyle \sqrt{2}\,P_-\varphi_0-2im\varphi_1&=&0\,.
\end{array}
$$
Similarly, for $n=1$ we get
$$
\begin{array}{lll}
P_-\varphi_1-i2\sqrt{2}\,m \varphi_2 &=&0\,,
\\[8pt]
i\displaystyle\frac{\partial}{\partial t}\varphi_1 -i
    \sqrt{2}\,P_+\varphi_2&=&0\,.
 \end{array}
$$
The last equation here is readily seen to follow from the first
three, however, and can therefore be dropped. For $n=2$ and then
for any $n>2$, the system (\ref{2HP1})-(\ref{2HP2}) yields
$\phi_n=0$. In conclusion, the non-relativistic limit of the
 Deser-Jackiw-Templeton equations (\ref{DJT}) reads,
\begin{equation}\begin{array}{lll}
    \displaystyle
    i\frac{\partial}{\partial t}\varphi_0 -
    \frac{i}{\sqrt{2}}\,P_+\varphi_1&=&0\,,
     \\[12pt]
     i\sqrt{2}\,m\varphi_1 -P_-\varphi_0 &=&0\,,
      \\[10pt]
      2\sqrt{2}\,m \varphi_2 + iP_-\varphi_1&=&0\,.
     \end{array}
     \label{NRDJT}
\end{equation}
$\varphi_1$ and $\varphi_2$ play the role of auxiliary
fields, and may be expressed in terms of $\varphi_0$. The
dynamics of each of these free fields is governed by the
Schr\"odinger equation.

The spin contribution to the boost, (\ref{simpleboost}),
and to the angular momentum, (\ref{simplerot}), read
\begin{equation}\label{DeltaDJT}
    \Delta_+=\left(%
\begin{array}{ccc}
  0 & 0 & 0\\
  -\sqrt{2} & 0 & 0\\
  0  & -\sqrt{2} & 0
\end{array}
\right)\,,
\qquad
\mathcal{J}_0=\left(\barray{ccc}
    -1&0&0\\0&0&0\\
    0&0&1\earray\right).
\end{equation}

%%%%%%%%%%%%%%%%%%%%%%%%%%%%%%%
%\subsection{The supermultiplet}
%%%%%%%%%%%%%%%%%%%%%%%%%%%%%%%

The L\'evy-Leblond (spin $s=-1/2$) and
non-relativistic DJT  (spin $s=-1$) systems
can now be unified into an $N=1$
supermultiplet following the general recipe of Section
\ref{secNRSUSY}.
The non-relativistic spin one half, (\ref{s12LL}),
and spin one, (\ref{phipsi}), multiplets are thus unified
into
\begin{equation}
    \mathbf{\Phi} = \left(
    \begin{array}{c}
    \varphi_0   \\
    {\phi}_0  \\
    \varphi_1   \\
    {\phi}_1  \\
    \varphi_2  %
    \end{array}
\right).\label{superDDJTvector}
\end{equation}
The Galilei generators  $P_i$ and \eqref{simplerot} with ${\cal
J}_0=diag\, (-1,-1/2,0,1/2,1)\,$
[cf. (\ref{finitecase20})]  act diagonally on the
vectors (\ref{superDDJTvector}). The corresponding
supercharges are  \eqref{q+-} with $\nu=-5$, i.e.
\begin{equation}
    \mathcal{Q}_+=\sqrt{m}\,a^+\Pi_+\,,\qquad
    \mathcal{Q}_-=-\sqrt{m}\,\left(a^-+\frac{1}{2m}P_-a^+\right)\Pi_-
    \,,\label{sQDDJT}
\end{equation}
where the odd $\mathfrak{osp}(1|2)$ generators $a^\pm$  and
projectors $\Pi_\pm$ are those matrices \eqref{5a+a-} and
\eqref{RPI}. Note the asymmetry between $\mathcal{Q}_+$ and
$\mathcal{Q}_-$, which is consistent with the one between
$\mathcal{K}_+$ and $\mathcal{K}_-$ in Eqns.
(\ref{simple1/2boost}), (\ref{DeltaDJT}).

The Dirac -- DJT supermultiplet, in particular its
 superGalilei symmetry, are further discussed in \cite{HPVLett}. See also \cite{3TMGGD}.

%\vskip30mm
%%%%%%%%%%%%%%%%%%%%%%%%%%%%%%%%%%%%%%%%%%%%%%%%%%%%%%%%%%%%%%%%%%%%%%%%%%%%%%%%%
\section{Extended $N=2$ non-relativistic
supersymmetry}\label{secN2NRSUSY}
%%%%%%%%%%%%%%%%%%%%%%%%%%%%%%%%%%%%%%%%%%%%%%%%%%%%%%%%%%%%%%%%%%%%%%%%%%%%%%%%%
Here we investigate the non-relativistic limit of the
relativistic  equations
\eqref{Vchi} with $N=2$ supersymmetry. The symmetries
of the corresponding four-particle
non-relativistic supermultiplet are obtained  by
In\"on\"u-Wigner contraction of the extended superPoincar\'e
algebra \eqref{SuperPoi1}, \eqref{QQant}-\eqref{QQanti}.
The limit can be usual, or exotic, or a mixture of both
kinds, since there are two independent spin parameters.

%%%%%%%%%%%%%%%%%%%%%%%%%%%%%%%%%%%%%%%%%
\subsection{Usual non-relativistic extended supersymmetry}

First we consider the usual non-relativistic limit of
the four-anyon system in  Eqns.\eqref{Vchi},
$$\barray{lll}
    V_\mu^{(\chi)}\psi^{++}(x)&=&0\,,\\[4pt]
    V_\mu^{(\chi+1/2)}\psi^{+-}(x)&=&0\,,\\[4pt]
    V_\mu^{(\chi+1/2)}\psi^{-+}(x)&=&0\,, \\[4pt]
    V_\mu^{(\chi+1)}\psi^{--}(x)&=&0\,.
    \earray
$$
 For $\psi^{++}(x)$, for instance, the simple limit yields
\begin{eqnarray}
    \sqrt{n+2\chi} \left(i\frac{\partial}{\partial t}
    \right) \Phi^{++}_n+\sqrt{n+1}\,P_+\Phi^{++}_{n+1}&=&0\,,
    \\[6pt]
    \sqrt{n+2\chi}\,P_-\Phi^{++}_n+ 2m \sqrt{n+1}\,
    \Phi^{++}_{n+1}&=&0\,,
\end{eqnarray}
obtained as in \eqref{HP+-1}, \eqref{HP+-2}. Identical
equations can be derived for the three remaining fields,
taking into account that the spin parameter in
$V_\mu^{(\hat{\emph{\ss}\;})}$ takes the values,
\begin{equation}
    \hat{\emph{\ss}\;}=diag\left(\chi,\chi+\frac{1}{2},
     \chi+\frac{1}{2},\chi+1\right),
     \quad \chi = \alpha_{\sc 1 +} +
     \alpha_{\sc 2 +} = \frac{1}{4}(1+\nu_{\sc 1})
     +\frac{1}{4}(1+\nu_{\sc 2}),
\end{equation}
as in \eqref{ssN2}.
The index $n$ of $\Phi$ refers to the lowest
weight representations of the Lorentz algebra of
spin $\emph{\ss}$, from which the nonrelativistic
limit was taken. These lowest weight representations
correspond to linear combinations of the Fock basis
\eqref{psidecomp3} (see also \eqref{psidecomp} and \eqref{psidecomp2}).
Hence,
\begin{equation}
\hat{\emph{\ss}\;}=\hat{\alpha}_{\sc 1 +}
    +\hat{\alpha}_{\sc 2 +}=\chi + \frac{1}{2}\left(\Pi_{\sc 1 -} +
     \Pi_{\sc 2 -}\right), \quad \chi = \alpha_{\sc 1 +} +
     \alpha_{\sc 2 +} = \frac{1}2
     + \frac{1}{4}(\nu_{\sc 1}+\nu_{\sc 2}).%\label{}
\end{equation}

The nonrelativistic limit of time translations, Galilei
 boosts and rotation generators  is obtained as in
Section \ref{NRlimit}.

 This representation is, of course, reducible,
 since that of the Poincar\'e algebra was reducible too.
In particular, rotations also involve the number operators
of the lowest-weight representations of the Lorentz algebra
as well as  $\hat{\emph{\ss}\;}$,
%%%%%%%%%%%%%%%%%%%%%%%%%%%%%%%%%%%%%%%%%%%%%%%%%%%%%%%%%%%%%%%%%%%%%%%%%%%%%%%%%
\footnote{ Here $\Pi_{\sc A \pm}=(1\pm R_{\sc A})/ 2$ are the projection
operators on Fock space of every oscillator labeled by ${\un A}$.
 They satisfy the relations \eqref{projector} for every index, and commute.
 The numbers operators satisfy,
$$\mathbf{N}_{\sc A}|n_1,n_2\rangle_{\pm \pm}=n_A |n_1,n_2\rangle_{\pm \pm},\qquad
 \mathbf{N}_{\sc A}|n_1,n_2\rangle_{\pm \mp}=n_A |n_1,n_2\rangle_{\pm \mp},\qquad A=1,2.
$$
}
%%%%%%%%%%%%%%%%%%%%%%%%%%%%%%%%%%%%%%%%%%%%%%%%%%%%%%%%%%%%%%%%%%%%%%%%%%%%%%%%%
\begin{eqnarray}
    &{\rm J}=\epsilon_{ij}x_iP_j+\mathbf{N}_{\sc 1}+\mathbf{N}_{\sc 2}+
    \hat{\emph{\ss}\;}.&\label{betahat}
\end{eqnarray}
The operator $\mathcal{C}_1$ in \eqref{Casvl} is still zero,
while $\mathcal{C}_2$ becomes $\mathcal{C}_2=m\hat{\emph{\ss}\;}$,
\textit{cf.} Eq. \eqref{Casvl}. Their eigenvalues are,
\begin{equation}\barray{llcllc}
    \mathcal{C}_2 \Phi^{++} &=& \chi \Phi^{++}\ ,\;
    &\mathcal{C}_2 \Phi^{+-} &=&
    (\chi+1/2) \Phi^{+-}\ ,
    \\[8pt]
    \mathcal{C}_2 \Phi^{-+} &=& (\chi+1/2) \Phi^{-+}, \quad
    &\mathcal{C}_2 \Phi^{--} &=& (\chi+1) \Phi^{--}\ .
     \earray%\label{}
\end{equation}
Each field $\Phi$ carries therefore a representation of the Galilei algebra.

The supercharges associated to this supermultiplet are the
nonrelativistic limits of the superchages \eqref{Q_1N=2} and
\eqref{Q_2N=2}, namely,
\begin{equation}
    \mathcal{Q}_{\sc 1+}=2i\,\sqrt{\frac{m}{1+\nu_{\sc 1}}}\,a^+_{\sc 1}\Pi_{\sc
    1+}\,,\qquad
    \mathcal{Q}_{\sc 1-}=-2i\,\sqrt{\frac{m}{1+\nu_{\sc 1}}}\left(a^-_{\sc 1}
    +\frac{1}{2m}P_-a^+_{\sc 1}\right)\Pi_{\sc 1-}
    \,,
\label{2q'+-}
\end{equation}
\begin{equation}
    \mathcal{Q}_{\sc 2+}=2i\,\sqrt{\frac{m}{1+\nu_{\sc 2}}}\,R_{\sc 1}a^+_{\sc 2}
    \Pi_{\sc 2+} \,,\qquad
    \mathcal{Q}_{\sc 2-}=-2i\,\sqrt{\frac{m}{1+\nu_{\sc 2}}}\,R_{\sc 1}\left(a^-_{\sc 2}
    +\frac{1}{2m}P_-a^+_{\sc 2}\right)\Pi_{\sc 2-} \,,
    \label{2q'+-2}
\end{equation}
cf. \eqref{q+-}. The algebra generated by the supercharges
\eqref{2q'+-},  (\ref{2q'+-2}) augmented with the Galilei
generators is analogous to \eqref{JQQ1}, \eqref{JQQ2}, with the
additional relations
\begin{equation}
    \{\mathcal{Q}_{\sc A +},\, \mathcal{Q}_{\sc B -}\} =
    \delta_{\sc{A} \sc{B}}\,4m,\qquad
    \{\mathcal{Q}_{\sc A \pm},\, \mathcal{Q}_{\sc B \pm}\} = 0.
    \label{QQAB}
\end{equation}
The supercharges with different indices anticommute, due to the
 presence of the reflection operator, $R_{\sc 1}$.

%%%%%%%%%%%%%%%%%%%%%%%%%%%%%%%%%%%%%%%%%%%%%%%%%%%%%%
\subsection{Exotic nonrelativistic extended supersymmetry}
%%%%%%%%%%%%%%%%%%%%%%%%%%%%%%%%%%%%%%%%%%%%%%%%%%%%

The exotic limit \eqref{exolim} is generalized to
\begin{equation}
    \label{kappa3}
     c\rightarrow \infty\,,\qquad
      \nu \rightarrow \infty\,,\qquad
       \frac{\hat{\emph{\ss}\;}}{c^2}=\frac{\chi}{c^2}=\kappa=const\,,
\end{equation}
cf. \eqref{kappa2}. Since $\chi$ depends on both of the deformation  parameters
(see \eqref{betahat}), the exotic limit can be carried
out varying only one, or both, parameters simultaneously.

%%%%%%%%%%%%%%%%%%%%%%%%%%%%%%%%%%%%
%\subsubsection
\kikezd{Simple exotic limit}
%%%%%%%%%%%%%%%%%%%%%%%%%%%%%%%%%%%%

Here we take the limit
\begin{equation}
    \label{exolim2++}
     c\rightarrow \infty\,,\qquad
      \nu_1 \rightarrow \infty\,,\qquad
      \nu_2 =const\,,\qquad
       \frac{\nu_1/4}{c^2}=\kappa=const\,.
\end{equation}
The rotation generator is obtained by removing the divergent
$c$-number term $\nu_1/4$ from $\hat{\emph{\ss}\;}$   in \eqref{betahat},
\begin{eqnarray}
    {\rm J}=\epsilon_{ij}x_iP_j+\mathbf{N}_{\sc
    1}+\mathbf{N}_{\sc 2}+ \alpha_{\sc 2
    +}+\frac{1}{4}+\frac{1}{2}\left(\Pi_{\sc 1 -} + \Pi_{\sc 2
    -}\right).
\end{eqnarray}
Galilei boosts are obtained from the non-relativistic limit of
Sec.  \ref{NRlimit},
\begin{equation}\label{K2Jv}
    \mathcal{K}_+= mx_+-tP_++i\left(\mathcal{J}_{\sc 2+}-\kappa v_+\right)\,,
    \qquad
    \mathcal{K}_-= mx_--tP_-+i\kappa v_-\,,
\end{equation}\label{K2}
\begin{equation}
    v_\pm =v_1\pm iv_2=-\sqrt{\frac{2}{\kappa}}b^\pm\,,
 \qquad
    b^\pm=\lim_{\nu_{\sc 1}\rightarrow \infty}\sqrt{\frac{2}{\nu_{\sc 1}}}\,\mathcal{J}_{\sc 1
    \pm}\,.
\end{equation}
Observe that $v_j$ and $b^\pm$ only involve the oscillator with
label ${\un 1}$.

 Galilei boosts satisfy the exotic
 commutation relation \eqref{EG1}. The $b^\pm$ and
 $\mathbf{N}_{\sc 1}$ operators
  satisfy relations like in \eqref{HANb}, \textit{i.e.}
  $\mathbf{N}_{\sc 1}=b^+b^-$.
 Due to the asymmetry of the limit \eqref{exolim2++},
  the supercharges are of different types,
\begin{eqnarray}
    & \mathcal{Q}_{\sc 1 \pm}={\rm lim}_{c,\nu\rightarrow \infty}\,
    \frac{1}{\sqrt{c}}Q_{\sc 1 \pm}\,,\qquad
    \mathcal{Q}_{\sc 2 \pm}={\rm lim}_{c\rightarrow
      \infty} \frac{1}{c}\check{Q}_{\sc 2
      \pm}\,,&\label{Q12mixture}\\[6pt]
    & \mathcal{Q}_{\sc 1 \pm}= \pm 2i \sqrt{m} \left(
     {\rm lim}_{\nu_{\sc 1} \rightarrow \infty}\, \Pi_{\sc 1 \mp}
     \frac{a_{\sc 1}^\pm}{\sqrt{\nu_{\sc 1}}}  \right)=\pm 2i
     \sqrt{m} \,c_{\sc 1}^\pm ,&\label{Q1E}\\[6pt]
     & \mathcal{Q}_{\sc 2+}=2i\,\sqrt{\frac{m}{1+\nu_{\sc 2}}}\,R_{\sc 1}a^+_{\sc 2}
    \Pi_{\sc 2+}\,,\quad
     \mathcal{Q}_{\sc 2-}=-2i\,\sqrt{\frac{m}{1+\nu_{\sc 2}}}\,R_{\sc
    1}\left(a^-_{\sc 2}
    +\frac{1}{2m}P_-a^+_{\sc 2}\right)\Pi_{\sc 2-}
    \,,  &\label{Q1U}
\end{eqnarray}
which are those obtained as in Eqns. \eqref{q+-E} (``exotic
supercharges'')
 and \eqref{q+-} (``usual supercharges'') respectively.
These supercharges yield commutation relation as in \eqref{QQAB}.

%%%%%%%%%%%%%%%%%%%%%%%%%%%%%%%%%%%
%\subsubsection
\kikezd{Double exotic limit}
%%%%%%%%%%%%%%%%%%%%%%%%%%%%%%%%%%%

Here we take the limit,
\begin{equation}
    \label{exolim2+++}
     c\rightarrow \infty\,,\qquad
      \nu_1 \rightarrow \infty\,,\qquad
      \nu_2 \rightarrow \infty,\qquad
       \frac{\nu_1/2}{c^2}=\frac{\nu_2/2}{c^2}=\kappa=const\,.
\end{equation}
The finite part of angular momentum is now,
\begin{eqnarray}
    {\rm J}=\epsilon_{ij}x_iP_j+\mathbf{N}_{\sc 1}+
    \mathbf{N}_{\sc 2}+ \frac{1}{2}+\frac{1}{2}
    \left(\Pi_{\sc 1 -} + \Pi_{\sc 2 -}\right).
\end{eqnarray}
The Galilei boosts are,
\begin{equation}\label{K2bis}
     \mathcal{K}_i= mx_i-tP_i+\frac{\kappa}2
     \epsilon_{ij}v_{\sc 1 j}+
     \frac{\kappa}2 \epsilon_{ij}v_{\sc 2 j}\,
\end{equation}
where the $v$ operators are defined analogously to \eqref{K2}, and the
labels $\un A$ indicate the dependence on the corresponding oscillator.
In particular, here we have $b_{\sc A}^\pm$ and $\mathbf{N}_{\sc A}=
b_{\sc A}^+ b_{\sc A}^-$, as in \eqref{HANb}.
Note that this double exotic limit
is symmetric in both oscillators. Hence, we obtain,
\begin{eqnarray}
    \mathcal{Q}_{\sc 1 \pm}&=&{\rm lim}_{c,\nu_{\sc 1}\rightarrow \infty}\,
    \frac{1}{\sqrt{c}}Q_{\sc 1 \pm}=
     \pm 2i \sqrt{m} \,c_{\sc 1}^\pm ,
     \label{Q'1E}\\[6pt]
     \mathcal{Q}_{\sc 2 \pm}&=&{\rm lim}_{c,\nu_{\sc 2}
     \rightarrow \infty} \frac{1} {\sqrt{c}}Q_{\sc 2 \pm}=
     \pm 2i \sqrt{m} R_{\sc 1}\,c_{\sc 2}^\pm\,,  \label{Q'2E}
\end{eqnarray}
which are those in Eqns. \eqref{q+-E} and \eqref{q+-} respectively.

%%%%%%%%%%%%%%%%%%%%%%%%%%%%%%%%%%%%%%%%%%%%%%%%%%%%%%%
%%%%%%%%%%%%%%%%%%%%%%%%%%%%%%%%%%%%%%%%%%%%%
\section{Discussion and outlook}
%%%%%%%%%%%%%%%%%%%%%%%%%%%%%%%%%%%%%%%%%%%%%
%%%%%%%%%%%%%%%%%%%%%%%%%%%%%%%%%%%%%%%%%%%%%%%%%%%%%%%%

In conclusion, we review our main results and hint at some
open problems which could be worth of
further study.

Our universal covariant vector %set of first order wave
equations describe massive spinning particles in $2+1$
dimensions. The Poincar\'e spin, a pseudoscalar, can take
arbitrary real values. The equations fix themselves the
type of the  representation of the (2+1)D Lorentz
algebra\,: only  $\mathfrak{so}(2,1)$ representations
bounded from below or from above (or from both sides) are
allowed. Depending on the spin parameter, $\alpha$, our
equations describe three classes of particles, namely
\begin{enumerate}
 \item[a)] Bosons/Fermions : $\alpha = -j$,\quad
 $2j=1,2,\ldots$,
 \item[b)] ``Unitary" anyons  : $\alpha >0$,
 \item[c)] ``Non-unitary" anyons : $-j < \alpha < -j+1/2.$
\end{enumerate}

Case a), based on finite-dimensional non-unitary
representations of the $\mathfrak{so}(2,1)$, reproduces,
for $\alpha=-1/2,-1,-3/2$ the Dirac,
Deser-Jackiw-Templeton, and Rarita-Schwinger theories,
 respectively. For $\alpha=-2$, our theory
provides us with the linearized equations of massive
gravity. More generally,  complete correspondence with
the $2+1$ dimensional Dirac-Fierz-Pauli equations
\cite{DFP} is expected.

Then, we investigated two types of anyons. Both involve
infinite-dimensional half-bounded
 representations of the (2+1)D Lorentz algebra.
The first type, which corresponds to \emph{unitary}
representations [case b) above], has been widely used in
the anyon context since the beginning \cite{JN1,
Pl1991,Ptor}.
 The second type [case c)] is associated with
 \emph{non-unitary} representations. No
 attention was paid to them in earlier investigations on
 anyons.
  We argue, however, that precisely these new- [and not the conventional]
 types
 of fractional spin particles
 do correspond to the intuitive picture, in
 which anyons interpolate between bosons and fermions.

The essential difference is that when the
parameter $\alpha$ %which labels the representation
 tends to
a (half) integer $-j$, the higher field components
$\psi_n(x)$ with index $n>2j$ are, for the new type,
suppressed by the
 universal numerical factor $(\alpha+j)^{1/2}$. Note that no
such a suppression happens in the conventional, unitary
case. As a result, in the limiting case $\alpha=-j$, such
an anyonic field reduces to a usual $(2j+1)$-component
field of spin modulus $|s|=j$. The spin zero case then also
can be incorporated into the theory letting
$\alpha\rightarrow 0$ for non-unitary anyons.

The physical difference between the two types of anyons
could, perhaps, be revealed  when interactions are switched
on.

The extended formulation of the theory, presented in Section
\ref{solBFA}, generalizes the Jackiw-Nair \cite{JN1} and the
Majorana-Dirac \cite{Pl1991} descriptions of anyons. Our framework
allows us to combine several irreducible representations of the
Lorentz algebra, ``entangled'' by Eqn. (\ref{JJpsi}), whereas
spins behave additively. Usual bosons and fermions can be
obtained, in particular, as \emph{entangled} anyons. The
topologically massive vector gauge field of
 Jackiw-Deser-Templeton \cite{DJT} can be, for example, viewed as a pair of entangled
Dirac particles.

It would be interesting to apply  such a picture  to quantum
computing \cite{anyQC}.

%We constructed a supersymmetric generalization
Promoting the parameter $\alpha$  to a diagonal operator
$\hat{\alpha}=diag\,(\alpha,\alpha+1/2)$ and taking the
direct sum of two irreducible representations of
$\mathfrak{so}(2,1)$ %marked by the diagonal elements of $\hat{\alpha}$
 provided us with an $N=1$ supersymmetric
system. It is described, \emph{on-shell}, by
 the usual  super-Poincar\'e algebra.

The natural question is, then\,: what kind of symmetry will
appear if the elements of $\hat{\alpha}$ are shifted by
$n/2$ with $n$ integer greater than $1$~? One can expect
that such a theory will be characterized by a kind of
generalized supersymmetry, with spin-tensorial supercharges
\cite{exoSUSY,NirP,CasGom}, which, on-shell, would generate
a \emph{nonlinear} generalization of the
 usual $N=1$ super-Poincar\'e algebra
\cite{P_para_susy,nonSUSY}.

In the simplest form of the extended realization two Lorentz spin
vector generators  are added; this also allows us to generalize
the $N=1$ supersymmetry  to  $N=2$, see Section \ref{N2susy} for
details.

The extended formulation can further be generalized by
adding an arbitrary number of the vector spin generators,
${\cal J}_\mu=\sum_i J^{(i)}_\mu$, with each $J^{(i)}_\mu$
belonging  to one of the representations ${\cal
D}^+_{\alpha_{(i)}}$ (or ${\cal D}^-_{\alpha_{(i)}}$) from
(\ref{tilD}). Choosing $\emph{\ss}=\sum_i\alpha_{(i)}$, one
finds then  that the basic vector set of equations splits
into a set of vector equations,  while the irreducibility
equation (\ref{Cas+aux3}) produces an entangling equation
which generalizes (\ref{JJpsi}),
\begin{equation}\label{extendedN}
    V^{(\alpha_{(i)})}_\mu\psi=0\,,\qquad
    \sum_{i>j}(J^{(i)}J^{(j)}+
    \alpha_{(i)}\alpha_{(j)})\psi=0\,.
\end{equation}
Each $J^{(i)}_\mu$  can be promoted to an even generator of
the $\mathfrak{osp}(1|2)$ superalgebra, realized in terms
of an independent reflection-deformed oscillator with
appropriately chosen parameter $\nu_{(i)}$. In this way,
extended supermultiplets of anyons, or of bosons and
fermions could be obtained.

Considering two types of non-relativistic limits to the
relativistic theory, we observed how either the usual
centrally-mass-extended, or the doubly centrally  extended
``exotic'' Galilei symmetries appear.
 Their Schr\"odinger-type conformal extensions for
 arbitrary (including anyonic) spin,  and also for
 supersymmetric generalizations could also be studied.

It would also be interesting to derive our basic vector
equations from an action principle. A nontrivial point is
that the three equations of our system are dependent: any
two of them generate the third one as consistency
(integrability) equation. This indicates that the
corresponding action should have a kind of gauge symmetry,
analogous to the Chern-Simons formulation of topologically
massive gauge fields \cite{DJT}. An additional indication
in this direction is the appearance of an
infinite-dimensional null subspace in the extended
formulation. Analogous null subspaces also appear, in fact,
in covariant quantization schemes of relativistic strings,
and are associated with gauge (diffeomorphism) symmetries.

In our unified scheme, all irreducible massive
representations of the  2+1 D Poincar\'e group have been
obtained. Does this theory have an analog in higher
dimensions? This question directs us towards  theories with
 arbitrary spin fields. For instance, to strings, or to higher
 spin theories \cite{Hspin}.

\vskip 0.4cm\noindent {\bf Acknowledgements}. MP is
indebted to the {\it Laboratoire de Math\'ematiques et de
Physique Th\'eorique} of Tours University, and PAH is
indebted to the {\it Departamento de F\'{\i}sica,
Universidad de Santiago de Chile}, respectively, for
hospitality. Partial support by the FONDECYT (Chile) under
the grant 1095027 and by DICYT (USACH) is acknowledged. MV
was supported by CNRS postdoctoral grant (contract number
87366).

%%%%%%%%%%%%%%%%%%%%%%%%%%%%%%%%%%%%%%%%%%%%%%%%%%%%%%%%%%%%%%%%%%%%%%%
%%%%%%%%%%%%%%%%%%%%%%%%%%%%%%%%%%%%%%%%%%%%%%%%%%%%%%%%%%%%%%%%%%%%%%%%%%%%%%%%%

%%%%%%%%%%%%%%%%%%%%%%%%%%%%%%%%%%%%%%%%%%%%%%%%%%%%%%%%%%%%%%%%%%%%%%%%%%%%%%
\appendix
%%%%%%%%%%%%%%%%%%%%%%%%%%%%%%%%%%%%%%%%%%%%%%%%%%%%%%%%%%%%%%%%%%%%%%%%%%%%%%
\setcounter{equation}{0}
\renewcommand{\theequation}{A.\arabic{equation}}

\section{Finite-dimensional representations of $\mathfrak{osp}(1|2)$}\label{app}

%%%%%%%%%%%%%%%%%%%%%%%%%%%%%%%%%%%%%%%%%%%%%%%%%%%%%%%%%%%
\subsection{$\mathfrak{osp}(1|2)$-supermultiplet
($\alpha_+=-1/2$, $\alpha_-=0$)}

Here $\nu=-3$ and $r=1$. The finite-Fock space, $\mathcal{
F}=\mathcal{ F}_+\oplus\mathcal{ F}_-,$ is
\begin{eqnarray}
    \mathcal{ F}_+= \left\{ |0\rangle_+= \left(
    \begin{array}{c}
    1  \\
    0  \\
    0 %
    \end{array}
    \right),\; |1\rangle_+=\left(
    \begin{array}{c}
    0  \\
    0  \\
    1 %
    \end{array}
    \right) \right\}\,,\quad \mathcal{ F}_-=\left\{ |0\rangle_-=\left(
    \begin{array}{c}
    0  \\
    1  \\
    0 %
    \end{array}
    \right) \right\}.
\end{eqnarray}
The reflection operator and projectors take the form,
\begin{equation}
    R=diag\, (1,-1,1)\,,\quad
     \Pi_+=diag\, (1,0,1)\,,\quad \Pi_-=diag\,(0,1,0)\,.
\end{equation}
The matrix form of the $\mathfrak{osp}(1|2)$ algebra is obtained
from (\ref{CnJ+-})--(\ref{aaevod-}), (\ref{FFpm}) and (\ref{J^2}),
\begin{equation}a^{+}=\left(
\begin{array}{ccc}
0 & 0 & 0 \\
i\sqrt{2} & 0 & 0 \\
0 & \sqrt{2} & 0%
\end{array}%
\right) ,\qquad a^{-}=\left(
\begin{array}{ccc}
0 & i\sqrt{2} & 0 \\
0 & 0 & \sqrt{2} \\
0 & 0 & 0%
\end{array}%
\right),
\end{equation}
\begin{equation}\label{finiteJ}
{\cal J}_{0}=\left(
\begin{array}{ccc}
-1/2 & 0 & 0 \\
0 & 0 & 0 \\
0 & 0 & 1/2%
\end{array}%
\right),\qquad {\cal J}_+=\left(
\begin{array}{ccc}
0 & 0 & 0 \\
0 & 0 & 0 \\
i & 0 & 0%
\end{array}%
\right),\qquad {\cal J}_-=\left(
\begin{array}{ccc}
0 & 0 & i \\
0 & 0 & 0 \\
0 & 0 & 0%
\end{array}%
\right).
\end{equation}
The  representation (\ref{finiteJ}) of the Lorentz algebra is
reducible. The spin $-\alpha_+=j=1/2$ part is projected from
\eqref{finiteJ} to
\begin{equation}\label{Cliff:DHA}
     J^{(+)}_0=-\frac{1}{2}\sigma_3\,,\qquad
    J^{(+)}_1=\frac{i}{2}\sigma_1,\qquad J^{(+)}_2
    =-\frac{i}{2}\sigma_2\,, \qquad J^{(+)}_\mu
    J^{(+)}\,^\mu=-\frac{3}{4}\,,
\end{equation}
where $\sigma_a$, $a=1,2,3,$ are the Pauli matrices. These
operators act on the vector space $\mathcal{F}_+$ formed by
$$
|0\rangle_+=\left(
\begin{array}{c}
1  \\
0  \\
\end{array}
\right)_+ \,,\qquad |1\rangle_+= \left(
\begin{array}{c}
0  \\
1  \\
\end{array}
\right)_+.
$$
$-2J^{(+)}_\mu=\gamma'_\mu$ are Dirac matrices that satisfy
the Clifford algebra \eqref{Cliff}. They are related to the
Majorana representation (\ref{Dmatrices}) by the unitary
transformation,
\begin{equation}\label{gprimeg}
    \gamma'_\mu=U\gamma_\mu U^\dagger\, ,\qquad
    U=\frac{1}{\sqrt{2}}(\sigma_2+\sigma_3).
\end{equation}

The Lorentz algebra \eqref{finiteJ} projected to the spin
$\alpha_-=0$ subspace is trivial, $J^{(-)}_\mu={\cal
J}_\mu\Pi_-=0$. It leaves invariant the scalar,
one-dimensional subspace $\mathcal{ F}_-= \{|0\rangle_-\}$.

\subsection{$\mathfrak{osp}(1|2)$-supermultiplet
($\alpha_+=-1$, $\alpha_-=-1/2$)}\label{DDJTsusy}

Here $\nu=-5$ and $r=2$. The finite-Fock space is
$\mathcal{ F}=\mathcal{ F}_+\oplus\mathcal{ F}_-$,
\begin{equation}
\label{Fock5}\\[6pt]
\mathcal{ F}_+= \left\{ |0\rangle_+= \left(
\begin{array}{c}
1  \\
0  \\
0  \\
0  \\
0 %
\end{array}
\right),\, |1\rangle_+=\left(
\begin{array}{c}
0  \\
0  \\
1 \\
0  \\
0  %
\end{array}
\right),\, |2\rangle_+=\left(
\begin{array}{c}
0  \\
0  \\
0  \\
0  \\
1 %
\end{array}
\right)
 \right\}\,,
\end{equation}
\begin{equation}
\mathcal{ F}_-=\left\{ |0\rangle_-=\left(
\begin{array}{c}
0  \\
1  \\
0  \\
0  \\
0 %
\end{array}
\right),\, |1\rangle_-=\left(
\begin{array}{c}
0  \\
0  \\
0  \\
1  \\
0 %
\end{array}
\right)
 \right\}.\label{Fock5a}
\end{equation}
The reflection operator and projectors take the form
\begin{equation}\label{RPI}
    R=diag\,(1,-1,1,-1,1)\,,\quad
    \Pi_+=diag\, (1,0,1,0,1)\,,\quad
    \Pi_-=diag\, (0,1,0,1,0)\,.
\end{equation}
The  $\mathfrak{osp}(1|2)$ matrix algebra is generated by
\begin{equation}a^{+}=\left(
\begin{array}{ccccc}
0 & 0 & 0 & 0 & 0 \\
2i & 0 & 0 & 0 & 0 \\
0 & \sqrt{2} & 0 & 0 & 0 \\
0 & 0 & i\sqrt{2}& 0 & 0 \\
0 & 0 & 0 & 2 & 0 \\
\end{array}%
\right) ,\qquad a^{-}=\left(
\begin{array}{ccccc}
0 & 2i & 0 & 0 & 0 \\
0 & 0 & \sqrt{2} & 0 & 0 \\
0 & 0 & 0 & i\sqrt{2} & 0 \\
0 & 0 & 0 & 0 & 2 \\
0 & 0 & 0 & 0 & 0 \\
\end{array}%
\right),\label{5a+a-}
\end{equation}
\begin{equation}\label{finitecase20}
    {\cal J}_{0}=diag\, (-1,-1/2,0,1/2,1)\,,
\end{equation}
\begin{equation}
{\cal J}_+=\left(
\begin{array}{ccccc}
0 & 0 & 0 & 0 & 0 \\
0 & 0 & 0 & 0 & 0 \\
i\sqrt{2} & 0 & 0 & 0 & 0 \\
0 & i & 0 & 0 & 0 \\
0 & 0 & i\sqrt{2} & 0 & 0 \\
\end{array}\right)\,,\quad
{\cal J}_-=\left(
\begin{array}{ccccc}
0 & 0 & i\sqrt{2} & 0 & 0 \\
0 & 0 & 0 & i & 0 \\
0 & 0 & 0 & 0 & i\sqrt{2} \\
0 & 0 & 0 & 0 & 0 \\
0 & 0 & 0 & 0 & 0 \\
\end{array}%
\right).\label{finitecase2}
\end{equation}
Projection to the odd subspace of spin $-\alpha_-=j=1/2$ produces
a representation with the same matrix elements as in
\eqref{Cliff:DHA} but now acting on $\mathcal{ F}_-$,
\begin{equation}
|0\rangle_-=\left(
\begin{array}{c}
1  \\
0  \\
\end{array}
\right)_- ,\qquad |1\rangle_-= \left(
\begin{array}{c}
0  \\
1  \\
\end{array}
\right)_-\,.\label{F-}
\end{equation}
The spin $-\alpha_+=j=1$ part is projected to
\begin{equation}\label{finiteJ2}
J^{(+)}_0=\left(
\begin{array}{ccc}
-1 & 0 & 0 \\
0 & 0 & 0 \\
0 & 0 & 1%
\end{array}%
\right), \quad J^{(+)}_1=\frac{1}{\sqrt{2}}\left(
\begin{array}{ccc}
0 & i & 0 \\
i & 0 & i \\
0 & i & 0%
\end{array}%
\right), \quad J^{(+)}_2=\frac{1}{\sqrt{2}}\left(
\begin{array}{ccc}
0 & -1 & 0 \\
1 & 0 & -1 \\
0 & 1 & 0%
\end{array}%
\right),
\end{equation}
which act on the vector space $\mathcal{F}_+$,
\begin{equation}
|0\rangle_+=\left(
\begin{array}{c}
1  \\
0  \\
0
\end{array}
\right)_+ ,\quad |1\rangle_+= \left(
\begin{array}{c}
0  \\
1  \\
0
\end{array}
\right)_+ ,\quad |2\rangle_+= \left(
\begin{array}{c}
0  \\
0  \\
1
\end{array}
\right)_+.\label{F+}
\end{equation}
(\ref{finiteJ2}) is related to the adjoint representation
of $\mathfrak{so}(2,1)$,
\begin{equation}\label{s=1usual}
    (J_\mu)^\nu{}_\lambda=-i\epsilon^{\nu}{}_
    {\mu\lambda}, \qquad \mu,\nu,\lambda=0,1,2,\qquad
    \epsilon^{012}=1\,,
\end{equation}
by means of the unitary transformation
\begin{equation}\label{U1}
     J^{(+)}_\mu=U J_\mu U^\dagger,
     \qquad U=\frac{1}{\sqrt{2}}\left(
    \begin{array}{ccc}
    0 & 1 & i \\
    -i\sqrt{2} & 0 & 0 \\
    0 & -1 & i%
    \end{array}%
    \right).
\end{equation}
The explicit matrix form of (\ref{s=1usual}) is
\begin{equation}\label{usualspin1matrix}
(J_0)^\nu{}_\lambda=\left(
\begin{array}{ccc}
0 & 0 & 0 \\
0 & 0 & -i \\
0 & i & 0%
\end{array}%
\right), \quad (J_1)^\nu{}_\lambda =\left(
\begin{array}{ccc}
0 & 0 & -i \\
0 & 0 &  0\\
-i & 0 & 0%
\end{array}%
\right), \quad (J_2)^\nu{}_\lambda = \left(
\begin{array}{ccc}
0 & i & 0 \\
i & 0 & 0 \\
0 & 0 & 0%
\end{array}%
\right).
\end{equation}

%%%%%%%%%%%%%%%%%%%%

\end{document}